\documentstyle[aps,12pt,epsfig]{revtex}

\renewcommand{\arraystretch}{1.3}

\def\be{\begin{equation}}
\def\ee{\end{equation}}
\def\disp{\displaystyle}
\def\foot{\footnotesize}

\newtheorem{corollary}{Corollary}

\def\R{\rm {\sf R\hspace{-3.05mm}I\hspace{2.1mm}}}
\def\Z{\rm {\sf Z\hspace{-3.15mm}Z\hspace{0.6mm}}}

\newcounter{fig}

\begin{document}

\title{Statistics of knots and entangled random walks}
\bigskip

\author{Sergei Nechaev$^{1,2}$} 

\address{\rm e-mail: nechaev@ipno.in2p3.fr}
\bigskip

\address{$^{1}$ UMR 8626, CNRS-Universit\'e Paris XI, LPTMS, Bat.100,
Universit\'e Paris Sud, \\ 91405 Orsay Cedex, France}
\address{$^{2}$ L D Landau Institute for Theoretical Physics, 117940, 
Moscow, Russia}
\maketitle
\bigskip

\begin{center}
Extended version of lectures presented at \\ 
LES HOUCHES 1998 SUMMER SCHOOL (SESSION LXIX) \\ 
{\sl Topological Aspects of Low Dimensional Systems}, \\ 
July 7 - 31, 1998 
\end{center}

\bigskip

\begin{abstract}
The lectures review the state of affairs in modern branch of mathematical
physics called probabilistic topology. In particular we consider the
following problems: (i) We estimate the probability of a trivial knot
formation on the lattice using the Kauffman algebraic invariants and show
the connection of this problem with the thermodynamic properties of 2D
disordered Potts model; (ii) We investigate the limit behavior of random
walks in multi-connected spaces and on non-commutative groups related to the
knot theory. We discuss the application of the above mentioned problems in
statistical physics of polymer chains. On the basis of non-commutative
probability theory we derive some new results in statistical physics of
entangled polymer chains which unite rigorous mathematical facts with more
intuitive physical arguments.
\end{abstract}
\bigskip

\tableofcontents


\section{Introduction}
\setcounter{equation}{0}

It wouldn't be an exaggeration to say that contemporary physical science is
becoming more and more mathematical. This fact is too strongly manifested
to be completely ignored. Hence I would permit myself to bring forward two
possible conjectures:

(a) On the one hand there are hardly discovered any newly physical problem
which would be beyond the well established methods of the modern
theoretical physics. This leads to the fact that nowadays real physical
problems seem to be less numerous than mathematical methods of their
investigation.

(b) On the other hand the mathematical physics is a fascinating field
which absorbs new ideas from different branches of modern mathematics,
translates them into the physical language and hence fills the abstract
mathematical constructions by the new fresh content. This ultimately leads
to creating new concepts and stimulates seeking for newel deep conformities
to natural laws in known physical phenomena.

The penetration of new mathematical ideas in physics has sometimes rather
paradoxical character. It is not a secret that difference in means (in
languages) and goals of physicists and mathematicians leads to mutual
misunderstanding, making the very subject of investigation obscure. What is
true for general is certainly true for particular. To clarify the point,
let us turn to statistics of entangled uncrossible random walks---the
well-known subject of statistical physics of polymers. Actually, since
1970s, after Conway's works, when the first algebraic topological
invariants---Alexander polynomials---became very popular in mathematical
literature, physicists working in statistical topology have acquired a much
more powerful topological invariant than the simple Gauss linking number.
The constructive utilization of algebraic invariants in statistical physics
of macromolecules has been developed in the classical works of A.
Vologodskii and M. Frank-Kamenetskii \cite{volog}. However until recently
in overwhelming majority of works the authors continue using the
commutative Gauss invariant invariant just making references to its
imperfectness.

One of the reasons of such inertia consists in the fact that new
mathematical ideas are often formulated as ``theorems of existence" and it
takes much time to retranslate them into physically acceptable form which
may serve as a real computational tool.

We intend to use some recent advances in algebraic topology and theory of
random walks on non-commutative groups for reconsidering the old
problem---evaluating of the entropy of randomly generated knots and
entangled random walks in a given homotopic state. Let us emphasize that
this is a real physical paper, and when it is possible the rigorous
statements are replaced by some physically justified conjectures. Generally
speaking, the work is devoted to an analysis of probabilistic problems in
topology and their applications in statistical physics of polymer systems
with topological constraints.

Let us formulate briefly the main results of our work.

1. The probability for a long random walk to form randomly a knot with
specific topological invariant is computed. This problem is considered
using the Kauffman algebraic invariants and the connection with the
thermodynamic properties of 2D Potts model with ``quenched" and ``annealed"
disorder in interaction constants is discussed.

2. The limit behavior of random walks on the non-commutative groups related
to the knot theory is investigated. Namely, the connection between the
limit distribution for the Lyapunov exponent of products of non-commutative
random matrices---generators of ``braid group"---and the asymptotic of
powers (``knot complexity") of algebraic knot invariants is established.
This relation is applied for calculating the knot entropy. In
particular, it is shown that the ``knot complexity" corresponds to the well
known topological invariant, ``primitive path", repeatedly used is
statistics of entangled polymer chains.

3. The random walks on multi-connected manifolds is investigated using
conformal methods and the nonabelian topological invariants are
constructed. It is shown that many nontrivial properties of limit behavior
of random walks with topological constraints can be explained in context of
random walks on hyperbolic groups.

Usage of the limit behavior of entangled random paths established above
for investigation of the statistical properties of so-called ``crumpled
globule" (trivial ring without self-intersections in strongly contracted
state).

The connection between all these problems is shown in Table 1.

\begin{figure}
\centerline{\epsfig{file=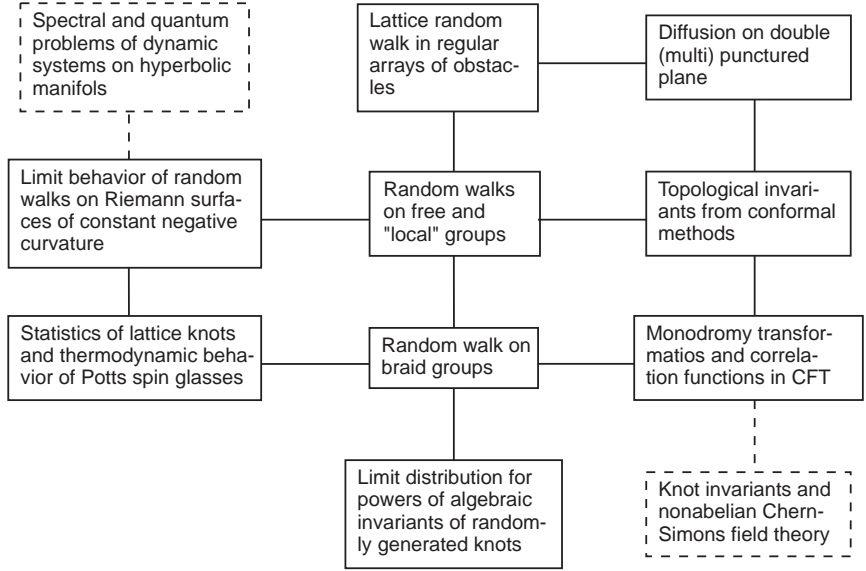,width=12cm}}
\caption{Links between topologically-probabilistic problems. Solid boxes
-- problems, discussed in the paper; dashed boxes -- problems not included
in the consideration.}
\label{tab:1}
\end{figure}

\section{Knot diagrams as disordered spin systems}
\subsection{Brief review of statistical problems in topology}
\setcounter{equation}{0}

The interdependence of such branches of modern theoretical and mathematical
physics as theory of integrable systems, algebraic topology and conformal
field theory has proved to be a powerful catalyst of development of the new
direction in topology, namely, of analytical topological invariants
construction by means of exactly solvable statistical models.

Today it is widely believed that the following three cornerstone findings
have brought the fresh stream in topology:

--- It has been found the deep relation between the Temperley-Lieb
algebra and the Hecke algebra representation of the braid group. This fact
resulted in the remarkable geometrical analogy between the Yang-Baxter
equations, appearing as necessary condition of the transfer matrix
commutativity in the theory of integrable systems on the one hand, and one
of Reidemeister moves, used in the knot invariant construction on the other
hand.

--- It has been discovered that the partition function of the Wilson loop
with the Chern-Simons action in the topological field theory coincides with
the representation of the known nonabelian algebraic knot invariants
written in terms of the time-ordered path integral.

--- The need for new solutions of the Yang-Baxter equations has given a
power impetus to the theory of quantum groups. Later on the related set of
problems was separated in the independent branch of mathematical physics.

Of course the above mentioned findings do not exhaust the list of all
brilliant achievements in that field during the last decade, but apparently
these new accomplishments have used profound ``ideological" changes in the
topological science: now we can hardly consider topology as an independent
branch of pure mathematics where each small step forward takes so much
effort that it seems incidental.

Thus in the middle of the 80s the ``quantum group" gin was released. It
linked by common mathematical formalism classical problems in topology,
statistical physics and field theory. A new look at the old problems and
the beauty of the formulated ideas made an impression on physicists and
mathematicians. As a result, in a few last years the number of works
devoted to the search of the new applications of the quantum group
apparatus is growing exponentially going beyond the framework of original
domains. As an example of persistent penetrating of the quantum group ideas
in physics we can name the works on anyon superconductivity \cite{wilczeck},
intensively discussing problems on ``quantum random walks" \cite{majid},
the investigation of spectral properties of ``quantum deformations" of
harmonic oscillators \cite{oscill} and so on.

The time will show whether such ``quantum group expansion" is physically
justified or it merely does tribute to today's fashion. However it is
clear that physics has acquired new convenient language allowing to
construct new ``nonabelian objects" and to work with them.

Among the vast amount of works devoted to different aspects of the theory
of integrable systems, their topological applications connected to the
construction of knot and link invariants and their representation in terms
of partition functions of some known 2D-models deserve our special
attention. There exist several reviews \cite{jones} and
books \cite{collect} on that subject and our aim by no means
consists in re-interpretation or compilation of their contents. We make an
attempt of consecutive account of recently solved {\it probabilistic}
problems in topology as well as attract attention to some interesting,
still unsolved, questions lying on the border of topology and the
probability theory. Of course we employ the knowledges acquired in the
algebraic topology utilizing the construction of new topological invariants
done by V.F.R. Jones \cite{jones} and L.H. Kauffman \cite{kauff2}.

Besides the traditional fundamental topological issues concerning the
construction of new topological invariants, investigation of homotopic
classes and fibre bundles we mark a set of adjoint but much less studied
problems. First of all, we mean the problem of so-called ``knot entropy"
calculation. Most generally it can be formulated as follows. Take the
lattice ${\Z}^3$ embedded in the space ${\R}^3$. Let $\Omega_N$ be the
ensemble of all possible closed nonselfintersecting $N$-step loops with one
common fixed point on ${\Z}^3$; by $\omega$ we denote the particular
trajectory configuration. The question is: what is the probability ${\cal
P}_N$ of the fact that the trajectory $\omega\in\Omega_N$ belongs to some
specific homotopic class. Formally this quantity can be represented in the
following way
\be \label{1:1.1}
\begin{array}{c} {\cal P}_N\{\mbox{Inv}\} = \disp
\frac{1}{\Omega_N}\sum_{\{\omega\}} \Delta
\left[\mbox{Inv}\{\omega\}-\mbox{Inv}\right] \equiv \medskip \\ \disp
\frac{1}{\Omega_N}\sum_{\{{\bf r}_1,\ldots,{\bf r}_N\}} \Delta
\left[\mbox{Inv}\{{\bf r}_1,\ldots,{\bf r}_N\}-\mbox{Inv}\right]
\Bigl(1-\Delta\left[{\bf r}_i-{\bf r}_j\right]\Bigr)
\Delta\left[{\bf r}_N\right]
\end{array}
\ee
where $\mbox{Inv}\{\omega\}$ is the functional representation of the knot
invariant corresponding to the trajectory with the bond coordinates $\{{\bf
r}_1\ldots,{\bf r}_N\}$; $\mbox{Inv}$ is the topological invariant
characterizing the knot of specific homotopic type and $\Delta(x)$ is the
Kronecker function: $\Delta(x=0)=1$ and $\Delta(x\neq 0)=0$. The first
$\Delta$-function in Eq.(\ref{1:1.1}) cuts the set of trajectories with the
fixed topological invariant while the second and the third
$\Delta$-functions ensure the $N$-step trajectory to be nonselfintersecting
and to form a closed loop respectively.

The distribution function ${\cal P}_N\{\mbox{Inv}\}$ satisfies the
normalization condition
\be
\sum_{\foot\mbox{all homotopic classes}}
P_N\{\mbox{Inv}\} = 1
\ee

The entropy $S_N\{{\rm Inv}\}$ of the given homotopic state of the knot
represented by $N$-step closed loop on ${\Z}^3$ reads
\be \label{1:entropy}
S_N\{{\rm Inv}\}=\ln\left[\Omega_N {\cal P}_N\{{\rm Inv}\}\right]
\ee

The problem concerning the knot entropy determination has been discussed
time and again by the leading physicists. However the number of new
analytic results in that field was insufficient till the beginning of the
80s: in about 90 percents of published materials their authors used
the Gauss linking number or some of its abelian modifications for
classification of a topological state of knots and links while the
disadvantages of this approach were explained in the rest 10 percent
of the works. We do not include in this list the celebrated investigations
of A.V. Vologodskii {\it et al} \cite{volog} devoted to the first fruitful
usage of the nonabelian Alexander algebraic invariants for the computer
simulations in the statistical biophysics. We discuss physical applications
of these topological problems at length in Section 5.

Despite of the clarity of geometrical image, the topological ideas are very
hard to formalize because of the non-local character of topological
constraints. Besides, the main difficulty in attempts to calculate
analytically the knot entropy is due to the absence of convenient analytic
representation of the complete topological invariant. Thus, to succeed, at
least partially, in the knot entropy computation we simplify the general
problem replacing it by the problem of calculating the distribution
function for the knots {\it with defined topological invariants}. That
problem differs from the original one because none of the known topological
invariants (Gauss linking number, Alexander, Jones, HOMFLY) are complete.
The only exception is Vassiliev invariants \cite{vassiliev}, which are
beyond the scope of the present book. Strictly speaking we are unable to
estimate exactly the correctness of such replacement of the homotopic class
by the mentioned topological invariants.  Thus under the definition of the
topological state of the knot or entanglement we simply understand the
determination of the corresponding topological invariant.

The problems where $\omega$ (see Eq.(\ref{1:1.1})) is the set of
realizations of the random walk, i.e. the Markov chain are of special
interest. In that case the probability to find a closed $N$-step random
walk in ${\R}^3$ in some prescribed topological state can be presented in
the following way
\be \label{1:1.2}
{\cal P}_N\{\mbox{Inv}\} = \int\ldots\int
\prod_{j=1}^{N}d{\bf r}_j \prod_{j=1}^{N-1}g\left({\bf r}_{j+1}-{\bf
r}_j\right) \delta\left[\mbox{Inv}\{{\bf r}_1\ldots,{\bf
r}_N\}-\mbox{Inv}\right] \delta\left[{\bf r}_N\right]
\ee
where
$g\left({\bf r}_{j+1}-{\bf r}_j\right)$ is the probability to find $j+1$th
step of the trajectory in the point ${\bf r}_{j+1}$ if $j$th step is in
${\bf r}_j$. In the limit $a\to 0$ and $N\to \infty$ ($Na=L=\mbox{const}$)
in three-dimensional space we have the following expression for
$g\left({\bf r}_{j+1}-{\bf r}_j\right)$
\be \label{1:1.2a}
g\left({\bf r}_{j+1}-{\bf r}_j\right) =
\left(\frac{3}{2\pi a^2}\right)^{3/2}\exp\left(-\frac{3({\bf r}_{j+1}-
{\bf r}_j)^2}{2a^2}\right)\simeq
\left(\frac{3}{2\pi a^2}\right)^{3/2}
\exp\left\{\frac{3}{2a}\left(\frac{d{\bf r}(s)}{ds}\right)^2\right\}
\ee
Introducing the ``time", $s$, along the trajectory we rewrite the
distribution function ${\cal P}_N\{\mbox{Inv}\}$ (Eq.(\ref{1:1.2})) in the
path integral form with the Wiener measure density
\be \label{1:1.2c}
{\cal P}_N\{\mbox{Inv}\} = \frac{1}{\cal Z}\int\ldots\int
{\cal D}\{{\bf r}\} \exp\left\{-\frac{3}{2a}\int_0^L
\left(\frac{d{\bf r}(s)}{ds}\right)^2 ds\right\}
\delta[\mbox{Inv}\{{\bf r}(s)\}-\mbox{Inv}]
\ee
and the normalization condition is as follows
$$
{\cal Z}=\sum_{\foot\begin{array}{c} \mbox{all different} \vspace{-3mm} \\
\mbox{knot invariants}\end{array}} {\cal P}_N\{\mbox{Inv}\}
$$

The form of Eq.(\ref{1:1.2c}) up to the Wick turn and the constants
coincides with the scattering amplitude $\alpha$ of a free quantum particle
in the multi-connected phase space. Actually, for the amplitude $\alpha$ we
have
\be \label{1:1.4}
\alpha \sim \sum_{\foot\begin{array}{c}\mbox{all paths from given} 
\vspace{-3mm} \\
\mbox{topological class}\end{array}} \exp\left\{\frac{i}{h}\int
\dot{\bf r}^2(s)ds\right\}
\ee
If phase trajectories can be mutually transformed by means of continuous
deformations, then the summation in Eq.(\ref{1:1.4}) should be extended to
all available paths in the system, but if the phase space consists of
different topological domains, then the summation in Eq.(\ref{1:1.4})
refers to the paths from the exclusively defined class and the ``knot
entropy" problem arises.

\subsection{Abelian problems in statistics of entangled random
walks and incompleteness of Gauss invariant}

As far back as 1967 S.F. Edwards had discovered the basis of the
statistical theory of entanglements in physical systems. In \cite{edwards}
he proposed the way of exact calculating the partition function of
self-intersecting random walk topologically interacting with the infinitely
long uncrossible string (in 3D case) or obstacle (in 2D-case). That problem
had been considered in mathematical literature even earlier---see the paper
\cite{yor} for instance---but S.F. Edwards was apparently the first to
recognize the deep analogy between abelian topological problems in
statistical mechanics of the Markov chains and quantum-mechanical problems
(like Bohm-Aharonov) of the particles in the magnetic fields. The review of
classical results is given in \cite{wiegel}, whereas some modern
advantages are discussed in \cite{com1}.

\begin{figure}
\centerline{\epsfig{file=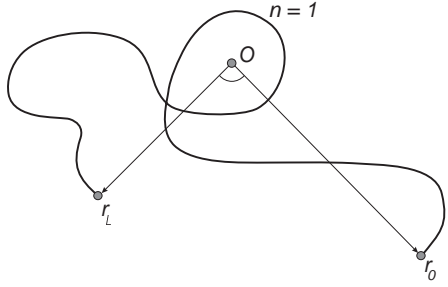,width=6cm}}
\caption{Random walk on the plane near the single obstacle.}
\label{1:fig:1}
\end{figure}

The 2D version of the Edwards' model is formulated as follows. Take a plane
with an excluded origin, producing the topological constraint for the
random walk of length $L$ with the initial and final points ${\bf r}_0$ and
${\bf r}_L$ respectively. Let trajectory make $n$ turns around the origin
(fig.\ref{1:fig:1}). The question is in calculating the distribution
function ${\cal P}_n({\bf r}_0,{\bf r}_L, L)$.

In the said model the topological state of the path $C$ is fully
characterized by number of turns of the path around the origin. The
corresponding abelian topological invariant is known as Gauss linking
number and when represented in the contour integral form, reads
\be \label{1:gauss}
\mbox{Inv}\{{\bf r}(s)\}\equiv G\{C\}=\int_C \frac{ydx-xdy}{x^2+y^2}=
\int_C {\bf A}({\bf r}) d{\bf r}\equiv 2\pi n+\vartheta
\ee
where
\be \label{1:abelian}
{\bf A}({\bf r})={\bf \xi}\times \frac{{\bf r}}{r^2}; \qquad
{\bf \xi}=(0,0,1)
\ee
and $\vartheta$ is the angle distance between ends of the random walk.

Substituting Eq.(\ref{1:gauss}) into Eq.(\ref{1:1.2c}) and using the
Fourier transform of the $\delta$-function, we arrive at
\be \label{1:edwards}
{\cal P}_n({\bf r}_0,{\bf r}_L, L)=
\frac{1}{\pi L a}\exp\left(\frac{r_0^2+r_L^2}{La}\right)
\int_{-\infty}^{\infty}I_{|\lambda|}\left(\frac{2r_0 r_L}{La}\right)
e^{i\lambda(2\pi n + \vartheta)}d\lambda
\ee
which reproduces the well known old result \cite{edwards} (some very
important generalizations one can find in \cite{com1}).

Physically significant quantity obtained on the basis of
Eq.(\ref{1:edwards}) is the entropic force
\be \label{1:force}
f_n(\rho)=-\frac{\partial}{\partial \rho} \ln {\cal P}_n(\rho, L)
\ee
which acts on the closed chain $({\bf r}_0={\bf r}_L=\rho,\; \vartheta=0)$
when the distance between the obstacle and a certain point of the
trajectory changes. Apparently the topological constraint leads to the
strong attraction of the path to the obstacle for any $n\neq 0$ and to the
weak repulsion for $n=0$.

Another exactly solvable 2D-problem closely related to the one under
discussion deals with the calculation of the partition function of a random
walk with given algebraic area. The problem concerns the determination of
the distribution function ${\cal P}_S({\bf r}_0,{\bf r}_L, L)$ for the
random walk with the fixed ends and specific algebraic area $S$.

As a possible solution of that problem, D.S. Khandekar and F.W. Wiegel
\cite{wiegel2} again represented the distribution function in terms of the
path integral Eq.(\ref{1:1.2c}) with the replacement
\be \label{1:area_inv}
\delta[\mbox{Inv}\{{\bf r}(s)\}-\mbox{Inv}]\to \delta[S\{{\bf r}(s)\}-S]
\ee
where the area is written in the Landau gauge:
\be \label{1:area}
S\{{\bf r}(s)\}=\frac{1}{2}\int_C ydx-xdy=\frac{1}{2}\int_C\tilde{\bf
A}\{{\bf r}\}\dot{\bf r}ds;\qquad \tilde{\bf A}=\xi\times {\bf r}
\ee
(compare to Eqs.(\ref{1:gauss})-(\ref{1:abelian})).

The final expression for the distribution function reads (\cite{wiegel})
\be \label{1:wiegel1}
{\cal P}_S({\bf r}_0,{\bf r}_L, L)=\frac{1}{2\pi}\int_{-\infty}^{\infty}
d g\; e^{iq S}\; {\cal P}_q({\bf r}_0,{\bf r}_L, L)
\ee
where
\be
\label{1:wiegel2}
\begin{array}{l}
\disp
{\cal P}_q({\bf r}_0,{\bf r}_L, L)=
\frac{\lambda}{4\pi\sin\frac{La\lambda}{4}} \\
\disp \qquad \times
\exp\left\{\frac{\lambda}{2}(x_0 y_L-y_0 x_L)-\frac{\lambda}{4}
\left((x_L-x_0)^2+(y_L-y_0)^2\right)\cot\frac{La\lambda}{4}\right\}
\end{array}
\ee
and $\lambda=-iq$.

For closed trajectories Eqs.(\ref{1:wiegel1})-(\ref{1:wiegel2}) can be
simplified essentially, giving
\be \label{1:closed}
{\cal P}_S^{cl}(N)=2La\cosh^2\left(\frac{2\pi S}{La}\right)
\ee
Different aspects of this problem have been extensively studied in
\cite{com1}.

There is no principal difference between the problems of random walk
statistics in the presence of a single topological obstacle or with a fixed
algebraic area---both of them have the ``abelian" nature. Nevertheless we
would like to concentrate on the last problem because of its deep
connection with the famous Harper-Hofstadter model dealing with spectral
properties of the 2D electron hopping on the discrete lattice in the
constant magnetic field \cite{harper}. Actually, rewrite Eq.(\ref{1:1.2})
with the substitution Eq.(\ref{1:area_inv}) in form of recursion relation
in the number of steps, $N$:
\be \label{1:recursion}
\begin{array}{c}
\disp {\cal P}_q({\bf r}_{N+1}, N+1)=\int d{\bf r}_N g\left({\bf r}_{N+1}-
{\bf r}_N\right) \exp\left(\frac{iq}{2} {\bf\xi}({\bf r}_N
\times{\bf r}_{N+1})\right) \medskip \\
\disp \times {\cal P}_q({\bf r}_N, N)
\end{array}
\ee
For the discrete random walk on ${\Z}^2$ we use the identity
\be
\int d{\bf r}_N g\left({\bf r}_{N+1}-{\bf r}_N\right) (\ldots) \to
\sum_{\{{\bf r}_N\}} w\left({\bf r}_{N+1}-{\bf r}_N\right) (\ldots)
\ee
where $w\left({\bf r}_{N+1}-{\bf r}_N\right)$ is the matrix of the local
jumps on the square lattice; $w$ is supposed to be symmetric:
\be
w=\left\{\begin{array}{cl} \frac{1}{4} & \mbox{for $(x,y)\to(x,y\pm 1)$ and
$(x,y)\to(x\pm 1,y)$} \medskip \\ 0 & \mbox{otherwise} \end{array}\right.
\ee
Finally, we get in the Landau gauge:
\be \label{1:harper}
\begin{array}{lll} \disp \frac{4}{\varepsilon} W(x,y,q,\varepsilon) & = &
\disp e^{\frac{1}{2}iq x} W(x,y-1,q) + \disp e^{-\frac{1}{2}iq x}
W(x,y+1,q) \;+ \medskip \\ & & \disp e^{\frac{1}{2}iq y}
W(x-1,y,q) + e^{-\frac{1}{2}iq y} W(x+1,y,q)
\end{array}
\ee
where $W(x,y,q,\varepsilon)$ is the generating function defined via
relation
$$
W(x,y,q,\varepsilon)=\sum_{N=0}^{\infty} \varepsilon^N {\cal
P}_S({\bf r}_N, N)
$$
and $q$ plays a role of the magnetic flux through the contour bounded by
the random walk on the lattice.

There is one point which is still out of our complete understanding. On the
one hand the continuous version of the described problem has very clear
abelian background due to the use of commutative ``invariants" like
algebraic area Eq.(\ref{1:area}). On the other hand it has been recently
discovered (\cite{zabrodin}) that so-called Harper equation, i.e.
Eq.(\ref{1:harper}) written in the gauge $S\{{\bf r}\}=\int_C ydx $,
exhibits the hidden quantum group symmetry related to the so-called
$C^{*}$--algebra (\cite{bellissard}) which is strongly nonabelian. Usually
in statistical physics we expect that the continuous limit (when lattice
spacing tends to zero with corresponding rescaling of parameters of the
model) of any discrete problem does not change the observed physical
picture, at least qualitatively. But for the considered model the spectral
properties of the problem are extremely sensitive to the actual physical
scale of the system and depend strongly on the lattice geometry.

The generalization of the above stated problems concerns, for instance,
the computation of the partition function for the random walk entangled
with $k>1$ obstacles on the plane located in the points $\{{\bf r}_1,
\ldots,{\bf r}_k\}$. At first sight, approach based on usage of Gauss
linking number as topological invariant, might allow us to solve such
problem easily. Let us replace the vector potential ${\bf A}({\bf r})$ in
Eq.(\ref{1:gauss}) by the following one
\be \label{1:many_gauss}
{\bf A}({\bf r}_1,\ldots,{\bf r}_k)={\bf \xi}\times
\sum_{j=1}^k\frac{{\bf r}-{\bf r}_j}{|{\bf r}-{\bf r}_j|^2}
\ee
The topological invariant in this case will be the algebraic sum of turns
around obstacles, which seems to be a natural generalization of the Gauss
linking number to the case of many-obstacle entanglements.

\begin{figure}
\centerline{\epsfig{file=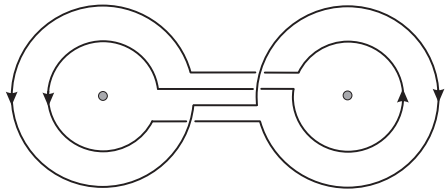,width=6cm}}
\caption{Pochammer contour entangled with two obstacles together but not
entangled with every one.}
\label{1:fig:3}
\end{figure}

However, the following problem is bound to arise: for the system with two
or more obstacles it is possible to imagine closed trajectories entangled
with a few obstacles together but not entangled with every one. In the
fig.\ref{1:fig:3} the so-called ``Pochhammer contour" is shown. Its
topological state with respect to the obstacles cannot be described using
any abelian version of the Gauss-like invariants.

To clarify the point we can apply to the concept of the homotopy group
\cite{dubrovin}. Consider the topological space ${\cal R}={\R}^2-\{{\bf
r}_1, {\bf r}_2\}$ where $\{{\bf r}_1, {\bf r}_2\}$ are the coordinates of
the removed points (obstacles) and choose an arbitrary reference point
${\bf r}_0$. Consider the ensemble of all directed trajectories starting
and finishing in the point ${\bf r}_0$. Take the {\it basis loops}
$\gamma_1(s)$ and $\gamma_2(s)$ $(0<s<L)$ representing the right-clock
turns around the points ${\bf r}_1$ and ${\bf r}_2$ respectively. The same
trajectories passed in the counter-clock direction are denoted by
$\gamma^{-1}_1(s)$ and $\gamma^{-1}_2(s)$.

The {\it multiplication} of the paths is their composition: for instance,
$\gamma_1\gamma_2=\gamma_1\circ\gamma_2$. The unit (trivial) path is the
composition of an arbitrary loop with its inverse:
\be
\label{1:unit} e=\gamma_i\gamma_i^{-1}=\gamma_i^{-1}\gamma_i
\qquad\qquad i=\{1,2\}.
\ee
The loops $\gamma_i(s)$ and $\tilde{\gamma}_i(s)$ are called equivalent if
one can be transformed into another by means of monotonic change of
variables $s=s(\tilde{s})$. The homotopic classes of directed trajectories
form the group with respect to the paths multiplication; the unity is the
homotopic class of the trivial paths. This group is known as the {\it
homotopy group} $\pi_1({\cal R},{\bf r}_0)$.

Any closed path on ${\cal R}$ can be represented by the ``word" consisting
of set of letters $\{\gamma_1,\gamma_2,\gamma_1^{-1},\gamma_2^{-1}\}$.
Taking into account Eq.(\ref{1:unit}), we can reduce each word to the
minimal irreducible representation. For example, the word
$W=\gamma_1\gamma_2^{-1}\gamma_1\gamma_1\gamma_1^{-1}\gamma_2^{-1}
\gamma_2\gamma_1^{-1}\gamma_2^{-1}$ can be transformed to the irreducible
form: $W=\gamma_1\gamma_2^{-1} \gamma_2^{-1}$. It is easy to understand
that the word $W \equiv e$ represents only the unentangled contours. The
entanglement in fig.\ref{1:fig:3} corresponds to the irreducible word
$W=\gamma_1^{-1}\gamma_2 \gamma_1\gamma_2^{-1}\equiv 1$. The non-abelian
character of the topological constraints is reflected in the fact that
different entanglements do not commute: $\gamma_1\gamma_2\neq
\gamma_2\gamma_1$. At the same time, the total algebraic number of turns
(Gauss linking number) for the path in fig.\ref{1:fig:3} is equal to zero,
i.e. it belongs to the trivial {\it class of homology}. Speaking more
formally, the mentioned example is the direct consequence of the well known
fact in topology: the classes of homology of knots (of entanglements) do
not coincide in general with the corresponding homotopic classes. The first
ones for the group $\pi_1$ can be distinguished by the Gauss invariant,
while the problem of characterizing the homotopy class of a knot
(entanglement) by an analytically defined invariant is one of the main
problems in topology.

The principal difficulty connected with application of the Gauss invariant
is due to its incompleteness. Hence, exploiting the abelian invariants for
adequate classification of topologically different states in the systems
with multiple topological constraints is very problematic.

\subsection{Nonabelian Algebraic Knot Invariants}

The most obvious topological questions concerning the knotting probability
during the random closure of the random walk cannot be answered using the
Gauss invariant due to its weakness.

The break through in that field was made in 1975-1976 when the algebraic
polynomials were used for the topological state identification of closed
random walks generated by the Monte-Carlo method \cite{volog}. It has been
recognized that the Alexander polynomials being much stronger invariants
than the Gauss linking number, could serve as a convenient tool for the
calculation of the thermodynamic properties of entangled random walks. That
approach actually appeared to be very fruitful and the main part of our
modern knowledge on knots and links statistics is obtained with the help of
these works and their subsequent modifications.

In the present Section we develop the analytic approach in statistical
theory of knots considering the basic problem---the probability to find a
randomly generated knot in a specific topological state. We would like to
reiterate that our investigation would be impossible without utilizing of
algebraic knot invariants discovered recently. Below we reproduce briefly
the construction of Jones invariants following the Kauffman approach in the
general outline.

\subsubsection{Disordered Potts model and generalized dichromatic
polynomials}

The graph expansion for the Potts model with the disorder in the
interaction constants can be defined by means of slight modification of the
well known construction of the ordinary Potts model \cite{wu,baxter}. Let
us recall the necessary definitions.

Take an arbitrary graph ${\cal L}$ with $N$ vertices. To each vertex of the
given graph we attribute the ``spin" variable ${\bf \sigma}_i$ $(i\in [1,N])$
which can take $q$ states labelled as $1,2,\ldots, q$ on the simplex.
Suppose that the interaction between spins belonging to the connected
neighboring graph vertices only contributes to the energy. Define the
energy of the spin's interaction as follows
\be
E_{kl}=J_{kl}\,\delta({\bf \sigma}_k,{\bf \sigma}_l)=
\cases{J_{kl} & ${\bf \sigma}_k={\bf \sigma}_l,\;
({\bf \sigma}_k,{\bf \sigma}_l)$ -- neighbors \medskip \cr
0 & otherwise}
\ee
where $J_{kl}$ is the interaction constant which varies for different
graph edges and the equality ${\bf \sigma}_k={\bf \sigma}_l$ means that the
neighboring spins take equal values on the simplex.

The partition function of the Potts model now reads
\be \label{1:potts}
Z_{potts}=\sum_{\{{\bf \sigma}\}}
\exp\left\{\sum_{\{kl\}}\frac{J_{kl}}{T}\delta({\bf \sigma}_k,{\bf
\sigma}_l)\right\}
\ee
where $T$ is the temperature.

Expression Eq.(\ref{1:potts}) gives for $q=2$ the well-known representation
of the Ising model with the disordered interactions extensively studied in
the theory of spin glasses \cite{mezard}. (Later on we would like to fill
in this old story by a new ``topological" sense.)

To proceed with the graph expansion of the Potts model \cite{baxter},
rewrite the partition function (\ref{1:potts}) in the following way
\be \label{1:pottsprod}
Z_{potts}=\sum_{\{{\bf \sigma}\}}\prod_{\{kl\}}
\left[1+v_{kl}\;\delta({\bf \sigma}_k,{\bf \sigma}_l)\right]
\ee
where
\be \label{1:interaction}
v_{kl}=\exp\left(\frac{J_{kl}}{T}\right)-1
\ee

If the graph ${\cal L}$ has $N$ edges then the product
Eq.(\ref{1:pottsprod}) contains $N$ multipliers. Each multiplier in that
product consists of two terms $\{1$ and $v_{kl}\,\delta({\bf \sigma}_k,{\bf
\sigma}_l)\}$. Hence the partition function Eq.(\ref{1:pottsprod}) is
decomposed in the sum of $2^N$ terms.

Each term in the sum is in one-to-one correspondence with some part of the
graph ${\cal L}$. To make this correspondence clearer, it should be that an
arbitrary term in the considered sum represents the product of $N$
multipliers described above in ones from each graph edge. We accept the
following convention:

(a) If for some edge the multiplier is equal to $1$, we remove the
corresponding edge from the graph ${\cal L}$;

(b) If the multiplier is equal to $v_{kl}\,\delta({\bf \sigma}_k,{\bf
\sigma}_l)$ we keep the edge in its place.

After repeating the same procedure with all graph edges, we find the unique
representation for all terms in the sum Eq.(\ref{1:pottsprod}) by
collecting the components (either connected or not) of the graph ${\cal
L}$.

Take the typical graph $G$ consisting of $m$ edges and $C$ connected
components where the separated graph vertex is considered as one
component. The presence of $\delta$-functions ensures the spin's
equivalence within one graph component. As a result after summation of
all independent spins and of all possible graph decompositions we get the
new expression for the partition function of the Potts system
Eq.(\ref{1:potts})
\be \label{1:graph}
Z_{potts}=\sum_{\{G\}} q^C \prod_{\{kl\}}^m v_{kl}
\ee
where the product runs over all edges in the fixed graph $G$.

It should be noted that the graph expansion Eq.(\ref{1:graph}) where
$v_{kl}\equiv v$ for all $\{k,l\}$ coincides with the well known
representation of the Potts system in terms of {\it dichromatic polynomial}
(see, for instance, \cite{wu,baxter}).

Another comment concerns the number of spin states, $q$. As it can be seen,
in the derivation presented above we did not account for the fact that $q$
has to take positive integer values only. From this point of view the
representation Eq.(\ref{1:graph}) has an advantage with respect to the
standard representation Eq.(\ref{1:potts}) and can be considered an
analytic continuation of the Potts system to the non-integer and even
complex values of $q$. We show in the subsequent sections how the defined
model is connected to the algebraic knot invariants.

\subsubsection{Reidemeister Moves and State Model for Construction of
Algebraic Invariants}

Let $K$ be a knot (or link) embedded in the 3D-space. First of all we
project the knot (link) onto the plane and obtain the 2D-knot diagram in
the so-called general position (denoted by $K$ as well). It means that only
the pair crossings can be in the points of paths intersections. Then for
each crossing we define the passages, i.e. parts of the trajectory on the
projection going ``below" and ``above" in accordance with its natural
positions in the 3D-space.

\begin{figure}
\centerline{\epsfig{file=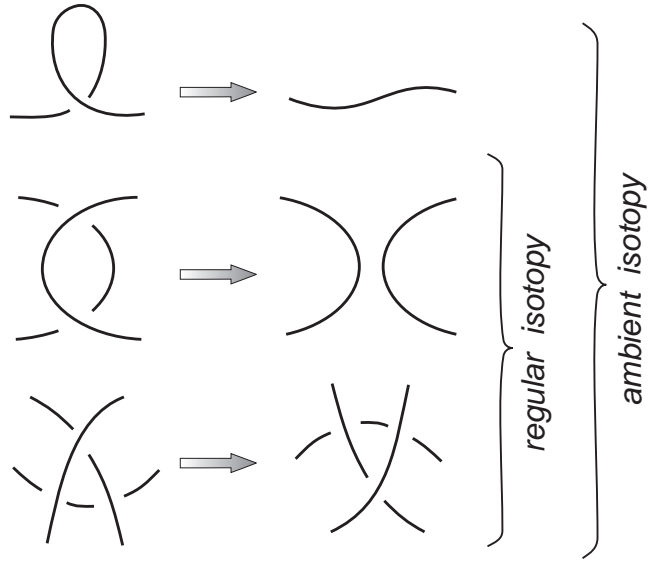,width=9cm}}
\caption{Reidemeister moves of types I, II and III.}
\label{1:fig:5}
\end{figure}

For the knot plane projection with defined passages the following theorem
is valid: (Reidemeister \cite{reid}):

{\it
Two knots embedded in ${\R}^3$ can be deformed continuously one into the
other if and only if the diagram of one knot can be transformed into the
diagram corresponding to another knot via the sequence of simple local
moves of types {\rm I, II} and {\rm III} shown in {\rm fig.\ref{1:fig:5}}.
}

The work \cite{reid} provides us with the proof of this theorem. Two knots
are called {\it regular isotopic} if they are isotopic with respect to two
last Reidemeister moves (II and III); meanwhile, if they are isotopic with
respect to all moves, they are called {\it ambient isotopic}. As it can be
seen from fig.\ref{1:fig:5}, the Reidemeister move of type I leads to the
cusp creation on the projection. At the same time it should be noted that
all real 3D-knots (links) are of ambient isotopy.

Now, after the Reidemeister theorem has been formulated, it is possible to
describe the construction of polynomial ``bracket" invariant in the way
proposed by L.H. Kauffman \cite{kauff2,kauff_bz}. This invariant
can be introduced as a certain partition function being the sum over the
set of some formal (``ghost") degrees of freedom.

Let us consider the 2D-knot diagram with defined passages as a certain
irregular lattice (graph). Crossings of path on the projection are the
lattice vertices. Turn all these crossings to the standard positions where
parts of the trajectories in each graph vertex are normal to each other and
form the angles of $\pm\pi/4$ with the $x$-axis. It can be proven that the
result does not depend on such standardization.

There are two types of vertices in our lattice---a) and b) which we label
by the variable $b_i=\pm 1$ as it is shown below:

\vspace*{-0.25in}
$$
\mbox{(a)} \quad
\unitlength=1mm
\;
\begin{picture}(10.00,10.00)
\linethickness{0.4pt}
\unitlength=.89mm
\put(0.00,-3.50){\line(1,1){4.00}}
\put(6.00,2.50){\line(1,1){4.00}}
\put(10.00,-3.50){\line(-1,1){10.00}}
\end{picture}
\quad b_i=+1 \quad\qquad \mbox{and}
\quad\qquad \mbox{(b)} \quad
\unitlength=1mm
\;
\begin{picture}(10.00,10.00)
\linethickness{0.4pt}
\unitlength=0.89mm
\put(0.00,-3.50){\line(1,1){10.00}}
\put(10.00,-3.50){\line(-1,1){4.00}}
\put(4.00,2.50){\line(-1,1){4.00}}
\end{picture}
\quad b_i=-1
$$
\vspace{-0.1in}

The next step in the construction of algebraic invariant is introduction
of two possible ways of {\it vertex splittings}. Namely, we attribute to
each way of graph splitting the following statistical weights: $A$ to the
horizontal splitting and $B$ to the vertical one for the vertex of type a);
$B$ to the horizontal splitting and $A$ to the vertical one for the
vertex of type b). The said can be schematically reproduced in the
following picture:
\vspace*{0.8in}

\hspace{2cm}
\begin{picture}(110.00,45.00)
\linethickness{0.4pt}
\unitlength=0.89mm
\put(-2.00,14.00){\line(1,1){4.00}}
\put(4.00,20.00){\line(1,1){4.00}}
\put(8.00,14.00){\line(-1,1){10.00}}
\put(11.00,21.00){\vector(3,2){9.00}}
\put(11.00,17.00){\vector(4,-3){9.00}}
\unitlength=1mm
\put(25.50,25.00){\oval(7.00,7.00)[t]}
\put(25.50,34.00){\oval(7.00,7.00)[b]}
\put(34.00,30.00){\makebox(0,0)[cc]{$A$}}
\put(21.00,6.00){\oval(7.00,7.00)[r]}
\put(30.00,6.00){\oval(7.00,7.00)[l]}
\put(34.00,6.00){\makebox(0,0)[cc]{$B$}}
\put(48.00,17.00){\makebox(0,0)[cc]{¨}}
\unitlength=0.89mm
\put(71.00,14.00){\line(1,1){10.00}}
\put(81.00,14.00){\line(-1,1){4.00}}
\put(75.00,20.00){\line(-1,1){4.00}}
\put(84.00,21.00){\vector(3,2){9.00}}
\put(84.00,17.00){\vector(4,-3){9.00}}
\unitlength=1mm
\put(90.50,25.00){\oval(7.00,7.00)[t]}
\put(90.50,34.00){\oval(7.00,7.00)[b]}
\put(99.00,30.00){\makebox(0,0)[cc]{B}}
\put(86.00,6.00){\oval(7.00,7.00)[r]}
\put(95.00,6.00){\oval(7.00,7.00)[l]}
\put(99.00,6.00){\makebox(0,0)[cc]{A}}
\end{picture}
\vspace{-0.05in}

\noindent the constants $A$ and $B$ to be defined later.

For the knot diagram with $N$ vertices there are $2^{N}$ different
microstates, each of them representing the set of splittings of all $N$
vertices. The entire microstate, $S$, corresponds to the knot (link)
disintegration to the system of disjoint and non-selfintersecting circles.
The number of such circles for the given microstate $S$ we denote as ${\cal
S}$. The following statement belongs to L. Kauffman (\cite{kauff2}).

{\it
Consider the partition function
\be \label{1:kauff}
\left<K\right> = \sum_{\{S\}}d^{{\cal S}-1} A^{i} B^{j},
\ee
where $\sum_{\{S\}}$ means summation over all possible $2^N$ graph
splittings, $i$ and $j=N-i$ being the numbers of vertices with weights $A$
and $B$ for the given realization of all splittings in the microstate $S$
respectively.

The polynomial in $A$, $B$ and $d$ represented by the partition function
{\rm Eq.(\ref{1:kauff})} is the topological invariant of knots of regular
isotopy if and only if the following relations among the weights $A$, $B$
and $d$ are fulfilled:
\be \label{1:abd}
\begin{array}{r} AB = 1 \medskip \\
ABd+A^{2}+B^{2} = 0
\end{array}
\ee
}

The sketch of the proof is as follows. Denote with $\left<\ldots\right>$
the statistical weight of the knot or of its part. The
$\left<K\right>$-value equals the product of all weights of knot parts.
Using the definition of vertex splittings, it is easy to test the following
identities valid for unoriented knot diagrams
\vspace{-0.25in}

\be \label{1:skein}
\begin{array}{ccl}
\left<
\unitlength=1mm
\;
\begin{picture}(10.00,10.00)
\linethickness{0.4pt}
\unitlength=.89mm
\put(0.00,-3.50){\line(1,1){4.00}}
\put(6.00,2.50){\line(1,1){4.00}}
\put(10.00,-3.50){\line(-1,1){10.00}}
\end{picture}
\right> & = & \left<
\unitlength=1.00mm
\;
\begin{picture}(7.00,12.50)
\linethickness{0.4pt}
\put(3.50,-3.30){\oval(7.00,7.00)[t]}
\put(3.50,5.50){\oval(7.00,7.00)[b]}
\end{picture}
\;
\right>\;
A\; +\; \left<
\unitlength=1.00mm
\;
\begin{picture}(12.50,6.50)
\linethickness{0.4pt}
\put(0.00,1.00){\oval(7.00,7.00)[r]}
\put(9.00,1.00){\oval(7.00,7.00)[l]}
\end{picture}
\hspace{-.1in}
\right>\; B \\
\left<
\unitlength=1mm
\;
\begin{picture}(10.00,10.00)
\linethickness{0.4pt}
\unitlength=0.89mm
\put(0.00,-3.50){\line(1,1){10.00}}
\put(10.00,-3.50){\line(-1,1){4.00}}
\put(4.00,2.50){\line(-1,1){4.00}}
\end{picture}
\right> & = & \left<
\unitlength=1.00mm
\;
\begin{picture}(7.00,12.50)
\linethickness{0.4pt}
\put(3.50,-3.30){\oval(7.00,7.00)[t]}
\put(3.50,5.50){\oval(7.00,7.00)[b]}
\end{picture}
\;
\right>\;
B\; +\; \left<
\unitlength=1.00mm
\;
\begin{picture}(12.50,6.50)
\linethickness{0.4pt}
\put(0.00,1.00){\oval(7.00,7.00)[r]}
\put(9.00,1.00){\oval(7.00,7.00)[l]}
\end{picture}
\hspace{-.1in}
\right>\; A
\end{array}
\ee
completed by the ``initial condition"
\be \label{1:skein_in}
\Bigl< K \bigcup O \Bigr> = d \Bigl<K\Bigr>; \qquad \mbox{$K$ is not empty}
\ee
where $O$ denotes the separated trivial loop.

The {\it skein relations} Eq.(\ref{1:skein}) correspond to the above
defined weights of horizontal and vertical splittings while the relation
Eq.(\ref{1:skein_in}) defines the statistical weights of the composition of
an arbitrary knot and a single trivial ring. These diagrammatic rules are
well defined only for fixed ``boundary condition" of the knot (i.e., for the
fixed part of the knot outside the brackets). Suppose that by convention
the polynomial of the trivial ring is equal to the unity;
\be
\label{1:trivial} \Bigl< O \Bigr> = 1
\ee

Now it can be shown that under the appropriate choice of the relations
between $A$, $B$ and $d$, the partition function Eq.(\ref{1:kauff})
represents the algebraic invariant of the knot. The proof is based on
direct testing of the invariance of $\left<K\right>$-value with respect to
the Reidemeister moves of types II and III. For instance, for the
Reidemeister move of type II we have:
\be \label{1:reid2}
\begin{array}{c} \left<
\unitlength=1.00mm
\;
\special{em:linewidth 0.4pt}
\linethickness{0.4pt}
\begin{picture}(10.00,10.00)
\put(10.00,1.00){\oval(10.00,12.00)[l]}
\put(8.00,1.00){\oval(4.00,8.00)[r]}
\put(4.00,5.00){\line(-1,0){3.00}}
\put(4.00,-3.00){\line(-1,0){3.00}}
\end{picture}
\;\;
\right> = \left<
\unitlength=1.00mm
\special{em:linewidth 0.4pt}
\linethickness{0.4pt}
\begin{picture}(11.00,12.00)
\put(6.00,-2.00){\line(-3,-2){5.00}}
\put(6.00,-2.00){\line(3,-2){5.00}}
\put(6.00,1.00){\circle{3.00}}
\put(11.00,7.00){\line(-3,-2){5.00}}
\put(1.00,7.00){\line(3,-2){5.00}}
\end{picture}
\;
\right>\;ABd\; +\; \left<
\unitlength=1.00mm
\;
\special{em:linewidth 0.4pt}
\linethickness{0.4pt}
\begin{picture}(12.00,10.00)
\put(6.00,0.50){\oval(4.00,3.00)[t]}
\put(7.00,0.50){\oval(2.00,2.00)[rb]}
\put(5.00,0.50){\oval(2.00,2.00)[lb]}
\put(5.00,-0.50){\line(-3,-2){5.00}}
\put(7.00,-0.50){\line(3,-2){5.00}}
\put(6.00,3.00){\line(2,1){6.00}}
\put(6.00,3.00){\line(-2,1){6.00}}
\end{picture}
\;
\right>\;B^2\; +\; \\
\left<
\unitlength=1.00mm
\;
\special{em:linewidth 0.4pt}
\linethickness{0.4pt}
\begin{picture}(12.00,10.44)
\put(0.00,-3.90){\line(2,1){6.00}}
\put(12.00,-3.90){\line(-2,1){6.00}}
\put(6.00,1.75){\oval(4.00,3.00)[b]}
\put(5.00,1.75){\oval(2.00,2.00)[lt]}
\put(7.00,1.75){\oval(2.00,2.00)[rt]}
\put(11.89,6.25){\line(-3,-2){5.00}}
\put(0.00,6.25){\line(3,-2){5.00}}
\end{picture}
\;
\right>\;A^2\; +\; \left<
\unitlength=1.00mm
\;
\special{em:linewidth 0.4pt}
\linethickness{0.4pt}
\begin{picture}(12.00,10.33)
\put(5.00,-1.00){\line(-3,-2){5.00}}
\put(7.50,-1.00){\line(3,-2){5.00}}
\put(5.00,1.00){\oval(4.00,4.00)[l]}
\put(7.50,1.00){\oval(3.00,4.00)[r]}
\put(0.00,6.30){\line(3,-2){5.00}}
\put(12.50,6.30){\line(-3,-2){5.00}}
\end{picture}
\;
\right>\;AB = \;\;\left<
\unitlength=1.00mm
\;
\begin{picture}(7.00,12.50)
\linethickness{0.4pt}
\put(3.50,-3.30){\oval(7.00,7.00)[t]}
\put(3.50,5.50){\oval(7.00,7.00)[b]}
\end{picture}
\;
\right>\;\left(ABd+A^2+B^2\right)\; +\; \left<
\unitlength=1.00mm
\;
\begin{picture}(12.50,6.50)
\linethickness{0.4pt}
\put(0.00,1.00){\oval(7.00,7.00)[r]}
\put(9.00,1.00){\oval(7.00,7.00)[l]}
\end{picture}
\hspace{-0.1in}
\right>\;AB
\end{array}
\ee

Therefore, the invariance with respect to the Reidemeister move of type II
can be obtained immediately if we set the statistical weights in the last
line of Eq.(\ref{1:reid2}) as it is written in Eq.(\ref{1:abd}). Actually,
the topological equivalence of two knot diagrams is restored with respect
to the Reidemeister move of type II only if the right- and left-hand sides
of Eq.(\ref{1:reid2}) are identical. It can also be tested that the
condition of obligatory invariance with respect to the Reidemeister move of
type III does not violate the relations Eq.(\ref{1:abd}).

The relations Eq.(\ref{1:abd}) can be converted into the form
\be \label{1:b_d}
B = A^{-1}, \qquad d = -A^{2} - A^{-2}
\ee
which means that the Kauffman invariant Eq.(\ref{1:kauff}) is the Laurent
polynomial in $A$-value only.

Finally, Kauffman showed that for oriented knots (links) the invariant of
ambient isotopy (i.e., the invariant with respect to all Reidemeister
moves) is defined via relation:
\be \label{1:ambient}
f[K] = (-A)^{3Tw(K)} \left<K\right>
\ee
here $Tw(K)$ is the twisting of the knot (link), i.e. the sum of signs of
all crossings defined by the convention:
\vspace{0.4in}

\hspace{2cm}
\begin{picture}(110.00,10.00)
\linethickness{0.4pt}
\unitlength=0.89mm
\put(5.00,5.00){\makebox(0,0)[cc]{( )}}
\put(12.00,0.00){\line(1,1){4.00}}
\put(18.00,6.00){\vector(1,1){4.00}}
\put(22.00,0.00){\vector(-1,1){10.00}}
\put(30.00,5.00){\makebox(0,0)[cc]{$+1$}}
\put(78.00,5.00){\makebox(0,0)[cc]{(¡)}}
\put(85.00,0.00){\vector(1,1){10.00}}
\put(95.00,0.00){\line(-1,1){4.00}}
\put(89.00,6.00){\vector(-1,1){4.00}}
\put(103.00,5.00){\makebox(0,0)[cc]{$-1$}}
\end{picture}
\vspace{.1in}

\noindent (not to be confused with the definition of the variable $b_i$
introduced above). Eq.(\ref{1:ambient}) follows from the following chain of
equalities
$$
\begin{array}{lll}
f\left[
\unitlength=1.00mm
\;\;
\linethickness{0.4pt}
\begin{picture}(14.00,6.00)
\put(13.00,6.00){\vector(-1,-1){10.00}}
\put(7.00,2.00){\line(-1,1){4.00}}
\put(9.00,0.00){\vector(1,-1){4.00}}
\put(3.00,1.00){\oval(9.00,10.00)[l]}
\end{picture}
\right] & = &
\left<
\unitlength=1.00mm
\;
\linethickness{0.4pt}
\begin{picture}(14.00,8.00)
\unitlength=0.83mm
\put(15.00,7.00){\line(-1,-1){5.00}}
\put(10.00,2.00){\line(-1,1){5.00}}
\put(15.00,-5.00){\line(-1,1){5.00}}
\put(10.00,0.00){\line(-1,-1){5.00}}
\put(5.00,1.00){\oval(11.00,12.00)[l]}
\end{picture}
\right>\;B\;+
\left<
\unitlength=1.00mm
\linethickness{0.4pt}
\begin{picture}(15.00,8.00)
\put(13.00,1.00){\oval(9.00,10.00)[l]}
\put(4.00,1.00){\oval(6.00,10.00)[]}
\end{picture}
\right>\;dA \\
& = & \left<
\unitlength=1.00mm
\linethickness{0.4pt}
\begin{picture}(8.00,8.00)
\put(5.00,1.00){\oval(9.00,10.00)[l]}
\end{picture}
\right>\;(B+dA)\; \equiv\;
\left<
\unitlength=1.00mm
\linethickness{0.4pt}
\begin{picture}(8.00,8.00)
\put(5.00,1.00){\oval(9.00,10.00)[l]}
\end{picture}
\right>(-A)^3
\end{array}
$$

The state model and bracket polynomials introduced by L.H. Kauffman seem to
be very special. They explore only the peculiar geometrical rules such as
summation over the formal ``ghost" degrees of freedom---all possible knot
(link) splittings with simple defined weights. But one of the main
advantages of the described construction is connected with the fact that
Kauffman polynomials in $A$-value coincide with Jones knot invariants in
$t$-value (where $t=A^{1/4}$).

Jones polynomial knot invariants were discovered first by V.F.R. Jones
during his investigation of topological properties of braids (see Section 3
for details). Jones' proposition concerns the establishment of the deep
connection between the braid group relations and the Yang-Baxter equations
ensuring the necessary condition of transfer matrix commutativity
\cite{collect}. The Yang-Baxted equations play an exceptionally important
role in the statistical physics of integrable systems (such as ice, Potts,
$O(n)$, 8-vertex, quantum Heisenberg models \cite{baxter}).

The attempt to apply Kauffman invariants of regular isotopy for
investigation of statistical properties of random walks with topological
constraints in a thin slit has been made recently \cite{grnech}.
Below we extend the ideas of the work \cite{grnech} considering the
topological state of the knot as a special kind of {\it a quenched
disorder}.

\subsection{Lattice knot diagrams as disordered Potts model}

\begin{figure}
\centerline{\epsfig{file=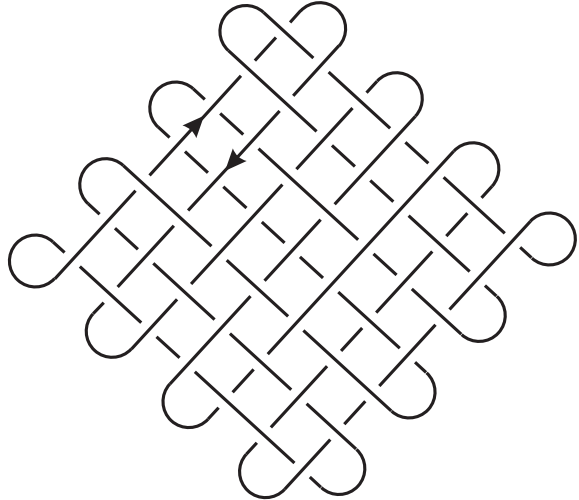,width=8cm}}
\caption{Lattice knot with topological disorder realized in a quenched random
pattern of passages.}
\label{1:fig:6}
\end{figure}

Let us specify the model under consideration. Take a square lattice ${\cal
M}$ turned to the angle $\pi/4$ with respect to the $x$-axis and project a
knot embedded in ${\R}^3$ onto ${\cal M}$ supposing that each crossing
point of the knot diagram coincides with one lattice vertex without fall
(there are no empty lattice vertices)---see fig.\ref{1:fig:6}. Define the
passages in all $N$ vertices and choose such boundary conditions which
ensure the lattice to form a single closed path; that is possible when
$\sqrt{N}$ (i.e. $N$) is an odd number. The {\it frozen pattern} of all
passages $\{b_i\}$ on the lattice together with the boundary conditions
fully determine the topology of some 3D knot.

Of course, the model under consideration is rather rough because we neglect
the ``space" degrees of freedom due to trajectory fluctuations and keep the
pure topological specificity of the system. Later on in Chapter 4 we
discuss the applicability of such model for real physical systems and
produce arguments in support of its validity.

The basic question of our interest is as follows: what is the probability
${\cal P}_N\{f[K]\}$ to find a knot diagram on our lattice ${\cal M}$ in a
topological state characterized by some specific Kauffman invariant $f[K]$
among all $2^N$ microrealizations of the disorder $\{b_i\}$ in the lattice
vertices. That probability distribution reads (compare to Eq.(\ref{1:1.1}))
\be \label{1:probab}
{\cal P}_N\{f[K]\}=\frac{1}{2^N}\sum_{\{b_i\}}
\Delta\Bigl[f[K\{b_1,b_2,\ldots,b_N\}]-f[K]\Bigr]
\ee
where $f[K\{b_1,\ldots,b_N\}]$ is the representation of the Kauffman
invariant as a function of all passages $\{b_i\}$ on the lattice ${\cal
M}$. These passages can be regarded as a sort of quenched ``external field"
(see below).

Our main idea of dealing with Eq.(\ref{1:probab}) consists in two steps:

(a) At first we convert the Kauffman topological invariant into the known
and well-investigated Potts spin system with the disorder in interaction
constants;

(b) Then we apply the methods of the physics of disordered systems to the
calculation of thermodynamic properties of the Potts model. It enables us
to extract finally the estimation for the requested distribution function.

Strictly speaking, we could have disregarded point (a), because it does not
lead directly to the answer to our main problem. Nevertheless we follow the
mentioned sequence of steps in pursuit of two goals: 1) we would like to
prove that the topologically-probabilistic problem can be solved within the
framework of standard thermodynamic formalism; 2) we would like to employ
the knowledges accumulated already in physics of disordered Potts systems
to avoid some unnecessary complications. Let us emphasize that the
mean--field approximation and formal replacement of the model with
short--range interactions by the model with infinite long--range ones
serves to be a common computational tool in the theory of disordered
systems and spin glasses.

\subsubsection{Algebraic invariants of regular isotopy}

The general outline of topological invariants construction deals with
seeking for the functional, $f[K\{b_1,...b_N\}]$, which is independent on
the knot shape i.e. is invariant with respect to all Reidemeister moves.

Recall that the Potts representation of the Kauffman polynomial invariant
Eq.(\ref{1:kauff}) of regular isotopy for some given pattern of
``topological disorder", $\{b_i\}$, deals with simultaneous splittings in
all lattice vertices representing the polygon decomposition of the lattice
${\cal M}$. Such lattice disintegration looks like a densely packed system
of disjoint and non-selfintersecting circles. The collection of all
polygons (circles) can be interpreted as a system of the so-called {\it
Eulerian circuits} completely filling the square lattice. Eulerian circuits
are in one-to-one correspondence with the graph expansion of some
disordered Potts system introduced in Section 2.3.1 (see details below and
in \cite{dupldav}).

Rewrite the Kauffman invariant of regular isotopy, $\left<K\right>$,
in form of disordered Potts model defined in the previous section.
Introduce the two-state ``ghost" spin variables, $s_i=\pm 1$ in each lattice
vertex independent on the crossing in the same vertex
\vspace{-0.3in}

$$
\unitlength=1.00mm
\;
\begin{picture}(7.00,12.50)
\linethickness{0.4pt}
\put(3.50,-3.30){\oval(7.00,7.00)[t]}
\put(3.50,5.50){\oval(7.00,7.00)[b]}
\end{picture}
\;
\quad s_i=+1 \quad\qquad \mbox{¨} \quad\qquad
\unitlength=1.00mm
\;
\begin{picture}(12.50,6.50)
\linethickness{0.4pt}
\put(0.00,1.00){\oval(7.00,7.00)[r]}
\put(9.00,1.00){\oval(7.00,7.00)[l]}
\end{picture}
\hspace{-0.1in}
\quad s_i=-1
$$
\vspace{-0.07in}

Irrespective of the orientation of the knot diagram shown in
fig.\ref{1:fig:6} (i.e. restricting with the case of regular isotopic
knots), we have
\be \label{1:splittings}
\left<K\{b_i\}\right>=\sum_{\{S\}}\left(A^2+A^{-2}\right)^{{\cal S}-1}
\exp\left(\ln A {\sum_{i=1}^Nb_i s_i}\right)
\ee

\begin{figure}
\centerline{\epsfig{file=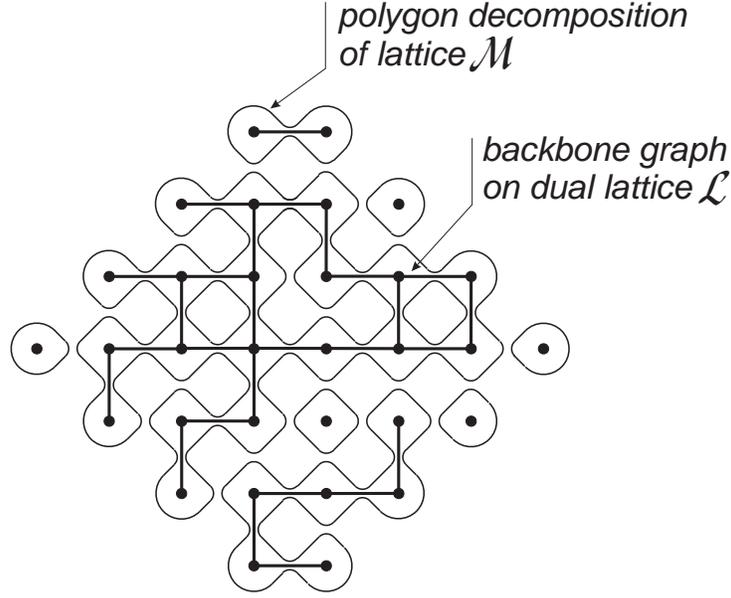,width=10cm}}
\caption{Disintegration of the knot diagram on the ${\cal M}$-lattice into
ensemble of nonselfintersecting loops (Eulerian circuits) and graph
representation of the Potts model on the dual ${\cal L}$-lattice.}
\label{1:fig:7}
\end{figure}

Written in such form the partition function $\left<K\{b_i\}\right>$
represents the weihgted sum of all possible {\it Eulerian circuits}
\footnote{Eulerian circuit is a trajectory on the graph which visits once
and only once all graph edges.} ~on the lattice ${\cal M}$. Let us show
explicitly that the microstates of the Kauffman system are in one-to-one
correspondence with the microstates of some disordered Potts model on a
lattice. Apparently for the first time the similar statement was expressed
in the paper \cite{kauff2}. To be careful, we would like to
use the following definitions:

(i) Let us introduce the lattice ${\cal L}$ dual to the lattice ${\cal M}$,
or more precisely, one of two possible (odd and even) diagonal dual
lattices, shown in fig.\ref{1:fig:7}. It can be easily noticed that the
edges of the lattice ${\cal L}$ are in one-to-one correspondence with the
vertices of the lattice ${\cal M}$. Thus, the disorder on the dual lattice
${\cal L}$ is determined on the {\it edges}. In turn, the edges of the
lattice ${\cal L}$ can be divided into the subgroups of vertical and
horisontal bonds. Each $kl$-bond of the lattice ${\cal L}$ carries the
``disorder variable" $b_{kl}$ being a function of the variable $b_i$ located
in the corresponding $i$-vertex of the lattice ${\cal M}$. The simplest and
most sutable choice of the function $b_{kl}(b_i)$ is as in Eq.(\ref{1:2.1})
(or vice versa for another choice of dual lattice); $i$ is the vertex of
the lattice ${\cal M}$ belonging to the $kl$-bond of the dual lattice
${\cal L}$.

(ii) For the given configuration of splittings on ${\cal M}$ and chosen
dual lattice ${\cal L}$ let us accept the following convention: we mark the
edge of the ${\cal L}$-lattice by the solid line if this edge is not
intersected by some polygon on the ${\cal M}$-latice and we leave the
corresponding edge unmarked if it is intersected by any polygon---as it is
shown in the fig.\ref{1:fig:7}. Similarly, the sum $\sum s_ib_i$ in
Eq.(\ref{1:splittings}) can be rewritten in terms of marked and unmarked
bonds on the ${\cal L}$-lattice
\be \label{1:2.2}
\begin{array}{ccl}
\disp \sum_is_ib_i & = & \disp \sum_{\rm mark}s_ib_i +
\sum_{\rm nonmark}s_ib_i \medskip \\
& = & \disp \sum_{\rm mark}^{\rm horiz}s_ib_i
+ \sum_{\rm mark}^{\rm vertic}s_ib_i + \sum_{\rm nonmark}^
{\rm horiz}s_ib_i+\sum_{\rm nonmark}^{\rm vertic}s_ib_i \medskip \\
& = & \disp -\sum_{\rm mark}^{\rm horiz}b_{kl} -
\sum_{\rm mark}^{\rm vertic}b_{kl} + \sum_{\rm nonmark}^{\rm horiz}b_{kl} +
\sum_{\rm nonmark}^{\rm vertic}b_{kl} \medskip \\
& = & \disp \sum_{\rm nonmark}b_{kl}-\sum_{\rm mark}b_{kl} =
\sum_{\rm all\; edges}b_{kl}-2\sum_{\rm mark}b_{kl}
\end{array}
\ee
where we used the relation $\sum\limits_{\rm
nonmark}b_{kl}+\sum\limits_{\rm mark}b_{kl} = \sum\limits_{\rm all\;
edges}b_{kl}$.

(iii) Let $m_s$ be the number of marked edges and $C_s$ be the number of
connected components of marked graph. Then the Euler relation reads:
\be \label{1:2.3}
{\cal S}=2C_s+m_s-N+\chi
\ee
The Eq.(\ref{1:2.3}) can be proved directly. The $\chi$-value depends on
the genus of the surface, which can be covered by the given lattice, (i.e.
$\chi$ depends on the boundary conditions). In the thermodynamic limit
$N\gg 1$ the $\chi$-dependence should disappear (at least for the flat
surfaces), so the standard equality ${\cal S}= 2C_s+m_s-N$ will be assumed
below.

By means of definitions (i)-(iii), we can easily convert
Eq.(\ref{1:splittings}) into the form:
\be \label{1:2.4}
\begin{array}{c}
\disp \left<K\{b_{kl}\}\right>=(A^2+A^{-2})^{-(N+1)}\prod_{\foot\mbox{all
edges}}^N \left[A^{b_{kl}}\right] \hspace{3cm} \\
\disp \hspace{2.7cm} \times\; \sum_{\{G\}}(A^2+A^{-2})^{2C_s}
\prod_{\foot\mbox{mark}}^{m_s}
\left[A^{-2b_{kl}}(-A^2-A^{-2})\right]
\end{array}
\ee
where we used Eq.(\ref{1:2.2}) and the fact that $N+1$ is even. Comparing
Eq.(\ref{1:2.4}) with Eq.(\ref{1:pottsprod}) we immediately conclude that
\be \label{1:2.4a}
\sum_{\{G\}}(A^2+A^{-2})^{2C_s}\prod_{\foot\mbox{mark}}^{m_s}
\left[A^{-2b_{kl}}(-A^2-A^{-2})\right]\equiv
\sum_{\{{\bf \sigma}\}}\prod_{\{kl\}}
\left(1+v_{kl}\delta({\bf \sigma}_k,{\bf \sigma}_l)
\right)
\ee
what coincides with the partition function of the Potts system written in
the form of dichromatic polynomial. Therefore, we have
\be \label{1:2.5}
\begin{array}{l}
v_{kl} \stackrel{def}{=} A^{-2b_{kl}}(-A^2-A^{-2})=-1-A^{-4b_{kl}}
\medskip \\ \hspace{0.8em} q = (A^2+A^{-2})^2
\end{array}
\ee
Since the ``disorder" variables $b_{kl}$ take the discrete values $\pm 1$
only, we get the following expression for the interaction constant $J_{kl}$
(see Eq.(\ref{1:interaction}))
\be \label{1:2.6}
\frac{J_{kl}}{T}=\ln\left[1-(A^2+A^{-2})A^{-2b_{kl}}\right]=
\ln [-A^{-4b_{kl}}]
\ee

Combining Eqs.(\ref{1:2.4})-(\ref{1:2.6}) we obtain the following
statement.

{\it
{\rm (a)} Take $N$-vertex knot diagram on the lattice ${\cal M}$ with
given boundary conditions and fixed set of passages $\{b_i\}$. {\rm (b)}
Take the dual lattice ${\cal L}$ in one-to-one correspondence with ${\cal
M}$ where one vertex of ${\cal M}$ belongs to one edge of ${\cal L}$.

The Kauffman topological invariant $\left<K(A)\right>$ of regular isotopy
for knot diagrams on ${\cal M}$ admits representation in form of {\rm 2D}
Potts system on the dual lattice ${\cal L}$:
\be \label{1:kauff_potts}
\left<K(A)\right>=H\bigl(A,\{b_{kl}\}\bigr)\;
Z_{potts}\bigl[q(A),\left\{J_{kl}\bigl(b_{kl},A\bigr)\right\}\bigr]
\ee
where:
\be
H\bigl(A,\{b_{kl}\}\bigr)=\left(A^2+A^{-2}\right)^{-(N+1)}
\exp\left(\ln A\sum_{\{kl\}}b_{kl}\right)
\ee
is the trivial multiplier {\rm (}$H$ does not depend on Potts spins{\rm
)};
\be
Z_{potts}\bigl[q(A),\left\{J_{kl}\bigl(b_{kl},A\bigr)\right\}\bigr]=
\sum_{\{{\bf \sigma}\}}\exp\left\{\sum_{\{kl\}}
\frac{J_{kl}\bigl(b_{kl},A\bigr)}{T}\delta({\bf \sigma}_k,{\bf
\sigma}_l)\right\}
\ee
is the Potts partition function with interaction constants, $J_{kl}$, and
number of spin states, $q$, defined as follows
\be
\frac{J_{kl}}{T}=\ln[-A^{-4b_{kl}}];\qquad q=\left(A^2+A^{-2}\right)^2
\ee
and the variables $b_{kl}$ play a role of disorder on edges of the lattice
${\cal L}$ dual to the lattice ${\cal M}$. The connection between $b_{kl}$
and $b_i$ is defined by convention
\be \label{1:2.1}
b_{kl}=\left\{\begin{array}{ll} -b_i\quad & \mbox{if $(kl)$-edge is
vertical} \medskip \\ b_i \quad & \mbox{if $(kl)$-edge is horizontal}
\end{array}\right.
\ee
}

Eq.(\ref{1:2.4a}) has the sense of partition function of the 2D disordered
Potts system with the random nearest-neighbor interactions whose
distribution remains arbitrary. The set of passages $\{b_{kl}\}$ uniquely
determines the actual topological state of the woven carpet for the
definite boundary conditions. Therefore the topological problem of the knot
invariant determination is reduced to usual statistical problem of
calculation of the partition function of the Potts model with the disorder
in the interaction constants. Of course, this correspondence is still
rather formal because the polynomial variable $A$ is absolutely arbitrary
and can take even complex values, but for some regions of $A$ that
thermodynamic analogy makes sense and could be useful as we shall see
below.

The specific feature of the Potts partition function which gives the
representation of the Kauffman algebraic invariant is connected with the
existence of the relation between the temperature $T$ and the number of
spin states $q$ (see Eq.(\ref{1:2.5})) according to which $T$ and $q$
cannot be considered anymore as independent variables.

\subsubsection{Algebraic invariants of ambient isotopy}

The invariance of the algebraic topological invariant, $f[K]$, with
respect to all Reidemeister moves (see Eq.(\ref{1:ambient})) for our system
shown in the fig.\ref{1:fig:6} is related to the oriented Eulerian circuits
called {\it Hamiltonian walks} \footnote{A Hamiltonian walk is a closed
path which visits once and only once all vertices of the given {\it
oriented} graph.}.

Let us suppose that the orientation of the knot diagram shown in
fig.\ref{1:fig:6} is chosen according to the natural orientation of the
path representing a knot $K$ in ${\R}^3$. For the defined boundary
conditions we get the so-called {\it Manhattan lattice} consisting of woven
threads with alternating directions.

It follows from the definition of twisting $Tw(K)$ (see the Section 1.3.2)
that $Tw(K)$ changes the sign if the direction of one arrow in the vertex is
changed to the inverse. Reversing the direction of any arrows in the given
vertex even times we return the sign of twisting to the initial value.

We define groups of ``even" and ``odd" vertices on the lattice ${\cal M}$ as
follows. The vertex $i$ is called {\it even} ({\it odd}) if it belongs to
the horizontal (vertical) bond $(kl)$ of the dual lattice ${\cal L}$. Now
it is easy to prove that the twisting of the knot on the Manhattan lattice
${\cal M}$ can be written in terms of above defined variables $b_{kl}$.
Finally the expression for the algebraic invariant of ambient isotopy
$f[K]$ on the lattice ${\cal L}$ reads
\be \label{1:2.8}
f[K]=\exp\left(3\ln [-A] \sum_{\{kl\}} b_{kl}\right)
\left<K\left(\{b_{kl}\},A\right)\right>
\ee
where $\left<K\left(\{b_{kl}\},A\right)\right>$ is defined by
Eq.(\ref{1:kauff_potts}).

\subsection{Notion about annealed and quenched realizations of topological
disorder}

Fixed topological structure of a trajectory of given length fluctuating in
space is a typical example of a quenched disorder. Actually, the knot
structure is formed during the random closure of the path and cannot be
changed without the path rupture. Because of the topological constraints
the entire phase space of ensemble of randomly generated closed loops is
divided into the separated domains resembling the multi-valley structure of
the spin glass phase space. Every domain corresponds to the sub-space of
the path configurations with the fixed value of the topological invariant.

The methods of theoretical description of the systems with quenched
disorder in interaction constants are rather well developed, especially in
regard to the investigation of {\it spin glass models} \cite{mezard}.
Central for these methods is the concept of {\it self-averaging} which can
be explained as follows. Take some additive function $F$ (the free energy,
for instance) of some disordered spin system. The function $F$ is the
self-averaging quantity if the observed value, $F_{obs}$, of any
macroscopic sample of the system coincides with the value $F_{av}$ averaged
over the ensemble of disorder realizations:
$$
F_{obs} = \left<F\right>_{av}
$$
The central technical problem is in calculation of the free energy $F=-T\ln
Z$ averaged over the randomly distributed quenched pattern in the
interaction constants. In this Section we show that this famous
thermodynamic problem of the spin glass physics is closely related to the
knot entropy computation.

Another problem arises when averaging the partition function $Z$ (but not
the free energy) over the disorder. Such problem is much simpler from
computational point of view and corresponds to the case of {\it annealed}
disorder. Physically such model corresponds to the situation when the
topology of the closed loop can be changed. It means that the topological
invariant, i.e. the Potts partition function, has to be averaged over all
possible realizations of the pattern disorder in the ensemble of open (i.e.
unclosed) loops on the lattice. It has been shown in \cite{grnech2} that
the calculation of the mean values of topological invariants allows to
extract rather rough but nontrivial information about the knot statistics.

\subsubsection{Entropy of knots. Replica methods}

Our main goal is the computation of the probability distribution ${\cal
P}_N\{f[K]\}$ (see Eq.(\ref{1:probab})). Although we are unable to evaluate
this function exactly, the representation of ${\cal P}_N\{f[K]\}$ in terms
of disordered Potts system enable us to give an upper estimation for the
fraction of randomly generated paths belonging to some definite topological
class (in particular, to the trivial one). We use the following chain of
inequalities restricting ourselves with the case of regular isotopic knots
for simplicity (\cite{grnech}):
\be \label{1:2.23}
\fbox{\parbox{3.7cm}{Probability ${\cal P}_N$ of knot formation in a 
given topological state}} 
\le 
\fbox{\parbox{5.2cm}{Probability ${\cal P}_N\{K(A)\}$ of knot 
formation with specific topological invariant
$\left<K(A)\right>$ for {\it all} $A$}} 
\le 
\fbox{\parbox{5.2cm}{Probability ${\cal P}_N\{K(A^{*})\}$ of knot
formation for {\it specific} va\-lu\-e of $A^{*}$ minimizing the free energy of
associated Potts system}}
\ee

The first inequality is due to the fact that Kauffman invariant of regular
isotopic knots is not a complete topological invariant, whereas the last
probability in the chain can be written as follows
\be \label{1:2.24}
{\cal P}_N\{K(A^*)\}=
\sum_{\{b_{kl}\}}\Theta\{b_{kl}\}\Delta\Bigl[\left<K\{b_{kl},A^*\}\right>
-\left<K(A^*)\right>\Bigr]
\ee
where $\sum$ means summation over all possible configurations of the
``crossing field" $\{b_{kl}\}$, $\Delta$-function cuts out all states of
the field $\{b_{kl}\}$ with specific value of Kauffman invariant
$\left<K\{b_{kl},A^*\}\right>\equiv \left<K(A^*)\right>$ and
$\Theta\{b_{kl}\}$ is the probability of realization of given crossings
configuration.

In principle the distribution $\Theta\{b_i\}$ depends on statistics of the
path in underlying 3D space and is determined physically by the process
of the knot formation. Here we restrict ourselves to the following simplest
suppositions:

(i) We regard crossings $\{b_i\}$ in different vertices of ${\cal
M}$-lattice as completely uncorrelated variables (or, in other words, we
assume that the variables $\{b_{kl}\}$ defined on the edges of the ${\cal
L}$-lattice are statistically independent):
\be \label{1:2.10}
\Theta\{b_i\}=\prod_i^N P(b_i)
\ee

(ii) We suppose variable $b_i$ (or $b_{kl}$) to take values $\pm 1$
with equal probabilities, i.e.:
\be \label{1:2.11}
P(b_i)=\frac{1}{2}\:\delta(b_i-1) + \frac{1}{2}\:\delta(b_i+1) \ee

The probability of trivial knot formation can be estimated now as follows
\be \label{1:2.25}
\begin{array}{lll}
\disp {\cal P}_N^{(0)}(A^*) & \le &
\disp \sum_{\{b_{kl}\}}\Theta\{b_{kl}\}
\Delta\Bigl[\ln\left<K\{b_{kl},A^*\}\right>\Bigr] \medskip \\
& \simeq & \disp
\frac{1}{2\pi}\int_{-\infty}^{\infty}dy\int\ldots\int \prod_{kl}
P(b_{kl})db_{kl}\left<K^{iy}\{b_{kl},A^*\}\right>
\end{array}
\ee
where $\left<K(A^*)\right>\equiv 1$ for trivial knots.

Thus our problem is reduced to the calculation of non-integer complex
moments of the partition function, i.e., values of the type
$\overline{\left<K^{iy}\{b_{kl},A^*\}\right>}$.  An analogous problem of
evaluation of non-integer moments is well known in the spin-glass theory.
Indeed, the averaging of the free energy of the system, $\overline{F}$,
over quenched random field is widely performed via so-called {\it
replica-trick} \cite{edw_an}. The idea of the method is as follows.
Consider the identity $Z^n\equiv e^{n\ln  Z}$ and expand the right-hand
side up to the first order in $n$. We get $Z^n=1+n\ln Z+O(n^2)$. Now we can
write
$$
F=-\ln Z=-\lim_{n\to 0}\frac{Z^n-1}{n}
$$

We proceed with the calculation of the complex moments of the partition
function $\left<K\{b_{kl}\}\right>$. In other words we would like to find
the averaged value $\overline{\left<K^n\right>}$ for integer values of $n$.
Then we put $n=iy$ and compute the remaining integral in Eq.(\ref{1:2.25})
over $y$-value. Of course, this procedure needs to be verified and it would
be of most desire to compare our predictions with the data obtained in
numerical simulations. However let us stress that our approach is no more
curious than replica one, it would be extremely desirable to test the
results obtained by means of computer simulations.

The outline of our calculations is as follows. We begin by rewriting the
averaged Kauffman invariant using the standard representation of the
replicated Potts partition function and extract the corresponding
free energy $\overline{F(A)}$ in the frameworks of the infinite--range
mean--field theory in two dimensions. Minimizing $\overline{F(A)}$ with
respect to $A$ we find the equilibrium value $A^*$. Then we compute the
desired probability of trivial knot formation ${\cal P}_N^{(0)}(A^*)$
evaluating the remaining Gaussian integrals.

Averaging the $n$th power of Kauffman invariant over independent values of
the ``crossing field" $b_{kl}=\pm 1$ we get
\be \label{1:15}
\begin{array}{lll}
\disp \overline{\left<K^n(A)\right>} & = & \disp
\int\ldots\int\prod_{kl}P(b_{kl})db_{kl}K^{2n}\{b_{kl}\} \medskip \\
& = & \disp
\left[2\cosh(2\beta)\right]^{-2n(N+1)} \medskip \\
& & \disp \hspace{-2cm}\times \sum_{\{{\bf \sigma}\}}
\prod_{kl}\exp\left\{i\pi\sum_{kl}\delta\bigl({\bf\sigma}_k^{\alpha},
{\bf\sigma}_l^{\alpha}\bigr)+\ln\cosh\left[\beta\sum_{\alpha=1}^n\Bigl(
4\delta\bigl({\bf\sigma}_k^{\alpha},{\bf\sigma}_l^{\alpha}\bigr)-1\Bigr)
\right]\right\}
\end{array}
\ee
where $\beta=\ln A$. Let us break for a moment the connection between the
number of spin states, $q$, and interaction constant and suppose
$|\beta|\ll 1$. Later on we shall verify the selfconsistency of this
approximation. Now the exponent in the last expression can be expanded as a
power series in $\beta$. Keeping the terms of order $\beta^2$ only, we
rewrite Eq.(\ref{1:15}) in the standard form of $n$-replica Potts
partition function
\be \label{1:16}
\begin{array}{l}
\disp
\overline{\left<K^n(A)\right>}=\left[2\cosh(2\beta)\right]^{-2n(N+1)}
\exp\left[N\left(\frac{1}{2}\beta^2n^2\right)\right] \medskip \\
\disp\times\sum_{\{{\bf\sigma}_1\ldots{\bf\sigma}_n\}}
\exp\Biggl\{\frac{J^2}{2}\sum_{kl}^N\sum_{\alpha\neq\beta}^n
{\bf\sigma}_{ka}^{\alpha}{\bf\sigma}_{kb}^{\beta}
{\bf\sigma}_{la}^{\alpha}{\bf\sigma}_{lb}^{\beta}
+\left(\frac{J^2}{2}(q-2) + \bar{J}_0\right)
\sum_{kl}^N\sum_{\alpha=1}^n{\bf\sigma}_{ka}^{\alpha}
{\bf\sigma}_{lb}^{\beta}\Biggr\}
\end{array}
\ee
where spin indexes $a,b$ change in the interval $[0,q-1]$, $\beta^2\ll 1$
and
\be \label{1:17}
\begin{array}{rll}
J^2 & = & 16\beta^2 \medskip \\
\bar{J}_0 & = & i\pi - 4\beta^2n \medskip \\
q   & = & 4 + 16\beta^2>4 \end{array}
\ee

According to the results of Cwilich and Kirkpatrick \cite{cw_kir} and later
works (see, for instance, \cite{pari}), the  spin-glass ordering takes
place and the usual ferromagnetic phase makes no essential contribution
to the free energy under the condition
\be \label{1:18}
\frac{\bar{J}_0}{J}<\frac{q-4}{2}
\ee
Substituting Eq.(\ref{1:17}) into Eq.(\ref{1:18}) it can be seen
that $\Re(\mbox{l.h.s.})<\Re(\mbox{r.h.s})$ in Eq.(\ref{1:18}) for all
$\beta$. Thus, we expect that the spin-glass ordering (in the
infinite-range model) corresponds to the solutions
$$
\begin{array}{lll} \disp m_a^{\alpha} & = & \disp
\Bigl<q\delta({\bf\sigma}_k^{\alpha},a)-1\Bigr>=0 \medskip \\
\disp Q_{ab}^{\alpha\beta} & = & \disp
\Bigl<q\delta({\bf\sigma}_k^{\alpha},a)-1\Bigr>
\Bigl<q\delta({\bf\sigma}_k^{\beta},b)-1\Bigr> \neq 0
\end{array}
$$
where $m_a^{\alpha}$ and $Q_{ab}^{\alpha\beta}$ are the ferromagnetic and
spin-glass order parameters respectively. If it is so, we can keep the
term in the exponent (Eq.(\ref{1:16})) corresponding to inter-replica
interactions only.

We follow now the standard scheme of analysis of Potts spin glasses
partition function exhaustively described in \cite{cw_kir,gross,pari}; main
steps of this analysis are shortly represented below. Performing the
Hubbard-Stratonovich transformation to the scalar fields
$Q^{\alpha\beta}_{iab}$ and implying the homogeneous isotropic solution of
the form $Q^{\alpha\beta}_{iab}=Q^{\alpha\beta}_i\delta_{ab}$, we can write
down the value $\overline{\left<K^n\right>}$ (Eq.(\ref{1:16})) as follows
(\cite{cw_kir}):
\be \label{1:28}
\begin{array}{c} \disp
\overline{\left<K^n\right>}=\exp\Biggl\{N\Biggl[\ln\frac{\pi}{J^2}n(n-1)
(q-1)^2-\ln\left(2\cosh{J\over 2}\right)+
\frac{J^2n^2}{32}\Biggr]\Biggr\} \medskip \\
\disp \times \sum_{\{{\bf\sigma}\}}\int\prod_i dQ_i^{\alpha\beta}
\exp\left\{-\int H\{Q_i^{\alpha\beta}\}d^2x\right\}
\end{array}
\ee
where
\be \label{1:29}
\begin{array}{l}
\disp
H\{Q^{\alpha\beta}\}=(q-1)\Biggl[{1\over 4}\left(\frac{2}{J^2}-1\right)
\sum_{\alpha\neq\beta}(Q^{\alpha\beta})^2-{1\over 6}
\sum_{\alpha\neq\beta\neq\gamma}Q^{\alpha\beta}Q^{\beta\gamma}
Q^{\gamma\alpha} \medskip \\ \disp \hspace{2cm}
-{{q-2}\over {12}}\sum_{\alpha\neq\beta}(Q^{\alpha\beta})^3-
{{q-2}\over 4}\sum_{\alpha\neq\beta\neq\gamma}(Q^{\alpha\beta})^2
Q^{\beta\gamma}Q^{\gamma\alpha} \medskip \\ \disp \hspace{2cm}
-{1\over 8} \sum_{\alpha\neq\beta\neq\gamma\neq\delta}Q^{\alpha\beta}
Q^{\beta\gamma}Q^{\gamma\delta}Q^{\delta\alpha}-{{q^2-6q+6}\over {48}}
\sum_{\alpha\beta}(Q^{\alpha\beta})^4\Biggr]
\end{array}
\ee

In \cite{gross,cw_kir} it was shown, that the mean-field replica symmetric
solution of the mean-field Potts spin glass is unstable for $q\ge 2$ and
the right ansatz of Eqs.(\ref{1:28})-(\ref{1:29}) corresponds to the first
level of Parisi replica breaking scheme for spin glasses. Hence, we have
\be \label{1:30}
\disp Q^{\alpha\beta}=\left\{\begin{array}{ll} Q & \mbox{if
$\alpha$ and $\beta$ belong to the same group of $m$ replicas} \medskip \\
0 & \mbox{otherwise}
\end{array}\right.
\ee

Analysis shows that for $q>4$ (our case) the transition to the glassy
state corresponds to $m=1$ which implies the accessory condition
$F_{pm}=F_{sg}$, where $F_{pm}$ and $F_{sg}$ are the free energies of
paramagnetic and spin-glass phases respectively. The transition occurs at
the point
\be \label{1:19}
1-{2\over {J^2}}={{(q-4)^2}\over {3(q^2-18q+42)}}
\ee
Substituting Eq.(\ref{1:17}) into Eq.(\ref{1:19}) we find the
self-consistent value of reverse temperature of a spin-glass transition,
$\beta_{tr}$:
\be \label{1:20}
\beta_{tr}\approx 0.35
\ee
This numerical value is consistent with the condition $\beta^2_{tr}\ll 1$
implied above in the course of expansion of Eq.(\ref{1:16}).

According to the results of the work \cite{cw_kir} the $n$-replica free
energy near the transition point has the following form
\be \label{1:21}
F\simeq\frac{1}{64}Nn(q-1)^2Q_{tr}
\left(\frac{1}{\beta^2}-\frac{1}{\beta^2_{tr}}\right)^2
\ee
with the following expression of the spin-glass order parameter
\be \label{1:21a}
Q_{tr}=\frac{2(4-q)}{q^2-18q+42}>0
\ee

From Eq.(\ref{1:21}) we conclude that the free energy $\overline{F}$
reaches its minimum as a function of $A=\exp{(\beta)}$ just at the point
$A^{*}= \exp(\beta_{tr})$. Using Eqs.(\ref{1:21}) and (\ref{1:21a}) we
rewrite the expression for the averaged $n$-replica Kauffman invariant
$\overline{\left<K^n\right>}$ in the vicinity of $\beta_{tr}$ as follows
(compare to \cite{cw_kir}):
\be \label{1:23}
\begin{array}{l}
\disp \overline{\left<K^{2n}\right>}\simeq
\exp\Biggl\{Nn^2\Biggl[\left(3+16\beta^2\right)^2\ln\frac{\pi}{16\beta^2}+
\frac{\beta^2}{2}\Biggr]-\medskip \\
\disp \hspace{1cm} Nn\Biggl[\left(3+16\beta^2\right)^2
\ln\frac{\pi}{16\beta^2}+\ln 2+\frac{\beta^2}{2} 
-\frac{\left(3+16\beta^2\right)^2
\left(\beta^{-2}-\beta_{tr}^{-2}\right)^2\beta_{tr}^2}
{\left(4+16\beta_{tr}^2\right)^2-18\left(4+16\beta_{tr}^2\right)+42}
\Biggr]\Biggr\}
\end{array}
\ee

Substituting Eq.(\ref{1:23}) into Eq.(\ref{1:2.25}) and bearing in mind,
that $n=iy$, we can easily evaluate the remaining Gaussian integral over
$y$-value and obtain the result for ${\cal P}^{(0)}_N(A)$. As it has been
mentioned above, to get the simplest estimation for probability of trivial
knot formation, we use the last inequality in the chain of equations
(\ref{1:2.23}) corresponding to the choice $A=A^{*}\equiv\exp(\beta_{tr})$:
\be
\label{1:24}
{\cal P}^{(0)}_N(A^{*})\simeq\exp(c\, N); \qquad c\approx 1
\ee

This dependence it is not surprising from the point of view of
statistical mechanics because the value $\eta={\cal P}^{(0)}_N(A^{*})$
is proportional to the free energy of the Potts system. But from the
topological point of view the value $\eta$ has the sense of typical
``complexity' of the knot (see also Section 3). The fact that $\eta$
grows linearly with $N$ means that the maximum of the distribution
function $P(\eta,N)$ is in the region of very ``complex" knots, i.e. knots
far from trivial. This circumstance directly follows from the
non-commutative nature of topological interactions.


\section{Random walks on locally non-commutative groups}
\setcounter{equation}{0}

Recent years have been marked by the emergence of more and more problems
related to the consideration of physical processes on non-commutative
groups. In trying to classify such problems, we distinguish between the
following categories in which the non-commutative origin of phenomena appear
with perfect clarity:

1. Problems connected with the spectral properties of the
Harper--Hof\-stad\-ter equation \cite{harper} dealing with the electron
dynamics on the lattice in a constant magnetic field. We mean primarily the
consideration of groups of magnetic translations and properties of quantum
planes \cite{belliss,zabrodin}.

2. Problems of classical and quantum chaos on hyperbolic manifolds:
spectral properties of dynamical systems and derivation of trace formulae
\cite{gutzw,terr,bog} as well as construction of probability measures for
random walks on modular groups \cite{chass}.

3. Problems giving rise to application of quantum group theory in physics:
deformations of classical abelian objects such as harmonic oscillators
\cite{oscill} and standard random walks \cite{majid}.

4. Problems of knot theory and statistical topology: construction of
nonabelian topological invariants \cite{jones,kauff_bz}, consideration of
probabilistic behavior of the words on the simplest non-commutative groups
related to topology (such as braid groups) \cite{nech}, statistical
properties of "anyonic" systems \cite{anyon}.

5. Classical problems of random matrix and random operator theory and
localization phenomena: determination of Lyapunov exponents for products of
random non-commutative matrices \cite{fuerst_tut,nesi1,nesi2}, study of the
spectral properties and calculation of the density of states of large
random matrices \cite{lifshits,yan}.

Certainly, such a division of problems into these categories is
very speculative and reflects to a marked degree the authors' personal
point of view. However, we believe that the enumerated items reflect, at
least partially, the currently growing interest in theoretical physics of
the ideas of non-commutative analysis. Let us stress that we do not touch
upon the pure mathematical aspects of non-commutative analysis in this paper
and the problems discussed in the present work mainly concern the points 4
and 5 of the list above.

In the present Section we continue analyzing the statistical problems in
knot theory, but our attention is paid to some more delicate matters
related to investigation of correlations in knotted random paths caused by
the topological constraints. The methods elaborated in Section 2 allow us
to discuss these questions but we find it more reasonable to take a look at
the problems of knot entropy estimation in terms of conventional random
matrix theory. We believe that many non-trivial properties of the knot
entropy problem can be clearly explained in context of the limit behavior
of random walks over the elements of some non-commutative (hyperbolic)
groups \cite{nevegr}.

Another reason which forces us to consider the limit distributions (and
conditional limit distributions) of Markov chains on {\it locally
non-commutative discrete groups} is due to the fact that this class of
problems could be regarded as the first step in a consistent harmonic
analysis on the multiconnected manifolds (like Teichm\"uller space); see
also the Section 4.

\subsection{Brownian bridges on simplest non-commutative groups and
knot statistics}

As it follows from the said above the problems dealing with the
investigation of the limit distributions of random walks on non-commutative
groups is not a new subject in the probability theory and statistical
physics.

However in the context of ``topologically-probabilistic" consideration the
problems dealing with distributions of non-commutative random walks are
practically out of discussion, except for very few special cases
\cite{nesi2,2:helf,khne}. Particularly, in these works
it has been shown that statistics of random walks with the fixed
topological state with respect to the regular array of obstacles on the
plane can be obtained from the limit distribution of the so-called
``Brownian bridges" (see the definition below) on the universal
covering---the graph with the topology of Cayley tree. The analytic
construction of nonabelian topological invariant for the trajectories on
the double punctured plane and statistics of simplest nontrivial random
braid $B_3$ was shortly discussed in \cite{never}.

Below we calculate the conditional limit distributions of the Brownian
bridges on the braid group $B_3$ and derive the limit distribution
of powers of Alexander polynomial of knots generated by random
$B_3$-braids. We also discuss the limit distribution of random walks on
locally free groups and express some conjectures about statistics of random
walks on the group $B_n$. More extended discussion of the results
concerning the statistics of Markov chains on the braid and locally free
groups one can find in \cite{2:debne,2:comtet,2:debne2}.

\subsubsection{Basic definitions and statistical model}

The braid group $B_n$ of $n$ strings has $n-1$ generators
$\{\sigma_1,\sigma_2,\ldots,\sigma_{n-1}\}$ with the following
relations:
\be \label{2:1}
\begin{array}{ll} \medskip
\sigma_i\sigma_{i+1}\sigma_i = \sigma_{i+1}\sigma_i\sigma_{i+1}
& \qquad (1\le i<n-1) \\
\sigma_i\sigma_j=\sigma_j\sigma_i & \qquad (|i-j|\ge 2) \medskip \\
\sigma_i\sigma_i^{-1}=\sigma_i^{-1}\sigma_i=e &
\end{array}
\ee

Any arbitrary word written in terms of ``letters"---generators
from the set $\{\sigma_1,\ldots, \break \sigma_{n-1},\sigma_1^{-1},\ldots,
\sigma_{n-1}^{-1}\}$---gives a particular {\it braid}. The geometrical
interpretation of braid generators is shown below:
\bigskip

\unitlength=0.8mm
\special{em:linewidth 0.4pt}
\linethickness{0.4pt}
\hspace{3cm}
\begin{picture}(100.00,50.00)
\put(5.00,50.00){\line(0,-1){15.00}}
\put(30.00,50.00){\line(1,-1){5.00}}
\put(40.00,40.00){\line(1,-1){5.00}}
\put(45.00,50.00){\line(-1,-1){15.00}}
\put(-10.00,50.00){\line(0,-1){15.00}}
\put(70.00,50.00){\line(0,-1){15.00}}
\put(85.00,50.00){\line(0,-1){15.00}}
\put(5.00,17.00){\line(0,-1){15.00}}
\put(-10.00,17.00){\line(0,-1){15.00}}
\put(70.00,17.00){\line(0,-1){15.00}}
\put(85.00,17.00){\line(0,-1){15.00}}
\put(30.00,17.00){\line(1,-1){15.00}}
\put(45.00,17.00){\line(-1,-1){5.00}}
\put(35.00,7.00){\line(-1,-1){5.00}}
\put(-10.00,28.00){\makebox(0,0)[cc]{$1$}}
\put(5.00,28.00){\makebox(0,0)[cc]{$2$}}
\put(17.00,35.00){\makebox(0,0)[cc]{$\ldots$}}
\put(17.00,28.00){\makebox(0,0)[cc]{$\ldots$}}
\put(30.00,28.00){\makebox(0,0)[cc]{$i$}}
\put(45.00,28.00){\makebox(0,0)[cc]{$i+1$}}
\put(57.00,35.00){\makebox(0,0)[cc]{$\ldots$}}
\put(57.00,28.00){\makebox(0,0)[cc]{$\ldots$}}
\put(70.00,28.00){\makebox(0,0)[cc]{$n-1$}}
\put(85.00,28.00){\makebox(0,0)[cc]{$n$}}
\put(100.00,40.00){\makebox(0,0)[ll]{$=\sigma_i$}}
\put(-10.00,-5.00){\makebox(0,0)[cc]{$1$}}
\put(5.00,-5.00){\makebox(0,0)[cc]{$2$}}
\put(17.00,2.00){\makebox(0,0)[cc]{$\ldots$}}
\put(17.00,-5.00){\makebox(0,0)[cc]{$\ldots$}}
\put(30.00,-5.00){\makebox(0,0)[cc]{$i$}}
\put(45.00,-5.00){\makebox(0,0)[cc]{$i+1$}}
\put(57.00,2.00){\makebox(0,0)[cc]{$\ldots$}}
\put(57.00,-5.00){\makebox(0,0)[cc]{$\ldots$}}
\put(70.00,-5.00){\makebox(0,0)[cc]{$n-1$}}
\put(85.00,-5.00){\makebox(0,0)[cc]{$n$}}
\put(100.00,7.00){\makebox(0,0)[ll]{$=\sigma_i^{-1}$}}
\end{picture}
\vspace{0.3in}

The {\it length} of the braid is the total number of the used
letters, while the {\it minimal irreducible length} hereafter referred to as
the ``primitive word" is the shortest noncontractible length of a particular
braid which remains after applying all possible group relations
Eq.(\ref{2:1}). Diagrammatically the braid can be represented as a set of
crossed strings going from the top to the bottom appeared after subsequent
gluing the braid generators.

The closed braid appears after gluing the ``upper" and the ``lower"
free ends of the braid on the cylinder.

Any braid corresponds to some knot or link. So, it is feasible
principal possibility to use the braid group representation for the
construction of topological invariants of knots and links. However the
correspondence between braids and knots is not mutually single valued and
each knot or link can be represented by infinite series of different
braids. This fact should be taken into account in course of knot invariant
construction.

Take a knot diagram $K$ in general position on the plane. Let $f[K]$ be the
topological invariant of the knot $K$. One of the ways to construct the
knot invariant using the braid group representation is as follows.

\begin{figure}
\centerline{\epsfig{file=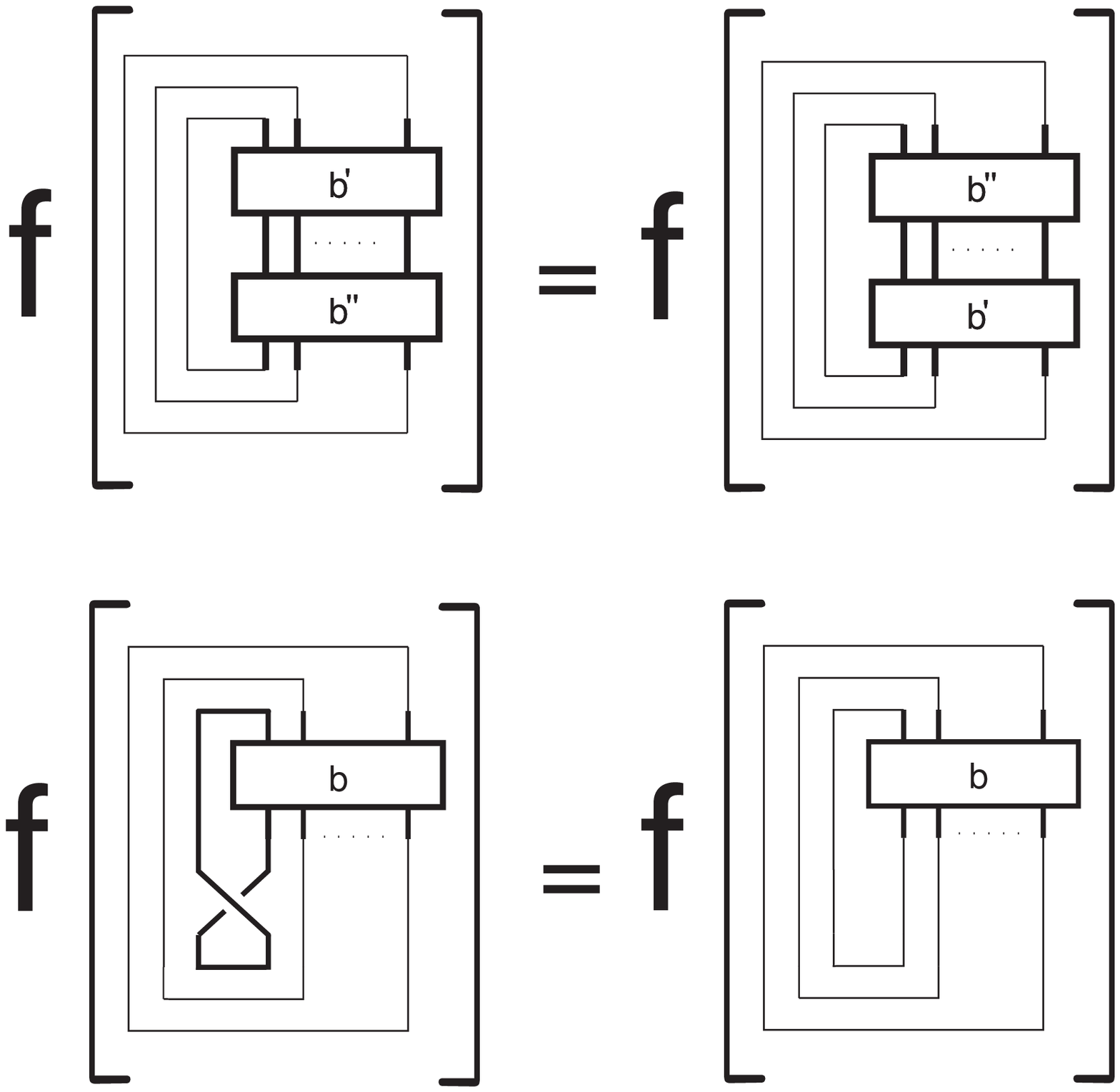,width=10cm}}
\caption{Geometric representation of Eqs.(\ref{2:2}).}
\label{2:fig:braid2}
\end{figure}

1. Represent the knot by some braid $b\in B_n$. Take the function $f$
$$
f: \; B_n \to {\sf C}
$$
Demand $f$ to take the same value for all braids $b$ representing the
given knot $K$. That condition is established in the well-known
Markov-Birman theorem (see, for instance, \cite{2:jones1}):

{\it
The function $f_K\{b\}$ defined on the braid $b\in B_n$ is the topological
invariant of a knot or link if and only if it satisfies the following
``Markov condition":
\be \label{2:2}
\begin{array}{lc} \medskip
f_K\{b'\,b''\}=f_K\{b''\,b'\} & \\
f_K\{b'\,\sigma_n\}=f_K\{\sigma_n\, b'\}=f_K\{b'\} & \quad b',b'' \in B_n
\end{array}
\ee
where $b'$ and $b''$ are two subsequent sub-words in the braid --- {\rm see
fig.\ref{2:fig:braid2}}.
}

2. Now the invariant $f_K\{b\}$ can be constructed using the linear
functional $\varphi\{b\}$ defined on the braid group and called {\it Markov
trace}. It has the following properties
\be \label{2:3}
\begin{array}{l}
\varphi\{b'\,b''\}=\varphi\{b''\,b'\} \medskip \\
\varphi\{b'\,\sigma_n\}=\tau\varphi\{b'\}  \medskip \\
\varphi\{b'\,\sigma_n^{-1}\}=\bar{\tau}\varphi\{b'\}
\end{array}
\ee
where
\be \label{3a}
\tau=\varphi\{\sigma_i\},\quad \bar{\tau}=
\varphi\{\sigma_i^{-1}\};\qquad i \in [1,n-1]
\ee

The invariant $f_K\{b\}$ of the knot $K$ is connected with the linear
functional $\varphi\{b\}$ defined on the braid $b$ as follows
\be \label{4}
f_K\{b\}= (\tau\bar{\tau})^{-(n-1)/2}
\left(\frac{\bar{\tau}}{\tau}\right)^{\frac{1}{2}\bigl(\#(+)-\#(-)\bigr)}
\varphi\{b\}
\ee
where $\#(+)$ and $\#(-)$ are the numbers of ``positive" and ``negative"
crossings in the given braid correspondingly.

The Alexander algebraic polynomials are the first well-known invariants of
such type. In the beginning of 1980s Jones discovered the new knot
invariants. He used the braid representation ``passed through" the Hecke
algebra relations, where the Hecke algebra, $H_n(t)$, for  $B_n$ satisfies
both braid group relations Eq.(\ref{2:1}) and an additional ``reduction"
relation (see the works \cite{2:jones1,2:veker})
\be \label{2:5}
\sigma_i^2 = (1-t)\sigma_i + t
\ee
Now the trace $\varphi\{b\}=\varphi (t)\{b\}$ can be said to take the value
in the ring of polynomials of one complex variable $t$. Consider the
functional $\varphi (t)$ over the braid $\{b'\,\sigma_i\,b''\}$.
Eq.(\ref{2:5}) allows us to get the recursion (skein) relations for
$\varphi(t)$ and for the invariant $f_K(t)$ (see for details
\cite{2:akde}):
\be
\varphi(t)\{b'\sigma_i b''\} =
(1-t)\varphi(t)\{b'b''\} + t\varphi(t)\{b'\sigma_i^{-1} b''\}
\ee
and
\be \label{2:6}
f_K^{+}(t) - t\left(\frac{\bar{\tau}}{\tau}\right)
f_K^{-}(t) = (1-t)\left(\frac{\bar{\tau}}{\tau}\right)^{1/2} f_K^{0}(t)
\ee
where $f_K^{+}\equiv f\{b'\,\sigma_i \,b''\}$;~ $f_K^{-}\equiv f\{b'\,
\sigma_i^{-1}\,b''\}$;~ $f_K^{0}\equiv f\{b'\,b''\}$ and the fraction
$\disp\frac{\bar{\tau}}{\tau}$ depends on the used representation.

3. The tensor representations of the braid generators can be written as
follows
\be
\sigma_i(u)=\lim_{u\to \infty}\sum_{klmn}R^{km}_{ln}(u)I^{(1)}\otimes
\cdots I^{(i-1)}\otimes E_{nk}^{i}\otimes E_{ml}^{i+1}\otimes
I^{(i+1)}\otimes\cdots I^{(n)}
\ee
where $I^{(i)}$ is the identity matrix acting in the position $i$;
$E_{nk}$ is a matrix with $(E_{nk})_{pq}=\delta_{np}\delta_{kq}$ and
$R^{km}_{ln}$ is the matrix satisfying the Yang-Baxter equation
\be \label{2:yb}
\sum_{abc}R^{bq}_{cr}(v)R^{ap}_{kc}(u+v)R^{ia}_{jb}(u)=
\sum_{abc}R^{ap}_{bq}(u)R^{ia}_{cr}(u+v)R^{jb}_{ka}(v)
\ee

In that scheme both known polynomial invariants (Jones and Alexander) ought
to be considered. In particular, it has been discovered in
\cite{2:kauf_sal,2:akde} that the solutions of Eq.(\ref{2:yb}) associated
with the groups $SU_q(2)$ and $GL(1,1)$ are linked to Jones and Alexander
invariants correspondingly.  To be more specific:

(a) $\disp\frac{\bar{\tau}}{\tau}=t^2$ for Jones invariants,
$f_K(t)\equiv V(t)$. The corresponding skein relations are
\be \label{2:7}
t^{-1} V^{+}(t) - t V^{-}(t) = (t^{-1/2} - t^{1/2})V^{0}(t)
\ee
and

(b) $\disp\frac{\bar{\tau}}{\tau}=t^{-1}$ for Alexander
invariants, $f_K(t) \equiv \nabla(t)$. The corresponding skein
relations\footnote{Let us stress that the standard skein relations for
Alexander polynomials one can obtain from Eq.(\ref{2:7a}) replacing
$t^{1/2}$ by $-t^{1/2}$.} ~are
\be \label{2:7a}
\nabla^{+}(t)-\nabla^{-}(t)=(t^{-1/2} - t^{1/2})\nabla^{0}(t)
\ee

To complete this brief review of construction of polynomial invariants
from the representation of the braid groups it should be mentioned that the
Alexander invariants allow also another useful description
\cite{2:birman}. Write the generators of the braid group in the so-called
Magnus representation
\be \label{2:braid}
\sigma_j\equiv\hat{\sigma}_j=\left(\begin{array}{ccccccccc}
1     & 0    &\cdots             &      &      \\
0     &\ddots&                   &      &      \\
\vdots&      &\framebox{\large A}&      &\vdots\\
&     &                          &\ddots& 0    \\
&     &\cdots                    & 0    & 1
\end{array}\right)\leftarrow j\mbox{th row}; \;
A=\left(\begin{array}{ccc}
1 & 0 & 0 \\ t & -t & 1 \\ 0 & 0 & 1 \end{array} \right)
\ee

Now the Alexander polynomial of the knot represented by the closed braid
$W=\prod_{j=1}^{N}\sigma_{{\alpha}_j}$ of the length $N$ one can
write as follows
\be \label{2:alex}
(1+t+t^2+\ldots+t^{n-1})\,\nabla(t)\{A\}=
\det\left[\prod_{j=1}^{N}\hat{\sigma}_{{\alpha}_j}-e\right]
\ee
where index $j$ runs ``along the braid", i.e. labels the number of used
generators, while the index $\alpha=\{1,\ldots,n-1,n,\dots,2n-2\}$ marks the
set of braid generators (letters) ordered as follows $\{\sigma_1,\ldots,
\sigma_{n-1}, \sigma_1^{-1},\ldots, \sigma_{n-1}^{-1}\}$. In our further
investigations we repeatedly address to that representation.

We are interested in the limit behavior of the knot or link invariants
when the length of the corresponding braid tends to infinity, i.e. when
the braid ``grows". In this case we can rigorously define some topological
characteristics, simpler than the algebraic invariant, which we call {\it
the knot complexity}.

{\it
Call the {\bf knot complexity}, $\eta$, the  power of some algebraic
invariant, $f_K(t)$ {\rm (Alexander, Jones, HOMFLY) (see also
\cite{grnech2})}
\be \label{1:complex}
\eta=\lim_{|t|\rightarrow\infty}\frac{\ln f_K(t)}{\ln |t|}
\ee
}

{\bf Remark.} By definition, the ``knot complexity" takes one and
the same value for rather broad class of topologically different knots
corresponding to algebraic invariants of one and the same power, being from
this point of view weaker topological characteristics than complete
algebraic polynomial. Let us summarize the advantages of knot complexity.

(i) One and the same value of $\eta$ characterizes a narrow class of
``topologically similar" knots which is, however, much broader than the
class represented by the polynomial invariant $X(t)$. This enables us to
introduce the smoothed measures and distribution functions for $\eta$.

(ii) The knot complexity $\eta$ describes correctly (at least from the
physical point of view) the limit cases: $\eta=0$ corresponds to ``weakly
entangled" trajectories whereas $\eta\sim N$ matches the system of
``strongly entangled" paths.

(iii) The knot complexity keeps all nonabelian properties of the polynomial
invariants.

(iv) The polynomial invariant can give exhaustive information about the
knot topology. However when dealing with statistics of randomly generated
knots, we frequently look for rougher characteristics of ``topologically
different" knots. A similar problem arises in statistical mechanics when
passing from the microcanonical ensemble to the Gibbs one: we lose some
information about details of particular realization of the system but
acquire smoothness of the measure and are able to apply standard
thermodynamic methods to the system in question.

The main purpose of the present section is the estimation of the limit
probability distribution of $\eta$ for the knots obtained by randomly
generated closed $B_n$-braids of the length $N$. It should be emphasized
that we essentially simplify the general problem ``of knot entropy". Namely,
we introduce an additional requirement that the knot should be represented
by a braid from the group $B_n$ without fail.

We begin the investigation of the probability properties of algebraic knot
invariants by analyzing statistics of the random loops (``Brownian bridges")
on simplest non-commutative groups. Most generally the problem can be
formulated as follows. Take the discrete group ${\cal G}_n$ with a fixed
finite number of generators $\{g_1,\ldots,g_{n-1}\}$. Let $\nu$ be the
uniform distribution on the set $\{g_1,\ldots, g_{n-1},
g_1^{-1},\ldots,g_{n-1}^{-1}\}$.  For convenience we suppose $h_j=g_i$ for
$j=i$ and $h_j=g_i^{-1}$ for $j=i+n-1$; $\nu(h_j)=\frac{1}{2n-2}$ for any
$j$. We construct the (right-hand) side random walk (the random word) on
${\cal G}_n$ with a transition measure $\nu$, i.e. the Markov chain
$\{\xi_n\}$, $\xi_0=e\in {\cal G}_n$ and Prob$(\xi_j=u|\xi_{j-1}=v)=
\nu(v^{-1}u)=\frac{1}{2n-2}$. It means that with the probability
$\frac{1}{2n-2}$ we add the element $h_{\alpha_N}$ to the given word
$h_{N-1}=h_{\alpha_1} h_{\alpha_2}\ldots h_{\alpha_{N-1}}$ from the
right-hand side \footnote {Analogously we can construct the left-hand side
Markov chain.}.

{\it
The random word $W$ formed by $N$ letters taken independently with the
uniform probability distribution $\nu=\frac{1}{2n-2}$ from the set
$\{g_1,\ldots,$ $g_{n-1}, g_1^{-1}, \ldots, g_{n-1}^{-1}\}$ is called
the {\bf Brownian bridge} {\rm (BB)} of length $N$ on the group ${\cal
G}_n$ if the shortest ({\bf primitive}) word of $W$ is identical to the
unity.
}

Two questions require most of our attention:

1. What is the probability distribution $P(N)$ of the Brownian bridge on
the group ${\cal G}_n$.

2. What is the conditional probability distribution $P(k,m|N)$ of the
fact that the sub-word $W'$ consisting of first $m$ letters of the
$N$--letter word $W$ has the primitive path $k$ under the condition that
the whole word $W$ is the Brownian bridge on the group ${\cal G}_n$.
(Hereafter $P(k,m|N)$ is referred to as the conditional distribution for
BB.)

It has been shown in the paper \cite{nesi2} that for the free group the
corresponding problem can be mapped on the investigation of the random
walks on the simply connected tree. Below we represent shortly some results
concerning the limit behavior of the conditional probability distribution
of BB on the Cayley tree. In the case of braids the more complicated group
structure does not allow us to apply the same simple geometrical image
directly. Nevertheless the problem of the limit distribution for the random
walks on $B_n$ can be reduced to the consideration of the random walk on
some graph $C(\Gamma)$. In case of the group $B_3$ we are able to construct
this graph evidently, whereas for the group $B_n$ ($n\ge 4$) we give upper
estimations for the limit distribution of the random walks considering the
statistics of Markov chains on so-called local groups.

\subsubsection{Random process on $PSL(2,{\Z})$, $B_3$ and limit
distribution of powers of Alexander invariant}

We begin with computing the distribution function for the conditional
random process on the simplest nontrivial braid group $B_3$. The group
$B_3$ can be represented by $2\times 2$ matrices. To be specific, the braid
generators $\sigma_1$ and $\sigma_2$ in the Magnus representation
\cite{2:birman} look as follows:

\be \label{2:2.1}
\sigma_1=\left(\begin{array}{cc} -t & 1 \\ 0 & 1 \end{array} \right);\qquad
\sigma_2=\left(\begin{array}{cc} 1 & 0 \\ t & -t
\end{array}\right),
\ee
where $t$ is ``the spectral parameter". It is well known that for $t=-1$ the
matrices $\sigma_1$ and $\sigma_2$ generate the group $PSL(2,\Z)$ in such a
way that the whole group $B_3$ is its central extension with the center
\be \label{2:2.2}
(\sigma_1\sigma_2\sigma_1)^{4\lambda}=
(\sigma_2\sigma_1\sigma_2)^{4\lambda}=
(\sigma_1\sigma_2)^{6\lambda}=(\sigma_2\sigma_1)^{6\lambda}=
\left(\begin{array}{cc} t^{6\lambda} & 0 \\ 0 & t^{6\lambda}
\end{array} \right)
\ee

First restrict ourselves with the examination of the group $PSL(2,\Z)$, for
which we define $\tilde{\sigma}_1=\sigma_1$ and $\tilde{\sigma}_2=\sigma_2$
(at $t=-1$).

The canonical representation of $PSL(2,\Z)$ is given by the unimodular
matrices $S,T$:
\be \label{2:2.3}
S=\left(\begin{array}{cc}
0 & 1 \\ -1 & 0 \end{array}\right);\qquad T=\left(\begin{array}{cc}
1 & 1 \\ 0 & 1 \end{array}\right)
\ee

The braiding relation $\tilde{\sigma}_1\tilde{\sigma}_2\tilde{\sigma}_1=
\tilde{\sigma}_2\tilde{\sigma}_1\tilde{\sigma}_2$ in the
$\{S,T\}$-representation takes the form
\be \label{2:2.4a}
S^2TS^{-2}T^{-1}=1
\ee
in addition we have
\be \label{2:2.4b}
S^4=(ST)^3=1
\ee

This representation is well known and signifies the fact that in terms of
$\{S,T\}$-generators the group $SL(2,\Z)$ is a free product $Z^2\otimes
Z^3$ of two cyclic groups of the 2nd and the 3rd orders correspondingly.

The connection of $\{S,T\}$ and $\{\tilde{\sigma}_1, \tilde{\sigma}_2\}$
is as follows
\be \label{2:2.5}
\begin{array}{ll} \tilde{\sigma}_1=T & (T=\tilde{\sigma}_1) \medskip \\
\tilde{\sigma}_2=T^{-1}ST^{-1} &
(S=\tilde{\sigma}_1\tilde{\sigma}_2\tilde{\sigma}_1) \end{array}
\ee

\begin{figure}
\centerline{\epsfig{file=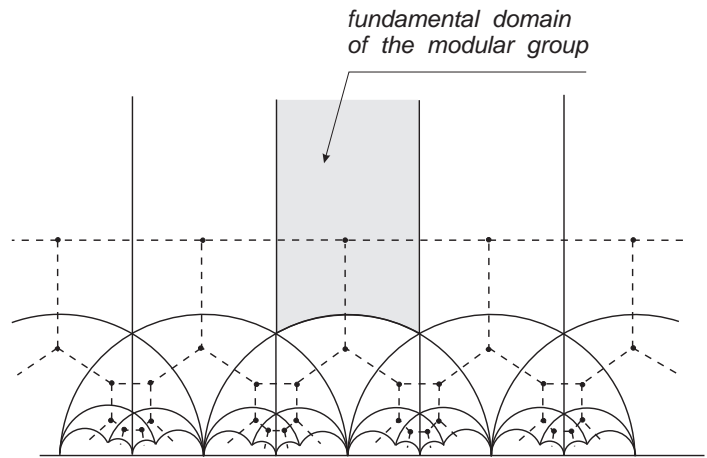,width=10cm}}
\caption{The Riemann surface for the modular group The graph $C(\Gamma)$
representing the topological structure of $PSL(2,~\Z)$ is shown by the
dashed line.}
\label{2:fig:2}
\end{figure}

The modular group $PSL(2,\Z)$ is a discrete subgroup of the group
$PSL(2,\R)$. The fundamental domain of $PSL(2,\Z)$ has the form of a
circular triangle $ABC$ with angles $\left\{0,\frac{\pi}{3},
\frac{\pi}{3}\right\}$ situated in the upper half-plane Im$\zeta>0$ of the
complex plane $\zeta=\xi+i\eta$ (see fig.\ref{2:fig:2} for details).
According to the definition of the fundamental domain, at least one element
of each orbit of $PSL(2,\Z)$ lies inside $ABC$-domain and two elements lie
on the same orbit if and only if they belong to the boundary of the
$ABC$-domain. The group $PSL(2,\Z)$ is completely defined by its basic
substitutions under the action of generators $S$ and $T$:
\be \label{2:2.6}
\begin{array}{l}
S:\quad\zeta\to -1/\zeta \medskip \\
T:\quad\zeta\to\zeta + 1
\end{array}
\ee

\begin{figure}
\centerline{\epsfig{file=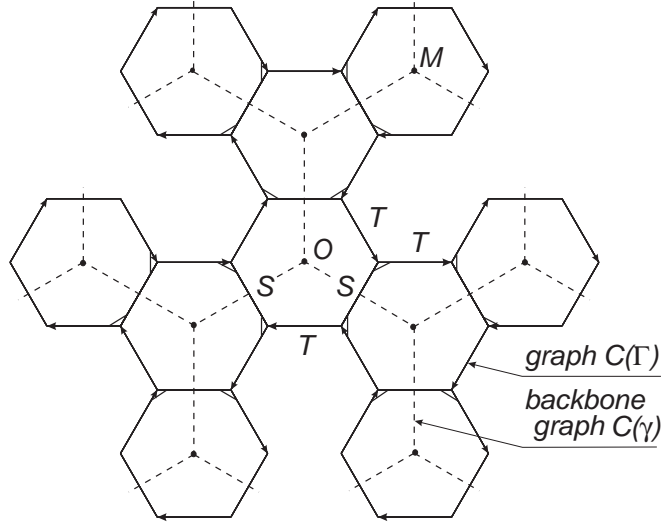,width=9cm}}
\caption{The graph $C(\Gamma)$ and its backbone graph $C(\gamma)$ (see the
explanations in the text).}
\label{2:fig:3}
\end{figure}

Let us choose an arbitrary element $\zeta_0$ from the fundamental domain
and construct a corresponding orbit. In other words, we raise a graph,
$C(\Gamma)$, which connects the neighboring images of the initial element
$\zeta_0$ obtained under successive action of the generators from the set
$\{S,T,S^{-1},T^{-1}\}$ to the element $\zeta_0$.  The corresponding graph
is shown in the fig.\ref{2:fig:2} by the broken line and its topological
structure is clearly reproduced in fig.\ref{2:fig:3}. It can be seen that
although the graph $C(\Gamma)$ does not correspond to the free group and
has local cycles, its ``backbone", $C(\gamma)$, has Cayley tree structure
but with the reduced number of branches as compared to the free group
$C(\Gamma_2)$.

Turn to the problem of limit distribution of a random walk on
the graph $C(\Gamma)$. The walk is determined as follows:

1. Take an initial point (``root") of the random walk on the graph
$C(\Gamma)$. Consider the discrete random jumps over the neighboring
vertices of the graph with the transition probabilities induced by the
uniform distribution $\nu$ on the set of generators $\{\tilde{\sigma}_1,
\tilde{\sigma}_2,\tilde{\sigma}_1^{-1},\tilde{\sigma}_2^{-1}\}$. These
probabilities are (see Eq.(\ref{2:2.5}))
\be \label{2:2.7}
\begin{array}{l} \medskip \mbox{Prob}(\xi_n=T\zeta_0 \;|\;
\xi_{n-1}=\zeta_0)=\frac{1}{4} \\ \medskip
\mbox{Prob}(\xi_n=(T^{-1}ST^{-1})\zeta_0 \;|\;
\xi_{n-1}=\zeta_0)=\frac{1}{4} \\ \medskip
\mbox{Prob}(\xi_n=T^{-1}\zeta_0 \;|\;
\xi_{n-1}=\zeta_0)=\frac{1}{4} \\ \medskip
\mbox{Prob}(\xi_n=(TS^{-1}T)\zeta_0 \;|\;
\xi_{n-1}=\zeta_0)=\frac{1}{4}
\end{array}
\ee
The following facts should be taken into account: the elements $S\zeta_0$
and $S^{-1}\zeta_0$ represent one and the same point, i.e. coincide (as it
follows from Eq.(\ref{2:2.6})); the process is Markovian in terms of
the alphabet $\{\tilde{\sigma}_1,\ldots, \tilde{\sigma}_2^{-1}\}$ {\it
only}; the total transition probability is conserved.

2. Define the shortest distance, $k$, along the graph between the root
and terminal points of the random walk. According to its construction, this
distance coincides with the length $|W_{\{S,T\}}|$ of the minimal
irreducible word $W_{\{S,T\}}$ written in the alphabet
$\{S,T,S^{-1},T^{-1}\}$. The link of the distance, $k$, with the
length $|W_{\{\tilde{\sigma}_1, \tilde{\sigma}_2\}}|$ of the minimal
irreducible word $W_{\{\tilde{\sigma}_1, \tilde{\sigma}_2\}}$ written in
terms of the alphabet $\{\tilde{\sigma}_1,\tilde{\sigma}_2,
\tilde{\sigma}_1^{-1}, \tilde{\sigma}_2^{-1}\}$ is as follows: (a)
$|W_{\{\tilde{\sigma}_1, \tilde{\sigma}_2\}}|=0$ if and only if $k=0$;
(b) for $k\gg 1$ the length $|W_{\{\tilde{\sigma}_1, \tilde{\sigma}_2\}}|$
has asymptotic: $|W_{\{\tilde{\sigma}_1,\tilde{\sigma}_2\}}|=k+o(k)$.

We define the ``coordinates" of the graph vertices in the following way
(see fig.\ref{2:fig:3}):

(a) We apply the arrows to the bonds of the graph $\Gamma$ corresponding to
$T$-generators. The step towards (backwards) the arrow means the
application of $T$ ($T^{-1}$).

(b) We characterize each elementary cell of the graph $\Gamma$ by its
distance, $\mu$, along the graph backbone $\gamma$ from the root cell.

(c) We introduce the variable $\alpha=\{1,2\}$ which numerates the vertices
{\it in each cell} only. We assume that the walker stays in the cell $M$
located at the distance $\mu$ along the backbone from the origin if and
only if it visits one of two in-going vertices of $M$. Such labelling gives
unique coding of the whole graph $C(\Gamma)$.

Define the probability $U_{\alpha}(\mu,N)$ of the fact that the $N$-step
random walk along the graph $C(\Gamma)$ starting from the root point is
ends in $\alpha$-vertex of the cell on the distance of $\mu$ steps along
the backbone. It should be emphasized that $U_{\alpha}(\mu,N)$ is the
probability to stay in {\it any} of ${\cal N}_{\gamma}(\mu)=3\cdot
2^{\mu-1}$ cells situated at the distance $\mu$ along the backbone.

It is possible to write the closed system of recursion relations for the
functions $U_{\alpha}(\mu,N)$. However, here we attend to rougher
characteristics of random walk. Namely, we calculate the ``integral"
probability distribution of the fact that the trajectory of the random walk
starting from an arbitrary vertex of the root cell $O$ has ended in an
arbitrary vertex point of the cell $M$ situated on the distance $\mu$ along
the graph backbone. This probability, $U(\mu,N)$, reads
$$
U(\mu,N)=\frac{1}{2}\sum_{\alpha=\{1,2\}}U_{\alpha}(\mu,N)
$$
The relation between the distances $k$, along the graph $\Gamma$, and $\mu$
along its backbone $\gamma$ is such: $k=\mu+o(\mu)$ for $\mu\gg 1$, what
ultimately follows from the constructions of the graphs $C(\Gamma)$ and
$C(\gamma)$.

Suppose the walker stays in the vertex $\alpha$ of the cell $M$
located at the distance $\mu>1$ from the origin along the graph backbone
$C(\gamma)$. The change in $\mu$ after making of one arbitrary step from
the set $\{\tilde{\sigma}_1, \tilde{\sigma}_2, \tilde{\sigma}_1^{-1},
\tilde{\sigma}_2^{-1}\}$ is summarized in the following table:
\bigskip

\begin{center}
{\renewcommand{\arraystretch}{1.3}
\begin{tabular}{||l|l||l|l||} \hline
\multicolumn{2}{||c||}{$\alpha=1$} &
\multicolumn{2}{c||}{$\alpha=2$} \\ \hline\hline
$\tilde{\sigma}_1=T$ & $\mu\rightarrow\mu+1$ &
$\tilde{\sigma}_1=T$ & $\mu\rightarrow\mu-1$ \\ \hline
$\tilde{\sigma}_2=T^{-1}ST^{-1}$ & $\mu\rightarrow\mu$ &
$\tilde{\sigma}_2=T^{-1}ST^{-1}$ & $\mu\rightarrow\mu+1$ \\ \hline
$\tilde{\sigma}_1^{-1}=T^{-1}$ & $\mu\rightarrow\mu-1$ &
$\tilde{\sigma}_1^{-1}=T^{-1}$   & $\mu\rightarrow\mu+1$ \\ \hline
$\tilde{\sigma}_2^{-1}=TS^{-1}T$ & $\mu\rightarrow\mu+1$ &
$\tilde{\sigma}_2^{-1}=TS^{-1}T$ & $\mu\rightarrow\mu$  \\ \hline
\end{tabular}
}
\end{center}
\bigskip

It is clear that for any value of $\alpha$ two steps increase the length of
the backbone, $\mu$, one step decreases it and one step leaves $\mu$
without changes.

Let us introduce the effective probabilities: $p_1$ -- to jump to some
specific cell among 3 neighboring ones of the graph $C(\Gamma)$ and $p_2$
-- to stay in the given cell. Because of the symmetry of the graph, the
conservation law has to be written as $3p_1+p_2=1$. By definition we have:
$p_1=\nu=\frac{1}{4}$. Thus we can write the following
set of recursion relations for the integral probability $U(\mu,N)$
\be \label{2:2.11}
\begin{array}{ll}
U(\mu,N+1)=\frac{1}{4}U(\mu+1,N)+\frac{1}{4}U(\mu,N)+
\frac{1}{2}U(\mu-1,N) & (\mu\ge 2) \medskip \\
U(\mu,N+1)=\frac{1}{4}U(\mu+1,N)+\frac{1}{2}U(\mu,N)
& (\mu=1) \medskip \\ U(\mu,N=0)=\delta_{\mu,1} &
\end{array}
\ee
The solution of Eq.(\ref{2:2.11}) gives the limit distribution for the
random walk on the group $PLS(2,\Z)$.

{\it
The probability distribution $U(k,N)$ of the fact that the randomly
generated $N$-letter word $W_{\{\tilde{\sigma}_1, \tilde{\sigma}_2\}}$ with
the uniform distribution $\nu=\frac{1}{4}$ over the generators
$\{\tilde{\sigma}_1,\tilde{\sigma}_2, \tilde{\sigma}_1^{-1},
\tilde{\sigma}_2^{-1}\}$ can be contracted to the minimal irreducible word
of length $k$, has the following limit behavior
\be \label{2:2.10}
U(k,N)\simeq\frac{h}{\sqrt{\pi}(4-h)}
\left(\frac{h}{4(h-2)}\right)^N
\left\{\begin{array}{lc} \disp \frac{1}{N^{3/2}} & k=0 \medskip \\
\disp \frac{k}{N^{3/2}}2^{k/2}\exp\left(-\frac{k^2 h}{4N}\right)
& 1\ll k \end{array} \right.
\ee
where $\disp h=2+\frac{\sqrt{2}}{2}$.
}

\begin{corollary}
The probability distribution $U(k,m|N)$ of the fact that in the randomly
generated $N$-letter {\bf trivial} word in the alphabet
$\{\tilde{\sigma}_1,\tilde{\sigma}_2, \tilde{\sigma}_1^{-1},
\tilde{\sigma}_2^{-1}\}$ the sub-word of first $m$ letters has a minimal
irreducible length $k$ reads \be \label{2:2.32} U(k,m|N)=
\frac{h}{\sqrt{\pi}(4-h)} \frac{k^2}{(m(N-m))^{3/2}}\exp \left\{\frac{k^2
h}{4}\left(\frac{1}{m}+\frac{1}{N-m}\right)\right\} \ee
\end{corollary}

Actually, the conditional probability distribution $U(\mu,m|N)$ that the
random walk on the backbone graph, $C(\gamma)$, starting in the
origin, visits after first $m$ ($\frac{m}{N}={\rm const}$) steps some graph
vertex situated at the distance $\mu$ and after $N$ steps returns to the
origin, is determined as follows
\be \label{2:2.23}
U(\mu,m|N)=\frac{U(\mu,m)U(\mu,N-m)}{U(\mu=0,N) {\cal N}_{\gamma}(\mu)}
\ee
where ${\cal N}_{\gamma}=3\cdot 2^{\mu-1})$ and $U(\mu,N)$ is given by
(\ref{2:2.10}).

The problem considered above helps us in calculating the conditional
distribution function for the powers of Alexander polynomial invariants of
knots produced by randomly generated closed braids from the group $B_3$.

\begin{figure}
\centerline{\epsfig{file=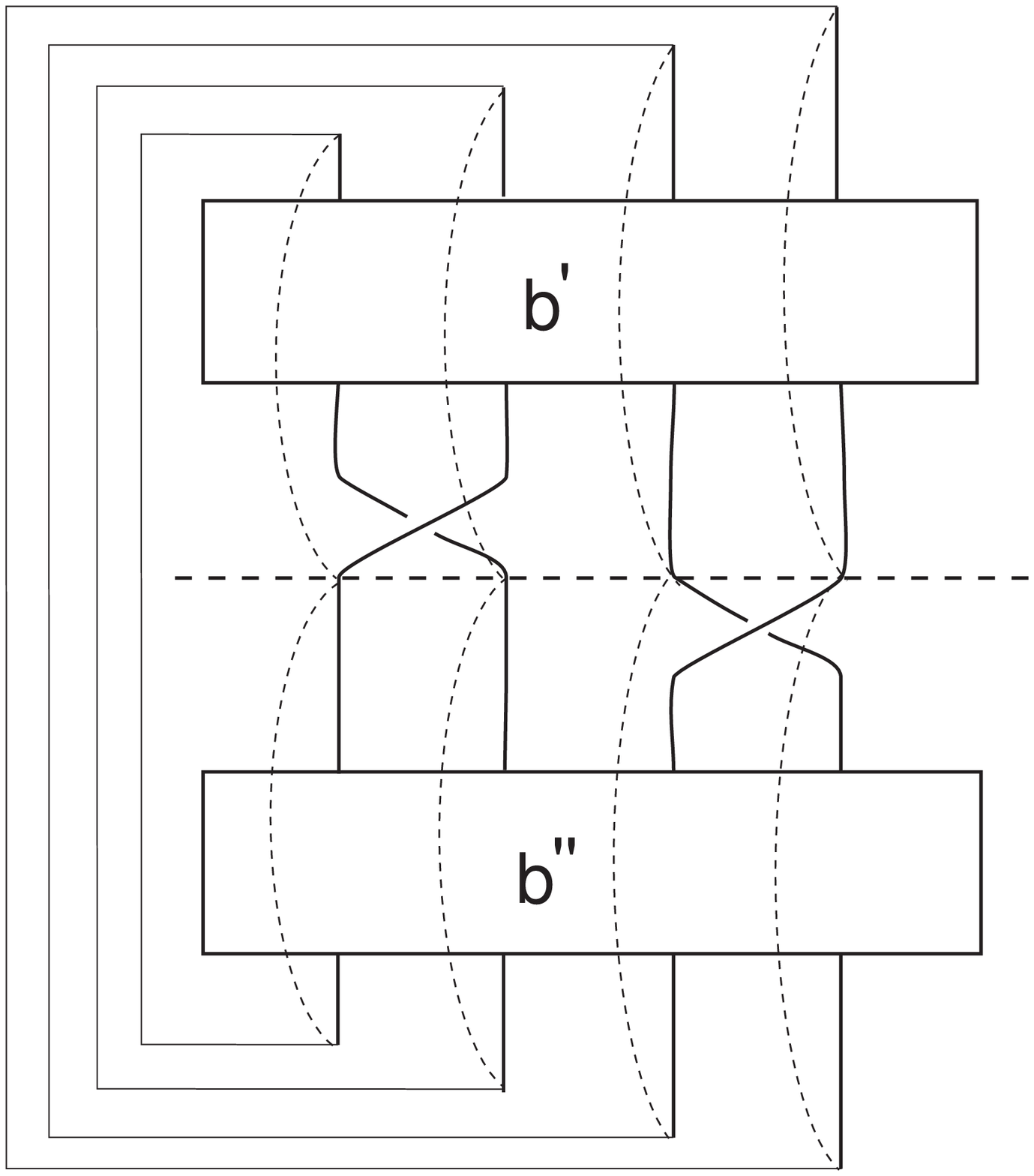,width=6cm}}
\caption{Construction of Brownian bridge for knots.}
\label{2:fig:closure}
\end{figure}

Generally the closure of an arbitrary braid $b\in B_n$ of the total length $N$ 
gives the knot (link) $K$. Split the braid $b$ in two parts $b'$ and $b''$ with
the corresponding lengths $m$ and $N-m$ and make the ``phantom closure" of
the sub-braids $b'$ and $b''$ as it is shown in fig.\ref{2:fig:closure}. The
phantomly closed sub-braids $b'$ and $b''$ correspond to the set of
phantomly closed parts (``sub-knots") of the knot (link) $K$. The next
question is what the conditional probability to find these sub-knots in the
state characterized by the complexity $\eta$ when the knot (link) $K$ as a
whole is characterized by the complexity $\eta=0$ (i.e. the topological
state of $K$ ``is close to trivial").

Returning the the group $B_3$, introduce normalized generators as follows
$$
||\sigma_j^{\pm 1}||=(\det \sigma_j^{\pm 1})^{-1} \sigma_j^{\pm 1}.
$$
To neglect the insignificant commutative factor dealing with norm of
matrices $\sigma_1$ and $\sigma_2$. Now we can rewrite the power of
Alexander invariant (Eq.(\ref{2:alex})) in the form
\be
\label{2:maxim} \eta=\left[\#(+)-\#(-)\right]+\overline{\eta}
\ee
where $\#(+)$ and $\#(-)$ are numbers of generators $\sigma_{{\alpha}_j}$
or $\sigma_{{\alpha}_j}^{-1}$ in a given braid and $\overline{\eta}$ is the
power of the normalized matrix product $\prod^{N}_{j=1}||
\sigma_{{\alpha}_j}||$. The condition of Brownian bridge implies $\eta=0$
(i.e. $\#(+)-\#(-)=0$ and $\overline{\eta}=0$).

Write
\be \label{2:nor}
||\sigma_1||=T(t); \qquad ||\sigma_2||=T^{-1}(t)S(t)T^{-1}(t)
\ee
where $T(t)$ and $S(t)$ are the generators of the ``$t$-deformed" group
$PSL_t(2,\Z)$
\be \label{2:gener}
\begin{array}{ccc}
\disp
T(t) & = & \left(\begin{array}{cc} (-t)^{1/2} & 0 \\ 0 & (-t)^{-1/2}
\end{array}\right) \left(\begin{array}{cc} 1 & (-t)^{-1} \\ 0 & 1
\end{array}
\right); \medskip \\
\disp
S(t) & = & \left(\begin{array}{cc} (-t)^{-1/2} & 0 \\ 0 & (-t)^{1/2}
\end{array}\right) \left(\begin{array}{rc} 0 & 1 \\ -1 & 0 \end{array}
\right)\end{array}
\ee
The group $PSL_t(2,\Z)$ preserves the relations of the group $PSL(2,\Z)$
unchanged, i.e., $(T(t)S(t))^3=S^4(t)=T(t)S^2(t)T^{-1}(t)S^{-2}(t)=1$
(compare to Eq.(\ref{2:2.4a})). Hence, if we construct the graph
$C(\Gamma_t)$ for the group $PSL_t(2,\Z)$ connecting the neighboring images
of an arbitrary element from the fundamental domain, we ultimately come to
the conclusion that the graphs $C(\Gamma_t)$ and $C(\Gamma)$
(fig.\ref{2:fig:3}) are topologically equivalent. This is the direct
consequence of the fact that group $B_3$ is the central extension of
$PSL(2,\Z)$. It should be emphasized that the metric properties of the
graphs $C(\Gamma_t)$ and $C(\Gamma)$ differ because of different
embeddings of groups $PSL_t(2,\Z)$ and $PSL(2,\Z)$ into the complex plane.

Thus, the matrix product $\prod^{N}_{j=1}||\sigma_{{\alpha}_j}||$ for the
uniform distribution of braid generators is in one-to-one correspondence
with the $N$-step random walk along the graph $C(\Gamma)$. Its power
coincides with the respective geodesics length along the backbone graph
$C(\gamma)$. Thus we conclude that limit distribution of random walks on
the group $B_3$ in terms of normalized generators (\ref{2:nor}) is given by
Eq.(\ref{2:2.10}) where $k$ should be regarded as the power of the product
$\prod^{N}_{\alpha=1} ||\sigma_{{\alpha}_j}||$. Hence we come to the
following statement.

{\it
Take a set of knots obtained by closure of $B_3$-braids of length $N$ with
the uniform distribution over the generators. The conditional probability
distribution $U(\overline{\eta},m|N)$ for the normalized complexity
$\overline{\eta}$ of Alexander polynomial invariant {\rm (see
(\ref{2:maxim}))} has the Gaussian behavior and is given by {\rm
Eq.(\ref{2:2.32})} where $k=\overline{\eta}$.
}

\subsection{Random walks on locally free groups}

We aim at getting the asymptotic of conditional limit distributions of BB
on the braid group $B_n$. For the case $n>3$ it presents a problem which is
unsolved yet. However we can estimate limit probability distributions of BB
on $B_n$ considering the limit distributions of random walks on the
so-called ``local groups"
(\cite{2:vershik,never,2:debne,2:comtet,2:debne2}).

The group ${\cal LF}_{n+1}(d)$ we call {\bf the locally free} if the
generators, $\{f_1,\ldots, f_n\}$ obey the following commutation relations:

(a) Each pair $(f_j, f_k)$ generates the free subgroup of the
group ${\cal F}_n$ if $|j-k|<d$;

(b) $f_j f_k=f_k f_j$ for  $|j-k|\ge d$.

(Below we restrict ourselves to the case $d=2$ where ${\cal
LF}_{n+1}(2)\equiv {\cal LF}_{n+1}$).

{\it
The limit probability distribution for the $N$-step random walk
{\rm(}$N\gg 1${\rm)} on the group ${\cal F}_{n+1}$ to have the minimal
irreducible length $\mu$ is
\be \label{2:16}
\begin{array}{ll}
\disp {\cal P}(\mu, N)\simeq\frac{\rm const}{N^{3/2}}e^{-N/6}\,
\mu\sinh\mu \exp\left(-\frac{3\mu^2}{2N}\right) & (n=3) \medskip \\ \disp
{\cal P}(\mu,N)\simeq\frac{1}{2\sqrt{14\pi N}}
\exp\left\{-\frac{8}{7N} \left(\mu-\frac{3}{4}N\right)^2\right\}
& (n\gg 1) \end{array}
\ee
}

We propose two independent approaches valid in two different cases:
(1) for $n=3$ and (2) for $n\gg 1$.

(1) The following geometrical image seems useful. Establish the one-to-one
correspondence between the random walk in some $n$-dimensional Hilbert
space ${\cal LH}^n (x_1,\ldots,x_n)$ and the random walk on the group
${\cal LF}_{n+1}$, written in terms of generators $\{f_1,\ldots,
f_n^{-1}\}$. To be more specific, suppose that when a generator, say,
$f_j$, (or $f_j^{-1}$) is added to the given word in ${\cal LF}_n$, the
walker makes one unit step towards (backwards for $f_j^{-1}$) the axis
$[0,x_j[$ in the space ${\cal LH}^n (x_1,\ldots,x_n)$.

Now the relations (a)-(b) of the definition of the locally free group
could be reformulated in terms of metric properties of the space ${\cal
LH}^n$.  Actually, the relation (b) indicates that successive steps along
the axes $[0,x_j[$ and $[0,x_k[$ $(|j-k|\ge 2)$ commute, hence the section
$(x_j,x_k)$ of the space ${\cal LH}^n$ is flat and has the Euclidean metric
$dx_j^2+dx_k^2$. Situation with the random trajectories in the sections
$(x_j,x_{j\pm 1})$ of the Hilbert space ${\cal LH}^n$ appears to be
completely different. Here the steps of the walk obey the free group
relations (a) and the walk itself is mapped to the walk on the Cayley tree.
It is well known that Cayley tree can be uniformly embedded (without
gaps and selfintersections) into the 3-pseudosphere which gives the
representation of the non-Euclidean plane with the constant negative
curvature. Thus, sections $(x_j,x_{j+1})$ have the metric of
Lobachevskii plane which can be written in the form $\frac{1}{x_j^2}
(dx_j^2+dx_{j+1}^2)$.

For the group ${\cal LF}_4$ these arguments result in the following metric
of appropriate space ${\cal LH}^{(3)}$
\be
\label{2:17} ds^2=\frac{dx_1^2+dx_2^2+dx_3^2}{x_2^2}
\ee
Actually, the space section $(x_1,x_3)$ is flat whereas the space sections
$(x_1,x_2)$ and $(x_2,x_3)$ have Lobachevskii plane metric. The
noneuclidean (hyperbolic) distance between two points $M'$ and $M''$ in the
space ${\cal H}^3$ is defined as follows
\be \label{2:lspace}
\disp \cosh\mu(M'M'')=1+\frac{1}{x_2(M')x_2(M'')}
\sum_{i=1}^3\left(x_i(M')-x_i(M'')\right)^2
\ee
where $\{x_1,x_2,x_3\}$ are the euclidean coordinates in the 3D-halfspace
$x_2>0$ and $\mu$ is regarded as geodesics on a 4-pseudosphere
(Lobachevskii space).

Some well known results concerning the limit behavior of random walks in
spaces of constant negative curvature are reviewed in the next Section
where solutions of the diffusion equations in the Lobachevskii plane and
space are given by Eq.(\ref{2:3pseud}) and Eq.(\ref{2:4pseud})
correspondingly.  Thus we can conclude that the distribution function for
random walk in Lobachevskii space ${\cal P}_s(\mu,N)$ defined by
Eqs.(\ref{2:4pseud})-(\ref{2:4area}) gives also the probability for the
$N$-letter random word (written in terms of uniformly distributed
generators on ${\cal F}_4$) to have the primitive word of length $\mu$ (see
Eq.(\ref{2:16})).

(2) For the group ${\cal LF}_{n+1}$ ($n\gg 1)$ we extract the limit
behavior of the distribution function evaluating the probabilities to increase
and to decrease the length of the primitive word if we randomly add one extra
letter to the given word. We follow below the line proposed by J. Desbois
\cite{2:debne,2:debne2}

Let us point out the main steps of our computations:
\begin{itemize}
\item[(a)] We generate randomly (with uniform probability distribution) the
words of lengths $N\in [1000; 20000]$, while the number of generators,
$n$, varies in the interval $[3; 200]$. The number of randomly generated
words is of order of 1000.
\item[(b)] We reduce the given word till the minimal irreducible
(primitive) word. This can be done by using the braid (or locally free)
group relations. The numerical procedure is as follows. First, we try to
push each braid generator in the word as far as possible to the left. Some
reductions can occur after that. Then, we play the same game but in the
opposite direction, pushing each braid generator to the right performing
possible reductions of the word, then---to the left again and so on... If
no reductions occur during two consecutive steps, we stop the process.
\end{itemize}

We compute the following quantities for braid and locally free groups:

\noindent The mean length of the shortest (primitive) word
$\left<\mu\right>$
\be \label{me}
\left<\mu\right>= \frac{\sum\limits_{\mu=0}^{\infty}\mu Z(\mu,N)}
{\sum\limits_{\mu=0}^{\infty} Z(\mu,N)}
\ee
and the variance $\mbox{Var}(\mu)$
\be \label{va}
\mbox{Var}(\mu)\equiv\bigl<\mu^2\bigr>-\bigl<\mu\bigr>^2 =
\frac{\sum\limits_{\mu=0}^{\infty}\mu^2 Z(\mu,N)}
{\sum\limits_{\mu=0}^{\infty} Z(\mu,N)}-\bigl<\mu\bigr>^2
\ee

The results of numerical simulations for the word statistics on braid
($B_n$) and locally free (${\cal LF}_n(d)$) groups are presented in
the Table 1.

{\small
\begin{center}
{\bf Table 1\footnote{\rm The groups ${\cal LF}_n(d)$ are completely free
when $d\ge n-1$.}.}
\bigskip

\renewcommand{\arraystretch}{1.3}

\begin{tabular}{|c||c|c||c|c||c|c||c|c||} \hline
Groups & \multicolumn{2}{|c||}{$B_n$} &
\multicolumn{2}{|c||}{${\cal LF}_n(2)$} &
\multicolumn{2}{|c||}{${\cal LF}_n(3)$} &
\multicolumn{2}{|c||}{${\cal LF}_n(4)$} \\ \hline\hline
& $\frac{\left<\mu\right>}{N}$ & $\frac{\mbox{\scriptsize Var}(\mu)}{N}$ &
$\frac{\left<\mu\right>}{N}$ & $\frac{\mbox{\scriptsize Var}(\mu)}{N}$ &
$\frac{\left<\mu\right>}{N}$ & $\frac{\mbox{\scriptsize Var}(\mu)}{N}$ &
$\frac{\left<\mu\right>}{N}$ & $\frac{\mbox{\scriptsize Var}(\mu)}{N}$
\\ \hline
n=3 & 0.29 & 0.85 & 0.50 & 0.76 & 0.50 & 0.76 & 0.50 & 0.75
\\ \hline
n=5 & 0.49 & 0.77 & 0.60 & 0.63 & 0.71 & 0.48 & 0.75 & 0.46
\\ \hline
n=10 &  0.56  &  0.63  &  0.65 & 0.56 & 0.77 & 0.40 & 0.82 & 0.34
\\ \hline
n=20 &  0.59  &  0.63  &  0.66  &  0.54 & 0.79 & 0.39 & 0.84 & 0.29
\\ \hline
n=50 & 0.61 & 0.61 & 0.67 & 0.56 & 0.80 & 0.38 & 0.85 & 0.27
\\ \hline
n=100 & 0.61 & 0.61 & 0.67 & 0.52 & 0.80 & 0.36 & 0.86 & 0.26
\\ \hline
n=200 & 0.61 & 0.60 & 0.67 & 0.53 & 0.80 & 0.35 & 0.86 & 0.26
\\ \hline
\end{tabular}
\end{center}
}

\renewcommand{\arraystretch}{1.0}

The maximal standard deviations in the Table 1 (and everywhere below) are:
$$
\left\{\begin{array}{ll}
\pm 0.01 & \mbox{for the mean value $\left<\mu\right>/N$} \\
\pm 0.05 & \mbox{for the variance $\mbox{Var}(\mu)/N$}
\end{array}\right.
$$

\subsection{Analytic Results for Random Walks on Locally Free Group} 

Let us estimate now the quantities $\left<\mu\right>/N$ and $\mbox{Var} 
(\mu)/N$ analytically. We present below two different approaches called 
"dynamical" and "statistical". The "dynamical" approach is based on simple 
estimation of the probability to reduce the primitive word by random adding 
one extra letter. The estimate obtained by this method is in very good 
agreement with corresponding numerical simulations. However the 
"statistical" approach dealing with rigorous enumeration of all 
nonequivalent primitive words in the locally free group ${\cal LF}_n(d)$ 
leads to another answer. The rest of this section is devoted to the 
explanation of the abovementioned discrepancy. 

\noindent{\sc Dynamical Consideration}. Under the conditions
\be \label{ineq}
\begin{array}{c}
n\gg 1 \\ N\gg n^2
\end{array}
\ee
we can easily develop the dynamical arguments which are in rather good
agreement with the results of numerical simulations presented above. The
last inequality in (\ref{ineq}) ensures the conditions, sufficient for
finding the limit probability distribution of Markov chains on the groups
of $n$ generators. Actually, the number of letters in the word, $N$ should
be much larger that the number of all possible pairs in the set of $2n$
letters\footnote{The total number of generators is $2n$ because each of $n$
generators has the inverse one.}. Only in this case the corresponding
Markov process has the reliable distribution function. The number of pairs
is of order $4n^2$, so we arrived at the inequality stated in (\ref{ineq}).

Take a randomly generated $N$--letter word $W$. This word is characterized
by the length of the primitive word $W_p$ (recall that $W_p$ is the length
of the word $W$ obtained after all possible contractions allowed by the
structure of the group ${\cal LF}_n(d)$\footnote{Our consideration is valid
for any values of $d$.}).

Let us compute the probability $\pi(d)$ of the fact that the primitive word
$W_p$ will be shortened in one letter after adding of the letter $f_i$
($i\in [1,n]$) to the word $W$ from the right-hand side. It is easy to
understand that the primitive word $W_p$ can be reduced if:
\begin{itemize}
\item[a)] The last letter in the word $W_p$ is just $f_i^{-1}$. The
probability of such event is $\frac{1}{2n}$;
\item[b)] The letter before the last in the word $W_p$ is $f_i^{-1}$ and
the last letter commutes with the letter $f_i$. The probability of such
event is $\frac{1}{2n}\left(1-\frac{4d-2}{2n}\right)$;
\item[c)] The third letter from the right end of the word $W_p$ is
$f_i^{-1}$ and two last letters commute with the letter $f_i$. The
probability of such event is $\frac{1}{2n}
\left(1-\frac{4d-2}{2n}\right)^2$;
\item[d)] ... and so on.
\end{itemize}

Finally we arrive at the following expression for the probability $\pi(d)$:
\be \label{sum}
\pi(d)=\frac{1}{2n}\sum_{l=0}^{\infty} \left(1-\frac{4d-2}{2n}\right)^l =
\frac{1}{4d-2}
\ee

The procedure described above assumes that the letters remaining in the
word $W_p$ are uniformly distributed---as in the initial (nonreduced word
$W$). The absence of "boundary effects" is ensured by the condition
(\ref{ineq}).

Once having the probability $\pi(d)$, we can  write down the master equation
for the probability $P(\mu,N)$ of the fact that in randomly generated
$N$--letter word the primitive path has the length $\mu$
\be \label{walk1}
P(\mu,N+1) = (1-\pi(d))\, P(\mu-1,N)+ \pi(d)\, P(\mu+1,N)
\qquad (\mu \ge 2)
\ee
where the relation between $P(\mu,N)$ and the partition function $Z(\mu,N)$
introduced above is as follows
$$
P(\mu,N)=\frac{Z(\mu,N)}{\sum\limits_{\mu=0}^{\infty} Z(\mu,N)}
$$

The recursion relation (\ref{walk1}) coincides with the equation describing
the random walk on the halfline with the drift from the origin or, what is the
same, with the equation describing the random walk on the simply Cayley tree
with the coordinational number
\be \label{eff}
z_{\rm eff}=\frac{1}{\pi(d)}=4d-2
\ee
Taking into account the last analogy we can complete the Eq.(\ref{walk1}) by
the boundary conditions
\be \label{walk2}
\begin{array}{l}
P(\mu=1,N+1) = P(\mu=0,N) + \pi(d)\, P(\mu=2,N) \\
P(\mu=0,N+1) = \pi P(\mu=1,N) \\
P(\mu,N=0) = \delta_{\mu,0}
\end{array}
\ee
It is noteworthy that these equations are written just for the Cayley tree
with $z_{\rm eff}$ branches. The actual structure of the
graph corresponding to the group ${\cal LF}_n(d)$ is much more complex,
thus Eqs.(\ref{walk2}) should be regarded as an aproximation. However the
exact form of boundary conditions does not influence the asymptotic
solution of Eq.(\ref{walk1}) in vicinity of the maximum of the distribution
function:
\be \label{prob}
P(\mu,N)\simeq\frac{1}{2\sqrt{2\pi (z_{\rm eff}-1)N}}
\exp\left\{-\frac{z_{\rm eff}^2}
{8(z_{\rm eff}-1)N}
\left(\mu-\frac{z_{\rm eff}-2}
{z_{\rm eff}}N\right)^2\right\}
\ee

Thus, we find
\be \label{aver}
\begin{array}{l}
\disp \frac{\left<\mu(d)\right>}{N}\simeq
\frac{z_{\rm eff}-2}{z_{\rm eff}}=
\frac{2d-2}{2d-1} \medskip \\
\disp \frac{\mbox{Var}(\mu,d)}{N}\simeq
\frac{4(z_{\rm eff}-1)}{z_{\rm eff}^2}=
\frac{4d-3}{(2d-1)^2}
\end{array}
\ee
Substituting in Eq.(\ref{aver}) $d=2,3,4$ we get the following numerical
values:
$$
\begin{array}{ll}
\disp \frac{\left<\mu(d)\right>}{N}=\frac{2}{3};\quad
\disp \frac{\mbox{Var}(\mu,d)}{N}=\frac{5}{9} &\quad \mbox{for $d=2$}
\medskip \\
\disp \frac{\left<\mu(d)\right>}{N}=\frac{4}{5};\quad
\disp \frac{\mbox{Var}(\mu,d)}{N}=\frac{9}{25} &\quad \mbox{for $d=3$}
\medskip \\
\disp \frac{\left<\mu(d)\right>}{N}=\frac{6}{7};\quad
\disp \frac{\mbox{Var}(\mu,d)}{N}=\frac{13}{49} &\quad \mbox{for $d=4$}
\end{array}
$$
what is in the excellent agreement with the asymptotic values ($n\gg 1$)
from the Table 1 for the same groups.

Another statistical problem appears when we are interested in the consideration
of the target space of the group ${\cal LF}_{n+1}$, i.e. in the evaluation of
the number of nonequivalent primitive words in the froup ${\cal LF}_{n+1}$ (see
for details \cite{2:comtet}).

Let $V_n(\mu)$ be the number of all nonequivalent primitive words of length
$\mu$ on the group ${\cal LF}_{n+1}$. When $\mu\gg 1$, $V_n(\mu)$ has the
following asymptotic:
\be \label{2:volume}
V_n(\mu)=\mbox{const}\left[1+2\left(3-
\frac{4\pi^2}{n^2}\right)\right]^{\mu}; \qquad n\gg 1
\ee

To get Eq.(\ref{2:volume}) we write each primitive word $W_p$ of length
$\mu$ in the group ${\cal LF}_{n+1}$ in the so-called {\it normal order}
(all $f_{\alpha_i}$ are different) similar to so-called "symbolic dynamics"
used in consideration of chaotic systems
\be \label{2:norm}
W_p=\left(f_{\alpha_1}\right)^{m_1}\left(f_{\alpha_2}\right)^{m_2}\ldots
\left(f_{\alpha_s}\right)^{m_s}
\ee
where $\disp \sum_{i=1}^s |m_i|=\mu\; (m_i\neq 0\; \forall\; i;\; 1\le s\le
\mu$) and sequence of generators $f_{\alpha_i}$ in Eq.(\ref{2:norm}) {\it
for all} $f_{{\alpha}_i}$ satisfies the following local rules:
\begin{itemize}
\item[(i)] If $f_{\alpha_i}=f_1$, then $f_{\alpha_{i+1}}\in
\left\{f_2,f_3,\ldots f_{n-1}\right\}$;
\item[(ii)] If $f_{\alpha_i}=f_k$ ($1<k\le n-1$), then $f_{\alpha_{i+1}}\in
\left\{f_{k-1},f_{k+1},\ldots f_{n-1}\right\}$;
\item[(iii)] If $f_{\alpha_i}=f_n$, then $f_{\alpha_{i+1}}=f_{n-1}$.
\end{itemize}
These local rules prescribe the enumeration of all distinct primitive
words. If the sequence of generators in the primitive word $W_p$ does not
satisfy the rules (i)-(iii), we commute the generators in the word $W_p$ up
the normal order is restored. Hence, the normal order representation
provides us with the unique coding of all nonequivalent primitive words in
the group ${\cal LF}_{n+1}$.

The calculation of the number of distinct primitive words,
$V_n(\mu)$, of the given length $\mu$ is rather straightforward:
\be \label{2:trace}
V_n(\mu)=\sum_{s=1}^{\mu} R(s)\mathop{{\sum}'}_{\{m_1,\ldots,m_s\}}
\Delta\left[\sum_{i=1}^s |m_i|-\mu\right]
\ee
where $R(s)$ is the number of all distinct sequences of $s$ generators
taken from the set $\{f_1,\ldots,f_n\}$ and satisfying the local rules
(i)-(iii) while the second sum gives the number of all possible
representations of the primitive path of length $\mu$ {\it for the fixed
sequence of generators} (``prime" means that the sum runs over all $m_i\neq
0$ for $1\le i\le s$; $\Delta$ is the Kronecker $\Delta$-function).

It should be mentioned that the local rules (i)-(iii) define the
generalized Markov chain with the states given by the $n\times n$
coincidence matrix $\hat{T}_n$ where the rows and columns correspond to the
generators $f_1,\ldots,f_n$:
\medskip

\be \label{matrix}
\hat{T}_n(d)=
{
\renewcommand{\arraystretch}{0}
\begin{tabular}{|c||c|c|c|c|c|c|c|} \hline
\strut \rule{0pt}{12pt} & $f_1$ & $f_2$ & $f_3$ & $f_4$ & $\ldots$ &
$f_{n-1}$ & $f_n$ \\ \hline \rule{0pt}{2pt} & & & & & & & \\ \hline
\strut \rule{0pt}{12pt} $f_1$ & 0 & 1 & 1 & 1 & $\ldots$ & 1 & 1 \\ \hline
\strut \rule{0pt}{12pt} $f_2$ & 1 & 0 & 1 & 1 & $\ldots$ & 1 & 1 \\ \hline
\strut \rule{0pt}{12pt} $f_3$ & 1 & 1 & 0 & 1 & $\ldots$ & 1 & 1 \\ \hline
\strut \rule{0pt}{12pt} $f_4$ & 0 & 1 & 1 & 0 & $\ldots$ & 1 & 1 \\ \hline
\strut $\vdots$ & $\vdots$ & $\vdots$ & $\vdots$ & $\vdots$ &
$\ddots$ & $\vdots$ & $\vdots$ \\ \hline
\strut \rule{0pt}{12pt} $f_{n-1}$ & 0 & 0 & 0 & 0 & $\ldots$ & 0 & 1 \\
\hline
\strut \rule{0pt}{12pt} $f_n$ & 0 & 0 & 0 & 0 & $\ldots$ & 1 & 0 \\ \hline
\end{tabular}
}
\ee
\medskip

The number of all distinct normally ordered {\bf sequences of words} of
length $s$ with allowed commutation relations is given by the following
partition function
\be \label{rs}
R_n(s,d)={\bf v}_{\mbox{\scriptsize in}}\left[\hat{T}_n(d)\right]^s
{\bf v}_{\mbox{\scriptsize out}}
\ee
where
\be \label{vec}
{\bf v}_{\mbox{\scriptsize in}}=(\;\overbrace{1\; 1\; 1\;\ldots\; 1}^{n}\;)
\qquad \mbox{and}\qquad {\bf v}_{\mbox{\scriptsize out}}=
\left.\left(
{\renewcommand{\arraystretch}{0.8} \begin{array}{c}1 \\ 1 \\ 1 \\ \vdots
\\ 1 \end{array}} \right)\right\}n
\ee

Supposing that the main contribution in Eq.(\ref{2:trace}) results from
$s\gg 1$ we take for $R_n(s)$ the following asymptotic expression
\be
R_n(s)\Big|_{s\gg 1}=\left(\lambda_n^{max}\right)^s; \qquad
\lambda_n^{max}=3-\frac{4\pi^2}{n^2}+ O\left(\frac{1}{n^3}\right)
\ee
where
$\lambda_n^{max}$ is the highest eigenvalue of the matrix $\hat{T}_n$
($n\gg 1$).

The remaining sum in Eq.(\ref{2:trace}) is independent of $R(s)$, so its
calculation is trivial:
\be \label{2:perm}
\mathop{{\sum}'}_{\{m_1,\ldots,m_s\}}
\Delta\left[\sum_{i=1}^s |m_i|-\mu\right] =
2^s\frac{(\mu-1)!}{(s-1)!(\mu-s)!}
\ee
Collecting all terms in Eq.(\ref{2:trace}) and evaluating the sum over $s$
we arrive at Eq.(\ref{2:volume}). The value $V_n(\mu,d)$ is growing
exponentially fast with $\mu$ and the "speed" of this grows is clearly
represented by the fraction
\be \label{fraction}
z_{\rm eff}-1=
\frac{V_n(\mu+1)}{V_n(\mu)}\bigg|_{\mu\gg 1}\simeq 7-\frac{8\pi^2}{n^2}
\ee
where $z_{\rm eff}$ is the coordinational number of effective tree
associated with the locally free group.

Thus, the random walk on the group ${\cal LF}_{n+1}$ can be viewed as
follows. Take the {\it free} group $\Gamma_n$ with generators
$\{\tilde{f}_1,\ldots, \tilde{f}_n\}$ where all $\tilde{f}_i$ ($1\le i\le
n$) do not commute. The group $\Gamma_n$ has a structure of $2n$-branching
Cayley tree, $C(\Gamma_n),$ where the number of distinct words of length
$\mu$ is equal to $\tilde{V}_n(\mu)$,
\be \label{2:vfree}
\tilde{V}_n(\mu)=2n(2n-1)^{\mu-1}
\ee
The graph $C({\cal LF}_{n+1})$ corresponding to the group ${\cal LF}_{n+1}$
can be constructed from the graph $C(\Gamma_n)$ in accordance with the
following recursion procedure: (a) Take the root vertex of the graph
$C(\Gamma_n)$ and consider all vertices on the distance $\mu=2$. Identify
those vertices which correspond to the equivalent words in group ${\cal
LF}_{n+1}$; (b) Repeat this procedure taking all vertices at the distance
$\mu=(1,2,\ldots)$ and ``gluing" them at the distance $\mu+2$ according to
the definition of the locally free group. By means of the described
procedure we raise a graph which in average has $z_{\rm eff}-1$ distinct
branches leading to the "next coordinational sphere". Thus this graph
coincides (in average) with $z_{\rm eff}$-branching Cayley tree.

Although the local structure of the graph $C({\cal LF}_{n+1})$ is very
complex, Eq.(\ref{fraction}) enables us to find the asymptotic of the
random walk on the graph $C({\cal LF}_{n+1})$. Once having $z_{\rm eff}$,
we can write down the master equation for the probability ${\cal P}(\mu,N)$
to find the walker at the distance $\mu$ from the origin after $N$ random
steps on the graph $C({\cal LF}_{n+1})$
\be \label{2:walk}
{\cal P}(\mu,N+1) = (1-\frac{1}{z_{\rm eff}} )\, P(\mu-1,N)+
\frac{1}{z_{\rm eff}} \, P(\mu+1,N) \qquad (\mu \ge 2)
\ee
The recursion relation (\ref{2:walk}) coincides with the equation
describing the random walk on the half-line with the drift from the origin.
Taking into account this analogy we can complete the Eq.(\ref{2:walk}) by
the boundary conditions \cite{2:debne}. However the exact form of boundary
conditions does not influence the asymptotic solution of Eq.(\ref{2:walk})
in vicinity of the maximum of the distribution function:
$$
{\cal P}(\mu,N)\simeq\frac{1}{2\sqrt{2\pi (z_{\rm eff}-1)N}}
\exp\left\{-\frac{z_{\rm eff}^2}
{8(z_{\rm eff}-1)N}
\left(\mu-\frac{z_{\rm eff}-2}
{z_{\rm eff}}N\right)^2\right\}
$$
Thus we obtain the desired distribution function (Eq.(\ref{2:16})) for the
primitive word length for the random walk on the group ${\cal LF}_{n+1}$.

{\it
The {\rm Eq.(\ref{2:16})} gives the estimation from below for the limit
distribution of the primitive words on the group $B_n$ for $n\gg 1$.
}

We find further investigation of the random walks on the groups ${\cal
LF}_{n+1}(d)$ for different values of $d$ very perspective. It should give
insight for consideration of random walk statistics on ``partially
commutative groups". Moreover, the set of problems considered there has
deep relation with the spectral theory of random matrices.

\subsection{Brownian bridges on Lobachevskii plane and products of
non-commutative random matrices}

The problem of word enumeration on locally non-commutative group has
evident connection with the statistics of Markov chains on graphs having
the Cayley tree--like structure and, hence, with random walk statistics on
the surfaces of a constant negative curvature. (We stressed once that the
Cayley tree--like graphs are isometrically embedded in the surfaces of a
constant negative curvature).

Recall that the distribution function, $P({\bf r},t)$, for the free random
walk in D-dimensional Euclidean space obeys the standard heat equation:
$$
\frac{\partial}{\partial t}P({\bf r},t) = {\cal D} \Delta P({\bf r},t)
$$
with the diffusion coefficient ${\cal D}=\frac{1}{2D}$ and appropriate
initial and normalization conditions
$$
\begin{array}{c} \disp P({\bf r},t=0)=\delta({\bf r}) \medskip \\ \disp
\int P({\bf r},t) d{\bf r} = 1
\end{array}
$$

Correspondingly, the diffusion equation for the scalar density $P({\bf
q},t)$ of the free random walk on a Riemann manifold reads (see
\cite{2:zinn} for instance)
\be\label{2:lob}
\frac{\partial}{\partial t} P({\bf q},t) = {\cal D}
\frac{1}{\sqrt{g}}\frac{\partial}{\partial q_i} \left(\sqrt{g}
\left(g^{-1}\right)_{ik}\frac{\partial}{\partial q_k}\right) P({\bf q},t)
\ee
where
\be
\begin{array}{c}
\disp P({\bf q},t=0)=\delta({\bf q}) \medskip \\
\disp \int \sqrt{g} P({\bf q},t) d{\bf q} = 1
\end{array}
\ee
and $g_{ik}$ is the metric tensor of the manifold; $g=\det g_{ik}$.

Eq.(\ref{2:lob}) has been subjected to thorough analysis for the manifolds
of the constant negative curvature. Below we reproduce the corresponding
solutions for the best known cases: for 2D-- and 3D--Lobachevskii spaces
(often referred to as 3-- and 4--pseudospheres) labelling them by indices
``$p\,$" and ``$s\,$" for 2D-- and 3D--cases correspondingly.

For the Lobachevskii plane one has
\be
||g_{ik}||=\left|\left| \begin{array}{cc}
1 & 0 \medskip \\ 0 & \sinh^2 \mu \end{array} \right|\right|
\ee
where $\mu$ stands for the geodesics length on 3-pseudosphere. The
corresponding diffusion equation now reads
\be \label{2:l2}
\frac{\partial}{\partial t}P_p(\mu,\varphi,t)=
{\cal D}\left(\frac{\partial^2}{\partial\mu^2}+\coth\mu
\frac{\partial}{\partial\mu} + \frac{1}{\sinh^2\mu}
\frac{\partial^2}{\partial\varphi^2}\right) P_p(\mu,\varphi,t)
\ee

The solution of Eq.(\ref{2:l2}) is believed to have the following form
\be \label{2:3pseud}
\begin{array}{lll}
P_p(\mu,t) & = &
\disp \frac{e^{-\frac{t{\cal D}}{4}}}{4\pi\sqrt{2\pi(t{\cal
D})^3}} \int_{\mu}^{\infty}\frac{\xi \exp\left(-\frac{\xi^2}{4t{\cal
D}}\right)} {\sqrt{\cosh \xi - \cosh \mu}} d\xi \medskip \\
& \simeq & \disp
\frac{e^{-\frac{t{\cal D}}{4}}}{4\pi t{\cal D}}\left(\frac{\mu}
{\sinh\mu}\right)^{1/2} \exp\left(-\frac{\mu^2}{4t{\cal D}}\right)
\end{array}
\ee

For the Lobachevskii space the corresponding metric tensor is
\be \label{2:4metr}
||g_{ik}||=\left|\left| \begin{array}{ccc}
1 & 0 & 0 \medskip \\ 0 & \sinh^2 \mu & 0 \medskip \\
0 & 0 & \sinh^2\mu \sin^2\theta
\end{array} \right|\right|
\ee

Substituting Eq.(\ref{2:lob}) for Eq.(\ref{2:4metr}) we have
\be \label{2:4pseud}
P_s(\mu,t) = \frac{e^{-t{\cal
D}}}{8\pi\sqrt{\pi(t{\cal D})^3}}\frac{\mu}{\sinh\mu}
\exp\left(-\frac{\mu^2}{4t{\cal D}}\right)
\ee
For the first time this spherically symmetric solution of the heat equation
(Eq.(\ref{2:lob})) in the Lobachevskii space was received in
\cite{2:ger_vas}.

In our opinion one fact must be given our attention. The distribution
functions $P_{\rm p}(\mu,t)$ and $P_{\rm s}(\mu,t)$ give the probabilities
to find the random walk starting at the point $\mu=0$ after time $t$ in
some {\it specific} point located at the distance $\mu$ in corresponding
noneuclidean space. The probability to find the terminal point of a random
walk after time $t$ {\it somewhere} at the distance $\mu$ is
\be \label{2:34lob}
{\cal P}_{\rm p,s}(\mu,t)=P_{\rm p,s}(\mu,t) {\cal N}_{\rm p,s}(\mu)
\ee where
\be \label{2:3perim}
{\cal N}_{\rm p}(\mu)=\sinh\mu
\ee
is the perimeter of circle of radius $\mu$ on the Lobachevskii plane
and
\be \label{2:4area}
{\cal N}_{\rm s}(\mu)=\sinh^2\mu
\ee
is the area of sphere of radius $\mu$ in the Lobachevskii space.

The difference between $P_{\rm p,s}$ and ${\cal P}_{\rm p,s}$ is
insignificant in euclidean geometry, whereas in the noneuclidean space it
becomes dramatic because of the consequences of the behavior of Brownian
bridges in spaces on constant negative curvature.

Using the definition of the Brownian bridge, let us calculate the
probabilities to find the $N$-step random walk (starting at $\mu=0$) after
first $t$ steps at the distance $\mu$ in the Lobachevskii plane (space)
under the condition that it returns to the origin on the last step. These
probabilities are ($N\to\infty$)
\be \label{2:gauss}
\begin{array}{l}
\disp {\cal P}_{\rm p}(\mu,t|0,N)=
\frac{P_{\rm p}(\mu,t){\cal P}_{\rm p}(\mu,N-t)}{P_{\rm p}(0,t)}=
\frac{N}{4\pi{\cal D}t(N-t)}\mu \exp\left\{-\frac{\mu^2}
{4{\cal D}}\left(\frac{1}{t}+\frac{1}{N-t}\right)\right\}
\medskip \\
\disp {\cal P}_{\rm s}(\mu,t|0,N)=
\frac{P_{\rm s}(\mu,t){\cal P}_{\rm s}(\mu,N-t)}{P_{\rm s}(0,t)}=
\frac{N^{3/2}}{8\pi t^{3/2}(N-t)^{3/2}}\mu^2 \exp\left\{-\frac{\mu^2}
{4{\cal D}}\left(\frac{1}{t}+\frac{1}{N-t}\right)\right\}
\end{array}
\ee
Hence we come to the standard Gaussian distribution function with zero
mean.

Equations (\ref{2:gauss}) describing the random walk on the Riemann surface
of constant negative curvature have direct application to the conditional
distributions of Lyapunov exponents for products of some non-commutative
matrices. Let us consider the first of Eqs.(\ref{2:gauss}). Changing
the variables $\mu=\ln\frac{1+|z|}{1-|z|};\quad \varphi=\arg z$ where
$z=x+iy;\, \bar{z}=x-iy$ we map the 3--pseudosphere $(\mu,\varphi)$
onto the unit disk $|z|<1$ known as the Poincare representation of the
Lobachevskii plane. The corresponding conformal metric reads $
dl^2=\frac{4\;dzd\bar{z}}{\left(1-|z|^2\right)^2}$. Using the conformal
transform $\disp z=\frac{1+iw}{1-iw}$ we recover the so-called Klein
representation of Lobachevskii plane, where
$dl^2=-\frac{4\;dwd\bar{w}}{\left(w-\bar{w}\right)^2}$
and the model is defined in Im$w>0$ ($w=u+iv;\, \bar{w}=u-iv$).

The following relations can be verified using conformal representations of
the Lobachevskii plane metric (see, for instance, \cite{dubrovin}). The
fractional group of motions of Lobachevskii plane is isomorphic to:

(i) the group $SU(1,1)/\pm 1\equiv PSU(1,1)$ in the Poincare model;
(ii) the group $SL(2,{\R})/\pm 1\equiv PSL(2,\R)$ in the Klein model.

Moreover, it is known (see, for example, \cite{terr}) that
the Lobachevskii plane $H$ can be identified with the group
$SL(2,{\R})/SO(2)$. This relation enables us to resolve (at least
qualitatively) the following problem. Take the Brownian bridge on the group
${\cal H}=SL(2,{\R})/SO(2)$, i.e. demand the products of $N$ independent
random matrices $\widehat{\cal M}_k\in {\cal H}\;(0\le k\le N)$ to be
identical to the unit matrix. Consider the limit distribution of the
Lyapunov exponent, $\hat{\delta}$, for the first $m$ matrices in that
products. To have a direct mapping of this problem on the random walk in
the Lobachevskii plane, write the corresponding stochastic recursion
equation for some vector $\disp {\bf W}_k={u_k \choose v_k}$
\be \label{2:recurs}
{\bf W}_{k+1}=\widehat{\cal M}_k {\bf W}_k;\qquad {\bf W}_0={1 \choose 1}
\ee
where ${\cal M}_k\in {\cal H}$ for all $k\in [1,N]$. The BB--condition
means that
\be \label{2:bb}
{\bf W}_N={\bf W}_1 \quad \mbox{for $N\gg 1$}
\ee
Let us consider the simplest case
\be
\label{2:norma}
\widehat{\cal M}_k=1+\widehat{M}_k;\qquad \mbox{norm}[\widehat{M}_k]\ll 1
\ee
In this case the discrete dynamic equation (\ref{2:recurs}) can be replaced
by the differential one. Its stationary measure is determined by the
corresponding Fokker-Plank equation (\ref{2:lob}). The Lyapunov exponent,
$\hat\delta$ of product of random matrices $\widehat{\cal M}$ coincides
with the length of geodesics in the Klein representation of the
Lobachevskii plane. Hence, under the conditions (\ref{2:bb}),
(\ref{2:norma}) we have for $\hat\delta$ the usual Gaussian distribution
coinciding with the first of Eq.(\ref{2:gauss}). Without the BB--condition
(i.e. for ``open walks") we reproduce the standard F\"urstenberg behavior
\cite{fuerst_tut}.

Although this consideration seems rather crude (for details see Appendix
A), it clearly shows the origin of the main result:

{\it
The ``Brownian bridge" condition for random walks in space of constant
negative curvature makes the space ``effectively flat" turning
the corresponding limit probability distribution for random walks to the
ordinary central limit distribution.
}

The question whether this result is valid for the case of the random walk
in noneuclidean spaces of non-constant negative curvature still remains.

Finally we would like to introduce some conjectures which naturally
generalize our consideration.

{\it
The complexity $\eta$ of any known algebraic invariants {\rm (Alexander,
Jones, HOMFLY)} for the knot represented by the $B_n$-braid of length $N$
with the uniform distribution over generators has the following limit
behavior:
\be \label{2:con2}
P(\eta,N)\sim
\frac{{\rm const}}{N^{3/2}}\eta\exp\left(-\alpha(n)N+\beta(n)\eta-
\frac{\eta^2}{\delta(n)N}\right)
\ee
where $\alpha(n),\;\beta(n),\;\delta(n)$ are numerical constants depending
on $n$ only.
}

{\it
The knot complexity $\eta$ in ensemble of Brownian bridges from the group
$B_n$ shown in {\rm fig.\ref{2:fig:closure}} has Gaussian distribution,
where
\be
\left<\overline{\eta}\right>=0; \qquad \left<\overline{\eta}^2 \right>=
\frac{1}{2}\delta(n)N
\ee
}

These conjectures are to be proven yet. The main idea is to employ the
relation between the knot complexity $\eta$, the length of the shortest
noncontractible word and the length of geodesics on some hyperbolic
manifold.


\section{Conformal methods in statistics of random walks with topological
constraints}
\setcounter{equation}{0}

The last few years have been marked by considerable progress in
understanding the relationship between Chern-Simons topological field
theory, construction of algebraic knot and link invariants and conformal
field theory (see, for review, \cite{3:polyakov}).

Although the general concepts have been well elaborated in the
field-theoretic context, their application in the related areas of
mathematics and physics, such as, for instance, probability theory and
statistical physics of chain-like objects is highly limited.

The present Section is mainly concerned with the conformal methods in
statistical analysis which allow us to correlate problems discussed in
Chapters 1 and 2 and the limit distributions of random walks on
multiconnected Riemann surfaces. To be more specific, we show on the level
of differential equations how simple geometrical methods can be applied to
construction of non-commutative topological invariants. The latter might
serve as nonabelian generalizations of the Gauss linking numbers for the
random walks on multi-punctured Riemann surfaces. We also study the
connection between the topological properties of random walks on the double
punctured plane and behavior of four-point correlation functions in the
conformal theory with central charge $c=-2$. The developed approach is
applied to the investigation of statistics of 2D--random walks with
multiple topological constraints. For instance, the methods presented here
allow us to extract nontrivial critical exponents for the contractible
(i.e., unentangled) random walks in the regular lattices of obstacles. Some
of our findings support conjectures of Sections 2 and 3 and have direct
application in statistics of strongly entangled polymer chains (see Section
5).

\subsection{Construction of nonabelian connections for $\Gamma_2$ and
$PSL(2,\Z)$ from conformal methods}

We analyze the random walk of length $L$ with the effective elementary
step $a$ ($a\equiv 1$) on the complex plane $z=x+iy$ with two points
removed. Suppose the coordinates of these points being $M_1$ ($z_1=(0,0)$)
and $M_2$ ($z_2=(c,0)$) ($c\equiv 1$). Such choice does not indicate the
loss of generality because by means of simultaneous rescaling of the
effective step, $a$, of the random walk and of the distance, $c$,
between the removed points we can always obtain of any arbitrary
values of $a$ and $c$.

\begin{figure}
\centerline{\epsfig{file=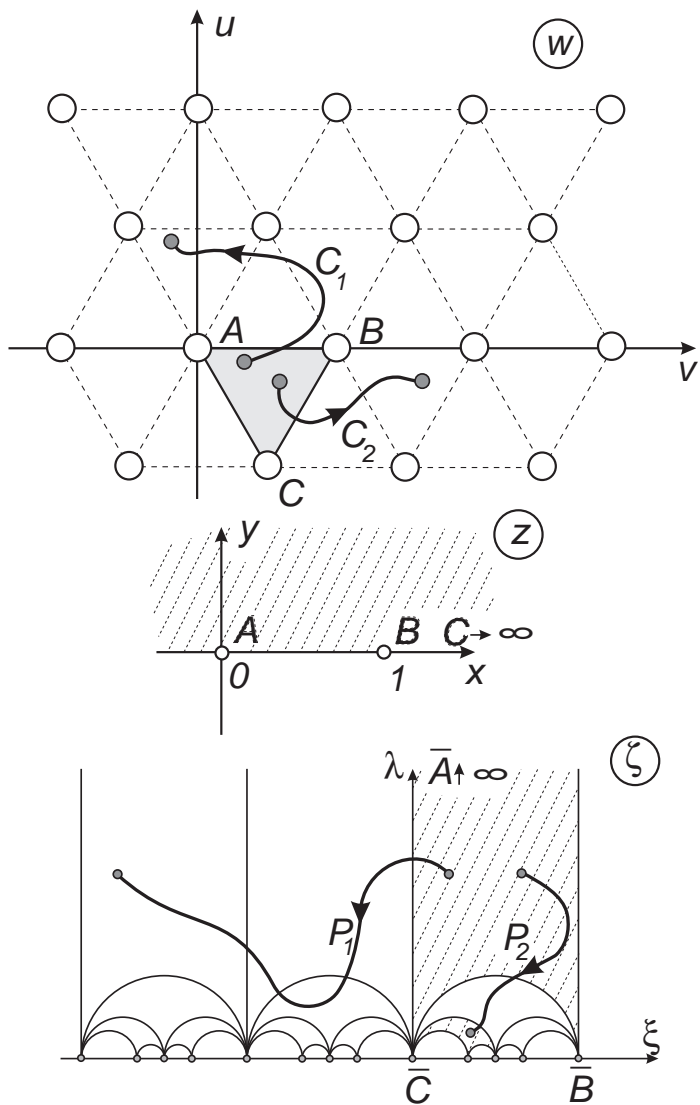,width=10cm}}
\caption{(a)---the double punctured complex plane $z$ with two basis loops
$C_1$ and $C_2$ enclosing points $M_1$ and $M_2$; (b)---the universal
covering $\zeta$ with fundamental domain corresponding to free group
$\Gamma_2$. The contours $P_1$ and $P_2$ are the images of the loops $C_1$
and $C_2$.}
\label{3:fig:1}
\end{figure}

Consider the closed paths on $z$ and attribute the generators $g_1,\, g_2$
of some group $G$ to the turns around the points $M_1$ and $M_2$ if we move
along the path in the clockwise direction (we apply $g_1^{-1},\, g_2^{-1}$
for counter-clockwise move)---see fig.\ref{3:fig:1}.

The question is: what is the probability $P(\mu,L)$ for the random walk of
length $L$ on the plane $z$ to form a closed loop with the shortest
noncontractible word written in terms of generators $\{g_1, g_2, g_1^{-1},
g_2^{-1}\}$ to have the length $\mu$ (see also Chapter 2).

Let the distribution function $P(\mu,L)$ be formally written as a path
integral with a Wiener measure
\be \label{2:pathint}
\begin{array}{ll}
\disp P(\mu,L)=\frac{1}{\cal Z} \int\ldots\int {\cal D}\{z\} &
\disp \hspace{-0.2cm} \exp\left\{-\frac{1}{a^2}
\int_0^L \left(\frac{dz(s)}{ds}\right)^2ds\right\}\medskip \\
& \times \delta\left[W\{g_1, g_2, g_1^{-1}, g_2^{-1}|z\}-\mu\right]
\end{array}
\ee
where ${\cal Z}=\int P(\mu,L)d\mu$ and $W\{\ldots|z\}$ is the length of the
shortest word on $G$ as a functional of the path on the complex plane.

Conformal methods enable us construct the connection and the topological
invariant $W$ for the given group as well as to rewrite
Eq.(\ref{2:pathint}) in a closed analytic form which is solvable at least
in the limit $L\to\infty$.

Let $\zeta(z)$ be the conformal mapping of the double punctured plane
$z=x+iy$ on the universal covering $\zeta=\xi+i\lambda$. The Riemann
surface $\zeta$ is constructed in the following way. Make three cuts on the
complex plane $z$ between the points $M_1$ and $M_2$, between $M_2$ and
$(\infty)$ and between $(\infty)$ and $M_1$ along the line Im$z=0$. These
cuts separate the upper (Im$z>0$) and lower (Im$z<0$) half-planes of $z$.
Now perform the conformal transform of the half-plane Im$z>0$ to the
fundamental domain of the group $G\{g_1,g_2\}$---the curvilinear triangle
lying in the half-plane Im$\zeta >0$ of the plane $\zeta$. Each fundamental
domain represents the Riemann sheet corresponding to the fibre bundle above
$z$. The whole covering space $\zeta$ is the unification of all such
Riemann sheets.

{\it
The coordinates of initial and final points of any trajectory on universal
covering $\zeta$ determine {\rm(\cite{3:nech}):} (a) The coordinates of
corresponding points on $z;$ (b) The homotopy class of any path on $z$. In
particular, the contours on $\zeta$ are closed if and only if
$W\{g_1,g_2|z\}\equiv 1$, i.e. they belong to the trivial homotopy class.
}

Coordinates of ends of the trajectory on universal covering $\zeta$ can
be used as the topological invariant for the path on double punctured plane
$z$ with respect to the action of the group $G$.

Thus, we characterize the topological invariant, Inv$(C)$, of some closed
directed path $C$ starting and ending in an arbitrary point $z_0 \neq
\{z_1,z_2,\infty\}$ on the plane $z$ by the coordinates of the initial,
$\zeta_{\rm in}(z_0)$, and final, $\zeta_{\rm fin}(z_0)$, points of the
corresponding contour $P$ in the covering space $\zeta$. The contour $P$
connects the images of the point $z_0$ on the different Riemann sheets.
Write Inv$_{(z)}(C)$ as a full derivative along the contour $C$:
\be \label{2:inv}
\mbox{Inv}_{(z)}(C)\stackrel{def}{=}\zeta_{\rm in}-
\zeta_{\rm fin} = \oint_C{{d\zeta(z)}\over {dz}}dz
\ee

The physical interpretation of the derivative $\disp \frac{d\zeta(z)}{dz}$
is very straightforward. Actually, the invariant, Inv$(C)$, can be
associated with the flux through the contour $C$ on the plane $(x,y)$:
\be \label{2:flux}
\mbox{Inv}(C)\equiv \mbox{Inv}_{(x,y)}(C)= \oint_C\nabla
\zeta(x,y){\bf n}d{\bf r}= \oint_C\nu\times \nabla \zeta(x,y){\bf v}(s)ds
\ee
where: ${\bf n}$ is the unit vector normal to the curve $C$, $d{\bf r}
={\bf e}_xdx+{\bf e}_ydy$ on the plane $(x,y)$; $\disp {\bf v}(s)=
\frac{d{\bf r}} {ds}$ denotes the "velocity" along the trajectory; and $ds$
stands for the differential path length. Simple transformations used in
Eq.(\ref{2:flux}) are: (a) ${\bf n}d{\bf r}={\bf e}_xdy-{\bf e}_ydx=d{\bf
r}\times\nu$; (b) $\nabla \zeta(x,y)(d{\bf r}\times{\bf \nu})=({\bf
\nu}\times \nabla \zeta(x,y)) d{\bf r}$, where ${\bf \nu} =(0,0,1)$ is the
unit vector normal to the plane $(x,y)$.

The vector product
\be \label{2:potential}
{\bf A}(x,y)={\bf \nu}\times\nabla \zeta(x,y)
\ee
can be considered a non-abelian generalization of the vector potential
of a solenoidal "magnetic field" normal to the plane $(x,y)$ and crossing
it in the points $(x_1,y_1)$ and $(x_2,y_2)$. Thus, ${\bf A}$ defines the
{\it flat connection} of the double punctured plane $z$ with respect to the
action of the group $G$.

It is easy to show how the basic formulae (\ref{2:inv}) and (\ref{2:flux})
transform in case of commutative group $G_{\rm comm}\{g_1,g_2\}$
which distinguishes only the classes of homology of the contour $C$ with
respect to the removed points on the plane. The corresponding conformal
transform is performed by the function $\zeta(z)=\ln(z-z_1)+\ln(z-z_2)$.
This immediately gives the abelian connection and the Gauss linking number
as a topological invariant:
$$
\begin{array}{c}
\disp {\bf A}({\bf r})={\bf \nu}\times\sum_{j=\{1,2\}} \frac{{\bf r}-{\bf
r}_j}{|{\bf r}-{\bf r}_j|^2}; \medskip \\
\disp \mbox{Inv}(C)=\oint_C A({\bf r})d{\bf r} =
\sum_{j=\{1,2\}} \oint_C
\frac{(y-y_j)dx-(x-x_j)dy}{(x-x_j)^2+(y-y_j)^2} = 2\pi (n_1+n_2)
\end{array}
$$
where $n_1$ and $n_2$ are the winding numbers of the path $C$ around the
points $M_1$ and $M_2$ of the plane $(x,y)$.

Substituting Eq.(\ref{2:pathint}) written in the Euclidean coordinates
$(x,y)$ for Eq.(\ref{2:flux}) and using the Fourier transform for the
$\delta$-function, we can rewrite equation (\ref{2:pathint}) as follows
\be \label{2:fourier2}
P(\mu,L)=\frac{1}{2\pi}\int_{-\infty}^{\infty}e^{-iq\mu}P(q,L)dq
\ee
where
\be \label{2:fourier1}
P(q,L)= \frac{1}{\cal Z} \int\ldots\int {\cal D}\{{\bf r}\}
\exp\left\{-\frac{1}{a^2}\int\limits_0^L
\left(\left(\frac{d{\bf r}(s)}{ds}\right)^2
-iq{\bf A}({\bf r})\frac{d{\bf r}(s)}{ds}\right)ds\right\}
\ee

The function $P(q,L)$ coincides with the Green function $P({\bf r}_0,{\bf
r}={\bf r}_0,q,L)$ of the non-stationary Schr\"odinger-like equation for
the free particle motion in a "magnetic field" with the vector potential
(\ref{2:potential}):
\be \label{2:diff}
\frac{\partial}{\partial L}P({\bf r}_0,{\bf r},q,L)-
\left(\frac{1}{2a}{\bf \nabla}-iq{\bf A}({\bf r})\right)^2
P({\bf r}_0,{\bf r},q,L)= \delta(L)\delta({\bf r}-{\bf r}_0)
\ee
where $q$ plays a role of a "charge" and the magnetic field is considered
transversal, i.e. $\mbox{rot}{\bf A}({\bf r})=0$.

Describe now the constructive way of getting the desired conformal
transform. The single-valued inverse function $z(\zeta)\equiv
\zeta^{-1}(z)$ is defined in the fundamental domain of $\zeta$---the
triangle $ABC$. The multivalued function $\phi(\zeta)$ is determined as
follows:

-- The function $\phi(\zeta)$ coincides with $z(\zeta)$ in the
basic fundamental domain;

-- In all other domains of the covering space $\zeta$ the
function $\phi(\zeta)$ is analytically continued through the boundaries of
these domains by means of fractional transformations consistent with the
action of the group $G$.

Consider two basic contours $P_1$ and $P_2$ on $\zeta$ being the
conformal images of the contours $C_1$ and $C_2$ (fig.\ref{3:fig:1}b). The
function $\phi(z)$ ($z\neq \{z_1,z_2,\infty\}$) obeys the following
transformations:
\be \label{2:monodr}
\phi\left[z\stackrel{C_1}{\rightarrow} z\right]\rightarrow
\tilde{\phi}_1(z)= {{a_1\phi(z)+b_1}\over {c_1\phi(z)+d_1}};\;
\phi\left[z\stackrel{C_2}{\rightarrow} z\right]\rightarrow
\tilde{\phi}_2(z)= {{a_2\phi(z)+b_2}\over {c_2\phi(z)+d_2}}
\ee
where
\be
\left(\begin{array}{cc}a_1 & b_1 \\ c_1 & d_1 \end{array}\right) = g_1;
\qquad \left(\begin{array}{cc}a_2 & b_2 \\ c_2 & d_2 \end{array}\right)
=g_2
\ee
are the matrices of basic substitutions of the group $G\{g_1,g_2\}$.

We assume $\zeta(z)$ to be a ratio of two fundamental solutions, $u_1(z)$,
and, $u_2(z)$, of some second order differential equation with peculiar
points $\{z_1=(0,0), z_2=(0,1), z_3=(\infty)\}$. As it follows from the
analytic theory of differential equations \cite{3:golubev}, the solutions
$u_1(z)$ and $u_2(z)$ undergo the linear transformations when the variable
$z$ moves along the contours $C_1$ and $C_2$:
\be
C_1:\,\left(\begin{array}{c}\tilde{u}_1(z) \\ \tilde{u}_2(z)
\end{array}\right) = g_1 \left(\begin{array}{c} u_1(z) \\
u_2(z) \end{array}\right);\quad
C_2:\,\left(\begin{array}{c}\tilde{u}_1(z) \\ \tilde{u}_2(z)
\end{array}\right) = g_2 \left(\begin{array}{c} u_1(z) \\
u_2(z) \end{array}\right)
\ee

The problem of restoring the form of differential equation knowing the
monodromy matrices $g_1$ and $g_2$ of the group $G$ known as
Riemann-Hilbert problem has an old history \cite{3:golubev}. In
our particular case we restrict ourselves with the well investigated groups
$\Gamma_2$ (the free group) and $PSL(2,\Z)$ (the modular group).
(\ref{2:2.1}). Thus, we have the following second-order differential
equations:
\be \label{2:free}
z(z-1){{d^2}\over {dz^2}}u^{(f)}(z)+(2z-1){{d}\over {dz}}
u^{(f)}(z)+{1\over 4}u^{(f)}(z)=0
\ee
for the free group and
\be \label{2:mod}
z(z-1){{d^2}\over {dz^2}}u^{(m)}(z) + \left({5\over 3}z-1\right)
{{d}\over {dz}}u^{(m)}(z)+{1\over {12}}u^{(m)}(z)=0
\ee
for the modular group.

The function which performs the conformal mapping of the upper half-plane
Im$z>0$ on the fundamental domain (the curvilinear triangle $ABC$) of the
universal covering $\zeta$ now reads
\be \label{2:mapping}
\zeta(z)=\frac{u_1^{\rm (f,m)}(z)}{u_2^{(f,m)}(z)}
\ee
where $u_{1,2}^{(f,m)}(z)$ and $u_{1,2}^{(f,m)}(z)$ are the basic
solutions of (\ref{2:free}) and (\ref{2:mod}) for $\Gamma_2$ and
$PSL(2,\Z)$ respectively.

As an example we give an explicit form of the complex potential $A(z)$ for
the free group $\Gamma_2$. Substituting Eq.(\ref{2:inv}) for
Eq.(\ref{2:mapping}), we get
\be \label{2:conn}
A(z)=
\frac{d\zeta(z)}{dz} = \frac{1}{2(z-1)}\left(\frac{F_1(z)F_4(z)}{F_2^2(z)}-
\frac{F_3(z)}{F_2(z)}\right)
\ee
where
$$
\begin{array}{rl}
\disp F_1(z)=\int\limits_1^{1/\sqrt z} \frac{d \kappa}
{\sqrt{(1-\kappa^2)(1-z\kappa^2)}}; & \disp
F_2(z)=\int\limits_0^1\frac{d\kappa}{\sqrt{(1-\kappa^2)(1-z\kappa^2)}}
\medskip \\
\disp F_3(z)=
\int\limits_1^{1/\sqrt z} \sqrt{\frac{1-\kappa^2}{1-z\kappa^2}}d\kappa;
& \disp F_4(z)=\int\limits_0^1 \sqrt{\frac{1-\kappa^2}{1-z\kappa^2}}d\kappa
\end{array}
$$
The asymptotic of (\ref{2:conn}) is as follows
$$
\frac{d\zeta(z)}{dz}\sim
\left\{\begin{array}{cl} \disp \frac{1}{z} & \qquad z\to 0 \medskip \\
\disp \frac{1}{z-1} & \qquad z\to 1
\end{array}\right.
$$
(compare to the abelian case).

\subsection{Random walk on double punctured plane and conformal field
theory}

The geometrical construction described in the previous section is evidently
related to the conformal field theory. In the most direct way this
relation could be understood as follows. The ordinary differential
equations Eq.(\ref{2:free}) and Eq.(\ref{2:mod}) can be associated with
equations on the four-point correlation function of some (still not
defined) conformal field theory. The question remains whether it is always
possible to adjust the central charge $c$ of the corresponding
Virosoro algebra and the conformal dimension $\Delta$ of the critical
theory to the coefficients in equations like (\ref{2:free}), (\ref{2:mod}).
The question is positive and we show that on the example of the random walk
on the double punctured plane with the monodromy of the free group.

We restrict ourselves to the "critical" case of infinite long trajectories,
i.e. we suppose $L\to\infty$. In the field-theoretic language that means
the consideration of the massless free field theory on $z$. Actually, the
partition function of the selfintersecting random walk on $z$ written in
the field representation is generated by the scalar Hamiltonian $H={1\over
2}(\nabla\varphi)^2+ m\varphi^2$ where the mass $m$ functions as the
"chemical potential" conjugated to the length of the path ($m\sim 1/L$).
Thus, for $L\rightarrow\infty$ we have $m_c=0$ which corresponds to the
critical point in conformal theory \cite{3:belpz}.

We introduce the conformal operator, $\varphi(z)$, on the complex plane
$z$. The dimension, $\Delta$, of this operator is defined from the
conformal correlator
\be \label{3:10}
\left<\varphi(z) \varphi(z')\right>\sim \frac{1}{\left| z-z'
\right|^{2\Delta}}
\ee
Let us suppose $\varphi(z)$ to be a primary field, then the four-point
correlation function $\left<\varphi(z_1)\varphi(z_2)\varphi(z_3)
\varphi(z_4)\right>$ satisfies the equation following from the conformal
Ward identity \cite{3:belpz,3:dots,3:kz}. In form of ordinary Riemann
differential equation, Eq.(\ref{3:10}) on the conformal correlator
$\psi(z|z_1,z_2,z_3)= \left<\varphi(z)\varphi(z_1)\varphi(z_2)
\varphi(z_3)\right>$ with the fixed points $\{z_1=(0,0), z_2=(1,0),
z_3=\infty\}$ reads \cite{3:belpz,3:dots}
$$
\begin{array}{ll}
\disp \left\{\frac{3}{2(2\Delta+1)}\frac{d^2}{dz^2} +
\frac{1}{z}\frac{d}{dz} + \frac{1}{z-1}\frac{d}{dz} - \frac{\Delta}{z^2} -
\frac{\Delta}{(z-1)^2} + \frac{2\Delta}{z(z-1)}\right\} & \medskip \\ &
\hspace{-2cm}\psi(z|z_1,z_2,z_3) = 0
\end{array}
$$
Performing the substitution
$$
\psi(z|z_1,z_2,z_3) = \left[ z(z-1) \right]^{-2\Delta} u(z)
$$
we get the equation
\be \label{3:11}
z(z-1)u''(z) - \frac{2}{3}(1-4\Delta)(1-2z)u'(z) - \frac{2}{3}(2\Delta -
8\Delta^2)u(z) = 0
\ee
which coincides with Eq.(\ref{2:free}) for one single value of $\Delta$
\be \label{3:12}
\Delta = - \frac{1}{8}
\ee

The conformal properties of the stress-energy tensor, $T(z)$, are defined
by the coefficients, $L_n$, in its Laurent expansion,
$$
T(z) = \sum_{n=-\infty}^{\infty} \frac{L_n}{z^{n+2}}
$$
These coefficients form the Virosoro algebra \cite{3:belpz}
$$
\left[ L_n, L_m \right]=(n-m) L_{n+m}+\frac{1}{12}C(n^3-n)\delta_{n+m,0}
$$
where the parameter, $c$, is the central charge of the theory. Using the
relation $c=\frac{2\Delta(5-8\Delta)}{(2\Delta+1)}$ established in
\cite{3:dots} and Eq.(\ref{3:12}) we obtain \be \label{3:13} c=-2
\ee

We find the following fact, mentioned by B. Duplantier, very intriguing. As
he has pointed out, the value $\Delta=-\frac{1}{8}$ (Eq.(\ref{3:12}))
coincides with the surface exponent (i.e. with the conformal dimension of
the two point correlator near the surface) for the dense phase of the
$O(n=0)$ lattice model (or, what is the same, for the Potts model with
$q=0$) describing statistics of the so-called "Manhattan random walks"
(known also as "dense polymers"---see the paper \cite{dupldav}). Recall
that Potts model has been already mentioned in the Chapter 1 in connection
with construction of algebraic knot invariants. It is hard to believe that
such coincidence is occasional and we hope that the relation between these
problems will be elucidated in the near future.

The conformal invariance of the random walk \cite{3:nech,3:itomc} together
with the geometrical interpretation of the monodromy properties of the
four-point conformal correlator established above enable us to express
the following assertion:

{\it
The critical conformal field theory characterized by the values $c=-2$ and
$\Delta=-\frac{1}{8}$ gives the field representation for the infinitely
long random walk on the double punctured complex plane.
}

With respect to the four-point correlation function, we could ask what
happens with the gauge connection $A_j(z)$ if the argument $z_j$ of the
primary field $\psi(z_j)$ moves along the closed contour $C$ around three
punctures on the plane. From the general theory it is known that $A_j(z)$
can be written as
\be \label{2:field}
A_j(z)=\frac{2}{k}\sum_{i\neq j}\frac{R_i R_j}{z-z_i} \ee
where $k$ is the level of the corresponding representation of the Kac-Moody
algebra and $R_i$, $R_j$ are the generators of representation of the
primary fields $\psi(z_i)$, $\psi(z_j)$ in the given group \cite{3:gmm}.

The holonomy operator $\chi(C)$ associated with $A_j(z)$ reads
\be \label{2:holon}
\chi(C)=P\exp\left(-\oint_C A_j(z)dz\right)
\ee
It would be interesting to compare Eq.(\ref{2:conn}) (with one puncture at
infinity) to Eq.(\ref{2:field}). Besides we could also expect that
Eq.(\ref{2:inv}) would allow us to rewrite the holonomy operator
(\ref{2:holon}) as follows
$$
\chi(C)=\exp\left(\zeta_{\rm in}-\zeta_{\rm fin}\right)
$$

At this point we finish the brief discussion of the field-theoretical
aspects of the geometrical approach presented above.

\subsection{Statistics of random walks with topological constraints in the
two--dimensional lattice of obstacles}

The conformal methods can be applied to the problem of calculating the
distribution function for random walks in regular lattices of topological
obstacles on the complex plane $w=u+iv$. Let the elementary cell of the
lattice be the equal-sided triangle with the side length $c$.

Introduce the distribution function $P(w_0,w,L|{\rm hom})$ defining the
probability of the fact that the trajectory of random walk starting at the
point $w_0$ comes after "time" $L$ to the point $w$ and {\it all paths
going from $w_0$ to $w$ belong to the same homotopy class with respect to
the lattice of obstacles}. Formally we can write the diffusion equation
\be \label{3:diffus}
\frac{a}{4}\Delta_{w}P(w,L|{\rm hom})=\frac{\partial}{\partial L}
P(w,L|{\rm hom})
\ee
with initial and normalization conditions:
$$
\begin{array}{c}
\disp P(w,L=0|{\rm hom})=\delta(z_0); \medskip \\ \disp
\sum_{\{{\rm hom}\}}P(w_0,w,L|{\rm hom})=\frac{1}{\pi aL}
\exp\left(-\frac{|w-w_0|^2}{aL}\right)
\end{array}
$$

The conformal methods can be used to find the asymptotic solution
of Eq.(\ref{3:diffus}) when $L\gg a$. Due to the conformal invariance of
the Brownian motion, the new random process in the covering space will be
again random but in the metric-dependent "new time". In particular, we are
interested in the probability to find the closed path of length $L$ to be
unentangled in the lattice of obstacles.

The construction of the conformal transformation $\zeta(w)$ (explicitly
described in \cite{3:nech}) can be performed in two steps---see
fig.\ref{3:fig:1}:

1. First, by means of auxiliary reflection $w(z)$ we transfer the
elementary cell of the $w$-plane to the upper half-plane of the Im$(z)>0$ of
the double punctured plane $z$. The function $w(z)$ is determined by the
Christoffel-Schwarts integral
\be \label{3a:9}
w(z)=\frac{c}{B\left(\frac{1}{3},\frac{1}{3}\right)} \int_0^z
\frac{d\tilde{z}}{\tilde{z}^{2/3}\left(1-\tilde{z}\right)^{2/3}}
\ee
where $B\left(\frac{1}{3},\frac{1}{3}\right)$ is the Beta-function. The
correspondence of the branching points is as follows:
$$
\begin{array}{lll}
A(w=0) & \to & \tilde{A}(z=0) \medskip \\
B(w=c) & \to & \tilde{B}(z=1) \medskip \\
C\left(w=c\; e^{-i\frac{\pi}{3}}\right) & \to & \tilde{C}(z=\infty)
\end{array}
$$

2. The construction of the universal covering $\zeta$ for the double
punctured complex plane $z$ is realized by means of automorphic functions.
If the covering space is free of obstacles, the corresponding conformal
transform should be as follows
\be \label{3:schwartz}
-\frac{1}{\bigl(z'(\zeta)\bigr)^2}\bigl\{z(\zeta)\bigr\}=
\frac{z^2-z+1}{2z^2(z-1)^2}
\ee
where $\bigl\{z(\zeta)\bigr\}$ is the
so-called Schwartz's derivative
$$
\bigl\{z(\zeta)\bigr\}=\frac{z'''(\zeta)}{z'(\zeta)}-
\frac{3}{2}\left(\frac{z''(\zeta)}{z'(\zeta)}\right)^2\;, \qquad
z'(z)=\frac{dz}{d\zeta}
$$

It is well known in the analytic theory of differential equations
\cite{3:golubev} that the solution of Eq.(\ref{3:schwartz}) can be
represented as ratio of two fundamental solutions of some second order
differential equation with two branching points, namely, of
Eq.(\ref{2:free}). The final answer reads
\be \label{3:ratio}
z(\zeta)\equiv k^2(\zeta)={{\theta_2^4 (0,e^{i\pi\zeta})}\over
{\theta_3^4 (0,e^{i\pi\zeta})}}
\ee
where $\theta_2(0,\zeta)$ and $\theta_3(0,\zeta)$ are the elliptic
Jacobi Theta-functions. We recall their definitions
\be \label{3:ell23}
\begin{array}{l}
\disp \theta_2\left(\chi,e^{i\pi\zeta}\right)=2e^{i{\pi \over 4}\zeta}
\sum_{n=0}^{\infty} e^{i\pi\zeta n(n+1)}\cos (2n+1)\chi \medskip \\
\disp \theta_3\left(\chi,e^{i\pi\zeta}\right)=1+2\sum_{n=0}^{\infty}
e^{i\pi\zeta n^2}\cos 2n\chi
\end{array}
\ee

The branching points $\tilde{A},\;\tilde{B},\;\tilde{C}$ have the images in
the vertex points of zero-angled triangle lying in the upper half-plane of
the plane $\zeta$. We have from Eq.(\ref{3:ratio}):
$$
\begin{array}{lll}
\tilde{A}(z=0) & \to & \bar{A}(\zeta=\infty) \medskip \\
\tilde{B}(z=1) & \to & \bar{B}(\zeta=0) \medskip \\
\tilde{C}(z=\infty) & \to & \bar{C}(\zeta=-1)
\end{array}
$$

The half-plane Im$(\zeta)>0$ functions as a covering space for the
plane $w$ with the regular array of topological obstacles. It does not
contain any branching point and consists of the infinite set of Riemann
sheets, each of them having form of zero-angled triangle. These Riemann
sheets correspond to the fibre bundle of $w$.

The conformal approach gives us a well defined nonabelian topological
invariant for the problem---the difference between the initial and final
points of the trajectory in the covering space (see Section 3.1). Thus, the
diffusion equation for the distribution function $P(\zeta,L)$ in the
covering space $\zeta$ with given initial point $\zeta_0$ yields
\be
\label{3:diff1}
\frac{a}{4}\frac{\partial^2}{\partial\zeta\partial\overline{\zeta}}
P(\zeta,\zeta_0,L)=\left|w'(\zeta)\right|^2
\frac{\partial}{\partial L}P(\zeta,\zeta_0,L)
\ee
where we took into account that under the conformal transform the Laplace
operator is transformed in the following way
$$
\Delta_w=\left|\frac{d\zeta}{dw}\right|^2\Delta_{\zeta}\qquad \mbox{and}
\qquad \left|\frac{d\zeta}{dw}\right|^2=\frac{1}{\left|w'(\zeta)\right|^2}
$$
In particular, the value $P(\zeta=\zeta_0,\zeta_0,L)$ gives the
probability for the path of length $L$ to be unentangled (i.e.
to be contractible to the point) in the lattice of obstacles.

The expression for the Jacobian $\left|w'(\zeta)\right|^2$ one can find
using the properties of Jacobi Theta-functions \cite{3:chandr}. Write
$w'(\zeta)=w'(z)\;z'(\zeta)$, where
$$
w'(z)=\frac{c}{B\left(\frac{1}{3},\frac{1}{3}\right)}
\frac{\theta_3^{16/3}}{\theta_2^{8/3}\;\theta_0^{8/3}}
$$
and
$$
z'(\zeta)=i\pi\frac{\theta_2^4\;\theta_0^4}{\theta_3^4}\;;\qquad
i\frac{\pi}{4}\theta_0^4=
\frac{d}{d\zeta}\ln\left(\frac{\theta_2}{\theta_3}\right)
$$
(we omit the arguments for compactness).

The identity
$$
\disp \theta'_1(0,e^{i\pi\zeta})\equiv
\left.\frac{d\theta_1(\chi,e^{i\pi\zeta})}{d\chi}\right|_{\chi=0}=
\pi\theta_0(\chi,e^{i\pi\zeta})\,\theta_2(\chi,e^{i\pi\zeta})\,
\theta_3(\chi,e^{i\pi\zeta})
$$
enables us to get the final expression
\be \label{3:jacobian}
\left|w'(\zeta)\right|^2=c^2 h^2
\left|\theta'_1\left(0, e^{i\pi\zeta}\right)\right|^{8/3}, \qquad
h=\frac{1}{\pi^{1/3}B\left(\frac{1}{3},\frac{1}{3}\right)}\simeq 0.129
\ee
where
\be \label{3:ell1}
\theta_1(\chi,e^{i\pi\zeta})=2e^{i\frac{\pi}{4}\zeta}
\sum_{n=0}^{\infty}(-1)^n e^{i\pi n(n+1)\zeta}\sin (2n+1)\chi
\ee

\begin{figure}
\centerline{\epsfig{file=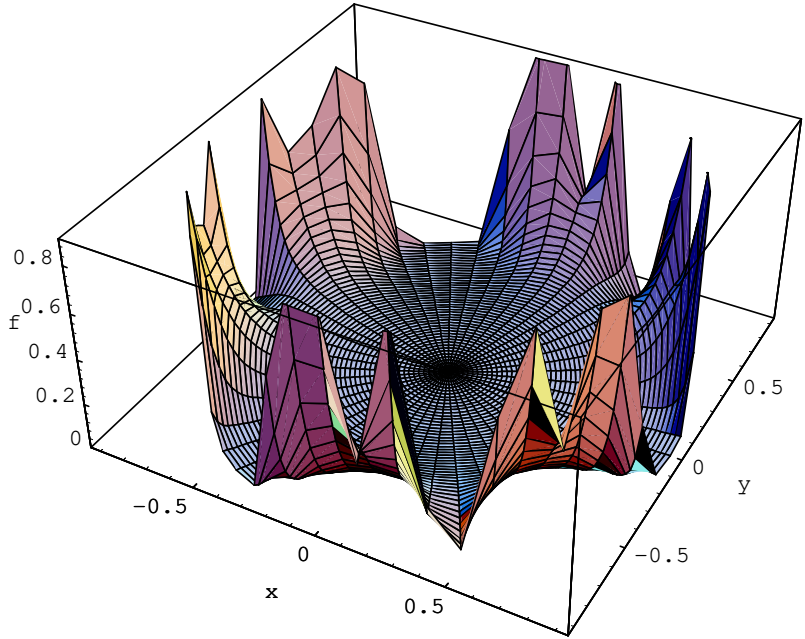,width=12cm}}
\caption{Relief of the function $g(r,\psi)$---see explanations in
the text.}
\label{3:fig:poincare}
\end{figure}

Return to Eq.(\ref{3:diff1}) and perform the conformal transform of the
upper half-plane Im$\zeta>0$ to the interior of the unit circle on the
complex plane $\tau$ in order to use the symmetry properties of
the system. It is convenient to choose the following mapping of the
vertices of the fundamental triangle $\bar{A}\bar{B}\bar{C}$
$$
\begin{array}{lll}
\bar{A}(\zeta=\infty) & \to & A'(\tilde{\zeta}=1) \medskip \\
\bar{B}(\zeta=0) & \to & B'(\tilde{\zeta}=e^{-i\frac{2\pi}{3}})
\medskip \\
\bar{C}(\zeta=-1) & \to & C'(\tilde{\zeta}=e^{i\frac{2\pi}{3}})
\end{array}
$$
The corresponding transform reads
\be \label{3:1.1}
\zeta(\tau)=e^{-i\frac{\pi}{3}}\frac{\tau-e^{i\frac{2\pi}{3}}}{\tau-1}-1
\ee
and the Jacobian $|w'(\tau)|^2$ takes the form
\be \label{3:1.2}
|w'(\tau)|^2=\frac{3c^2h^2}{|1-\tau|^4}
\left|\theta'_1\left(0,e^{i\pi\zeta(\tau)}\right)^{8/3}\right|
\ee
In fig.(\ref{3:fig:poincare}) we plot the function
$\disp g(r,\psi)=\frac{1}{c^2}|w'(\tau)|^2$ where $\tau=r\,e^{i\psi}$.
\bigskip

\begin{figure}
\centerline{\epsfig{file=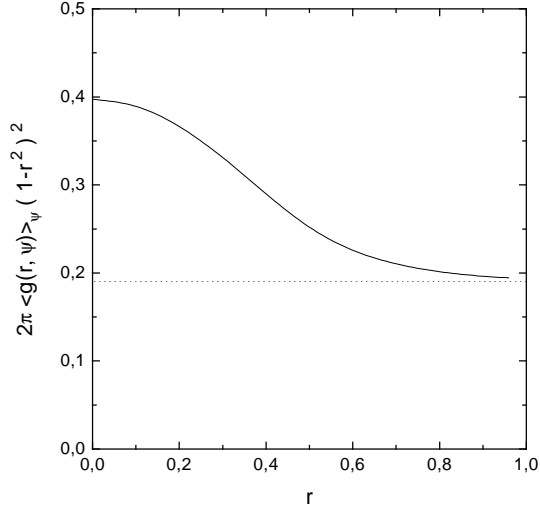,width=8cm}}
\caption{Plot of product $2\pi\left<g(r,\psi)\right>_{\psi}\times
(1-r^2)^2$ as a function of $r$.}
\label{3:fig:metric}
\end{figure}

The gain of such representation becomes clear if we average the function
$g(r,\psi)$ with respect to $\psi$. The numerical calculations give us:
\be \label{2:aver}
\lim_{r\to 1} \left<g(r,\psi)\right>_{\psi} \equiv
\lim_{r\to 1} \frac{1}{2\pi}\int_0^{2\pi} g(r,\psi)d\psi=
\frac{\varpi}{(1-r^2)^2}
\ee
where $\varpi\simeq 0.0309$ (see the fig.(\ref{3:fig:metric})).

Thus it is clear that for $r$ rather close to $1$ the diffusion is governed
by the Laplacian on the surface of the constant negative curvature (the
Lobachevskii plane). Representation of the Lobachevskii plane in the unit
circle and in the upper half-plane (i.e. Poincare and Klein models) has been
discussed in Section 2.4. Finally the diffusion equation (\ref{3:diff1})
takes the following form:
\be \label{2:diff2}
\frac{\partial}{\partial N}P(r,\psi,N)={\cal D}(1-r^2)^2
\Delta_{r,\psi}P(r,\psi,N)
\ee
where $\disp {\cal D}=\frac{a^2}{4\varpi c^2}$ is the "diffusion
coefficient" in the Lobachevskii plane and $N=L/a$ is the dimensionless
chain length (i.e. effective number of steps).

Changing the variables $(r,\psi)\to (\mu,\psi)$, where $\disp
\mu=\ln\frac{1+r}{1-r}$, we get the unrestricted random walk on the
3-pseudosphere (see Eq.(\ref{2:3pseud})). Correspondingly the distribution
function $P(\mu,N)$ reads
\be
P(\mu,N)=\frac{e^{-\frac{N{\cal D}}{4}}}{4\pi\sqrt{2\pi(N{\cal D})^3}}
\int_{\mu}^{\infty}\frac{\xi \exp\left(-\frac{\xi^2}{4N{\cal D}}
\right)}{\sqrt{\cosh\xi- \cosh\mu}}d\xi
\ee

The physical meaning of the geodesics length on 3-pseudosphere, $\mu$, is
straightforward: $\mu$ is the length of so-called "primitive path" in
the lattice of obstacles, i.e. length of the shortest path remaining after
all topologically allowed contractions of the random trajectory in the
lattice of obstacles. Hence, $\mu$ can be considered a nonabelian
topological invariant, much more powerful than the Gauss linking number.
This invariant is not complete except one point $\mu=0$ where it precisely
classifies the trajectories belonging to the trivial homotopic class. Let
us note that the length $\eta$ is proportional to the length of the
primitive (irreducible) word written in terms of generators of the free
group $\Gamma_2$.


\section{Physical applications. Polymer language in statistics of entangled
chain--like objects}
\setcounter{equation}{0}

Topological constraints essentially modify the physical properties of
statistical systems consisting of chain-like objects of completely
different nature. It should be said that topological problems
are widely investigated in connection with quantum field and string
theories, 2D-gravitation, statistics of vortexes in superconductors and
world lines of anyons, quantum Hall effect, thermodynamic properties of
entangled polymers etc. Modern methods of theoretical physics allow us to
describe rather comprehensively the effects of nonabelian statistics on
physical behavior for each particular referred system; however, in our
opinion, the following general questions remain obscure:

How does the changes in topological state of the system of entangled
chain-like objects effect their physical properties?

How can the knowledge accrued in statistical topology be applied to the
construction of the Ginzburg-Landau-type theory of fluctuating entangled
chain-like objects?

In order to have representative and physically clear image for the system
of fluctuating chains with the full range of nonabelian topological
properties it appears quite natural to formulate general topological
problems in terms of polymer physics. It allows us: to use a geometrically
clear image of polymer with topological constraints as a model
corresponding to the path integral formalism in the field theory; to
advance in investigation of specific physical properties of biological and
synthetical polymer systems where the topological constraints play a
significant role.

For physicists the polymer objects are attractive due to many reasons.
First of all, the adjoining of monomer units in chains essentially reduces
all equilibrium and dynamic properties of the system under consideration.
Moreover, due to that adjoining the behavior of polymers is determined by
the space-time scales larger than for low-molecular-weight substances. The
chain-like structure of macromolecules causes the following peculiarities
(see, for instance, \cite{4:grosbk}): the so-called "linear memory" (i.e.
fixed position of each monomer unit along the chain); the low translational
entropy (i.e. the restrictions on independent motion of monomer units due
to the presence of bonds); large space fluctuations (i.e. just a single
macromolecule can be regarded as a statistical system with many degrees of
freedom).

It should be emphasized that the above mentioned "linear memory" leads to
the fact that different parts of polymer molecules fluctuating in space can
not go one through another without the chain rupture. For the system of
non-phantom closed chains this means that only those chain conformations
are available which can be transformed continuously into one another, which
inevitably give rise to the problem of knot entropy determination (see
Section 2 for details).

\subsection{Polymer chain in 3D-array of obstacles}

The 3D-model "polymer chain in an array of obstacles" (PCAO) can be defined
as follows (\cite{2:helf,khne,3:nech}). Suppose a
polymer chain of length $L=Na$ is placed between the edges of the simple
cubic lattice with the spacing $c$, where $N$ and $a$ are the number of
monomer units in the chain and the length of the unit correspondingly. We
assume that the chain cannot cross ("pass through") any edges of the
lattice.

The PCAO-model can be considered as the basis for a mean-field-like
self-consistent approach to the major problem of entropy calculation of
ensembles of strongly entangled fluctuating chains. Namely, choose the test
chain, specify its topological state and assume that the lattice of
obstacles models the effect of entanglements with the surrounding chains
("background"). Neglecting the fluctuations of the background and the
topological constraints which the test chain produces for itself, we lose
information about the correlations between the test chain and the
background. However even in the simplest case we arrive at some nontrivial
results concerning statistics of the test chain caused by topological
interactions with the background. This means that for the investigation of
properties of real polymer systems with topological constraints it is not
enough to be able to calculate the statistical characteristics of chains in
lattices of obstacles, but it is also necessary to be able to adjust any
specific physical system to the unique lattice of obstacles, which is much
more complicated task.

So, let us take a closed polymer chain without volume interactions
(i.e. a chain with selfintersections) in the trivial topological state with
respect to the 3D lattice of obstacles. It means that the chain trajectory
can be continuously contracted to the point. It is clear that because of
the obstacles, the macromolecule will adopt more compact conformation than
the standard random walk without any topological constraints.

It is convenient to begin with the lattice realization of the problem. In
this case the polymer chain can be represented as a closed $N$-step random
walk on a cubic lattice with the length of elementary step $a$ being equal
to the spacing of the array of obstacles, $c$. The general case $a\neq c$
will be considered later.

{\it
The random walk on a 3D-cubic lattice in the presence of the
regular array of topological constraints produced by uncrossible strings on
the dual lattice is equivalent to the free random walk on the graph---the
Cayley tree with the branching number $z=6$.
}

The the average space dimension $R(N)\equiv \sqrt{R^2(N)}$ of the closed
unentangled $N$-step random walk is (\cite{khne}):
\be \label{4:r}
R\sim a N^{1/4}
\ee

The outline of the derivation of the result (\ref{4:r}) is as follows.
First of all note that the Cayley tree with $z$ branches (called latter as
$z$--tree with $z=2D$ branches) plays a role of the universal covering and
is just a visualization of the free group $\Gamma_{\infty}$ with the
infinite number of generators. At the same time $\Gamma_{\infty}/{\Z}^D=
\Gamma_{z/2}$, where $Gamma_{z/2}$ is the free group with $z$ generators.
Writing down the recursion relations for the probability $P(k,N)$ for the
$N$-step random walk on the $z$-tree (compare to
(\ref{2:2.11})--(\ref{2:walk})), we can easily find the conditional
limiting distribution for the function $\disp P(k,m|N) =
\frac{P(k,m)P(k,N-m)} {z(z-1)^{k-1}}$. Recall that $P(k,m|N)$ gives the
conditional probability distribution of the fact that two sub-chains $C_1$
and $C_2$ of lengths $m$ and $N-m$ have the common primitive path $k$ under
the condition that the composite chain $C_1\,C_2$ of length $N$ is closed
and unentangled in regard to the obstacles;
\be \label{4:3a}
P(k,m|N)\simeq \left(\frac{N}{2m(N-m)}\right)^{3/2}\,
k^2\,\exp\left(-\frac{k^2 N}{2m(N-m)}\right)
\ee
This equation enables us to get the following expressions for the mean
length of the primitive path, $\left<k(m)\right>$ of {\it closed
unentangled} $N$-link chain divided into two parts of the lengths $m$ and
$N-m$ correspondingly
\be \label{4:prim}
\left<k(m)\right>=\sum_{k=0}^N k P(k,m|N)\simeq
\frac{2}{\sqrt{\pi}}\sqrt{\frac{2m(N-m)}{N}} \qquad (N\gg 1)
\ee
The primitive path itself can be considered a random walk in a 3D space
with restriction that any step of the primitive path should not be strictly
opposite to the previous one. Therefore the mean-square distance in the
space $\left<({\bf r}_0-{\bf r}_m)^2\right>$ between the ends of the
primitive path of $k(m)$ steps is equal to
\be \label{4:dist}
\left<({\bf r}_0-{\bf r}_m)^2\right>=\frac{z}{z-2}k a^2
\ee
where ${\bf r}_m$ is the radius-vector of a link with the number $m$ and
the boundary conditions are: ${\bf r}_N={\bf r}_0=0$. The mean-square
gyration radius, $\left<R_g^2\right>$ of $N$-step closed unentangled random
walk in the regular lattice of obstacles reads
\be \label{4:1}
\begin{array}{lllll} \left<R_g^2\right> & = & \disp
\frac{1}{2N^2}\sum_{n\neq m} \left<({\bf r}_n-{\bf r}_m)^2\right> & = &
\disp \frac{1}{2N}\sum_{m=1}^{N} \left<({\bf r}_0-{\bf r}_m)^2\right>
\medskip \\ & & & = & \disp \frac{z}{z-2}\frac{\sqrt{2\pi}}{8}a^2\sqrt{N}
\end{array}
\ee
This result should be compared to the mean-square gyration radius of the
closed chain without any topological constraints,
$\left<R_{g,0}^2\right>=\frac{1}{12}a^2N$

The relation $R\sim N^{1/4}$ is reminiscent of the well-known expression
for the dimension of randomly branched ideal macromolecule. The gyration
radius of an ideal "lattice animal" containing $N$ links is proportional to
$N^{1/4}$. It means that both systems belong to the same universality
class.

Now we turn to the mean-field calculation of the critical exponent $\nu$ of
nonselfintersecting random walk in the regular lattice of obstacles
\cite{4:gr_gu_sh}. Within the framework of Flory-type mean-field theory
the nonequilibrium free energy, $F(R)$, of the polymer chain of size $R$
with volume interactions can be written as follows
\be \label{4:energy}
F(R)=F_{\rm int}(R)+F_{\rm el}(R)
\ee
where $F_{\rm int}(R)$ is the energy of the chain self-interactions and
$F_{\rm el}(R)$ is the "elastic", (i.e. pure entropic) contribution to the
total free energy of the system. Minimizing $F(R)$ with respect to $R$ for
fixed chain length, $L=Na$, we get the desired relation $R\sim N^{\nu}$.

Write the interacting part of the chain free energy written in the virial
expansion
\be \label{4:mean}
F_{\rm int}(R)=V \left( B\rho^2 + C\rho^3 \right)
\ee
where $V\sim R^d$ is the volume occupied by the chain in $d$-dimensional
space; $\rho=\frac{N}{V}$ is the chain density;
$B=b\frac{T-\theta}{\theta}$ and $C={\rm const}>0$ are the two-- and
three-- body interaction constants respectively. In the case $B>0$ third
virial coefficient contribution to Eq.(\ref{4:mean}) can be neglected
\cite{4:grosbk}.

The "elastic" part of the free energy $F_{\rm el}(R)$ of an unentangled closed
chain of size $R$ and length $Na$ in the lattice of obstacles can be
estimated as follows
\be \label{4:elastic}
F_{\rm el}(R)= {\rm const}+\ln P(R,m=N/2,N)= {\rm const}+\ln \int
dk\,P(k,N)\,P(R,k)
\ee
where the distribution function $P(k,m,N)$ is the same as in
Eq.(\ref{4:3a}) and $P(R,k)$ gives the probability for the primitive path
of length $k$ to have the space distance between the ends equal to $R$:
\be \label{4:space}
P(R,k)=\left(\frac{D}{2\pi ack}\right)^{d/2}
\exp\left(-\frac{DR^2}{2ack}\right)
\ee
Substituting Eq.(\ref{4:elastic}) for Eqs.(\ref{4:3a}) and (\ref{4:space})
we get the following estimate
\be \label{4:elastic1}
F_{\rm el}(R)=-\left(\frac{R^4}{a^2c^2N}\right)^{1/3} + o\left(R^{4/3}\right)
\ee

Equations (\ref{4:mean}) and (\ref{4:elastic1}) allow us to rewrite
Eq.(\ref{4:energy}) in the form
\be \label{4:energy1}
F(R)\simeq B\frac{N^2}{R^D}-\left(\frac{R^4}{a^2c^2N}\right)^{1/3}
\ee
Minimization of Eq.(\ref{4:energy1}) with respect to $R$ for fixed $N$
yields
\be \label{4:index}
R\sim B^{3/(4+3D)} (ac)^{2/(4+3D)} N^{\nu}; \qquad \nu=\frac{7}{4+3D}
\ee

The upper critical dimension for that system is $D_{cr}=8$. For $D=3$
Eq.(\ref{4:index}) gives
\be \label{4:index1}
R\sim N^{7/13}
\ee

It is interesting to compare Eq.(\ref{4:index}) to the critical exponent
$\nu_{\rm an}$ of the lattice animal with excluded volume in the
D--dimensional space, $\nu_{\rm an}=\frac{3}{4+D}$, which gives
$\nu_{\rm an}=\frac{3}{7}$ for $D=3$. The difference in exponents signifies
that the unentangled ring with volume interactions and the
nonselfintersecting "lattice animal" belong to {\it different universality
classes} (despite in the absence of volume interactions they belong to the
same class).

\subsection{Collapsed phase of unknotted polymer}

In this Section we show which predictions about the fractal structure of a
strongly collapsed phase of unknotted ring polymer can be made using the
concept of "polymer chain in array of obstacles".

\subsubsection{"Crumpled globule" concept in statistics of strongly
collapsed unknotted polymer loops}

Take closed nonselfintersecting polymer chain of length $N$ in the trivial
topological state \footnote{~The fact that the closed chain cannot
intersect itself causes two types of interactions: a) volume interactions
which vanish for infinitely thin chains and b) topological constraints
which remain even for chain of zero thickness.}. After a temperature
decrease the formation of the collapsed globular structure becomes
thermodynamically favorable \cite{4:lgkh}. Supposing that the globular
state can be described in the virial expansion we introduce as usual two--
and three--body interaction constants: $B=b\frac{T-\theta}{\theta}<0$ and
$C={\rm const}>0$. But in addition to the standard volume interactions we
would like to take into account the non-local topological constraints which
obviously have a repulsive character. In this connection we express our
main assertion \cite{4:grnesh}.

{\it
The condition to form a trivial knot in a closed polymer changes
significantly all thermodynamic properties of a macromolecule and leads
to specific non-trivial fractal properties of a line representing the chain
trajectory in a globule.
}
We call such structure {\bf crumpled (fractal) globule}. We prove this
statement consistently describing the given crumpled structure and
showing its stability.

It is well-known that in a poor solvent there exists some critical chain
length, $g^{*}$, depending on the temperature and energy of volume
interactions, so that chains which have length bigger than $g^{*}$
collapse. Taking long enough chain, we define these $g^{*}$-link parts as
new block monomer units (crumples of minimal scale).

\begin{figure}
\centerline{\epsfig{file=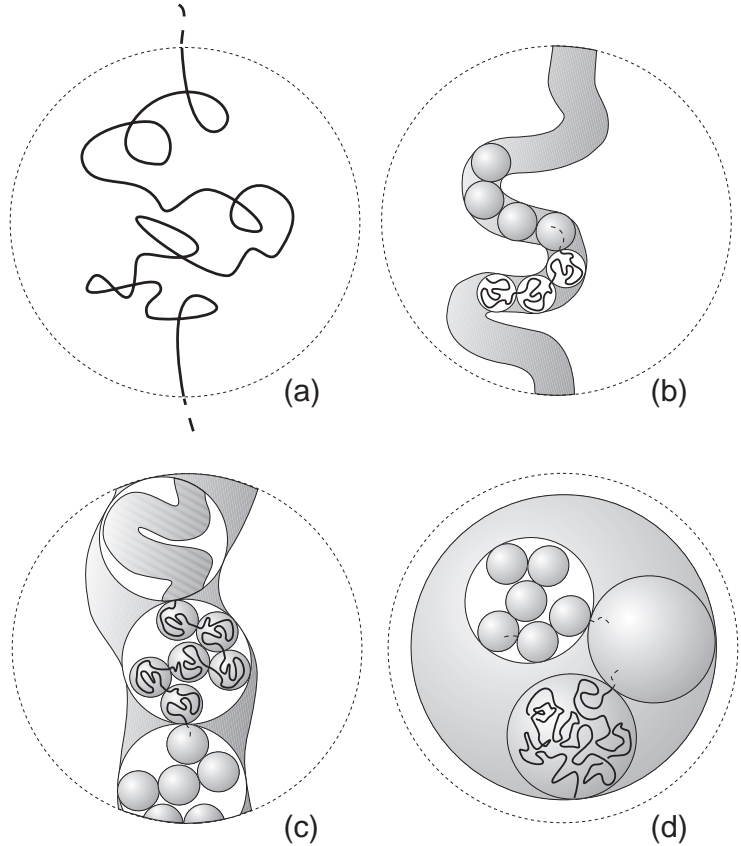,width=10cm}}
\caption{(a)--(c) Subsequent stages of collapse; (d) Self-similar structure
of crumpled globule segregated on all scales.}
\label{4:fig:crump}
\end{figure}

Consider now the part of a chain with several block monomers, i.e.
crumples of smallest scale. This new part should again collapse in itself,
i.e. should form the crumple of the next scale if other chain parts do not
interfere with it. The chain of such new sub-blocks (crumples of new scale)
collapses again and so on till the chain as a whole (see
fig.\ref{4:fig:crump}) forms the largest final crumple. 
Thus the procedure
is completed when all initial links are united into one crumple of the
largest scale. It should be noted that the line representing the chain
trajectory obtained through the procedure described above resembles the
3D-analogue of the well known self-similar {\it Peano curve}.

The specific feature of the crumpled globule is in the fact that different
chain parts are not entangled with each others, completely fill the allowed
volume of space and are "collapsed in themselves" starting from the
characteristic scale $g^*$. It may seem that due to space fluctuations of
the chain parts all that crumples could penetrate each others with the
loops, destroying the self-similar scale-invariant structure described
above. However it can be shown on the basis of PCAO--model that if the
chain length in a crumple of an arbitrary scale exceeds $N_{e}$ then the
crumples coming in contact do not mix with each other and remain segregated
in space. Recall that $N_e$ is the characteristic distance between
neighboring entanglements along the chain expressed in number of segments
and, as a rule, the values of $N_e$ lie in the range $30\div 300$
~\cite{4:grosbk}.

\begin{figure}
\centerline{\epsfig{file=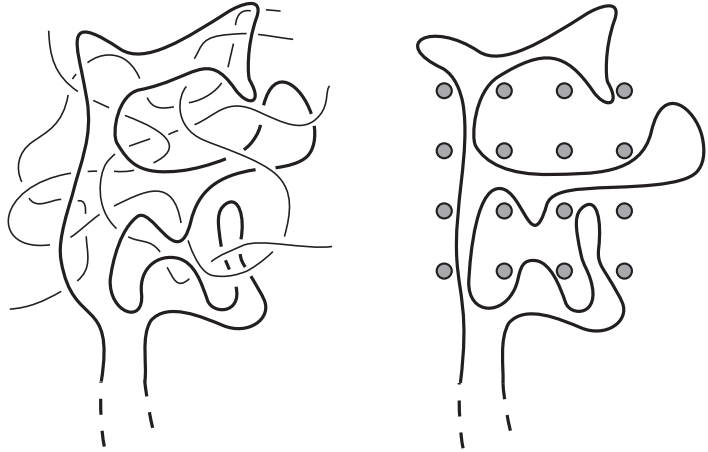,width=10cm}}
\caption{(a) Part of the closed unknotted chain surrounded by other parts
of the same chain; (b) Unentangled ring in lattice of obstacles. The
obstacles replace the effect of topological constraints produced by other
part of the same chain.}
\label{4:fig:loop}
\end{figure}

Since the topological state of the chain part in each crumple is fixed and
coincides with the state of the whole chain (which is trivial) this chain
part can be regarded as an unknotted ring. Other chain parts (other
crumples) function as effective lattice of obstacles surrounding the "test"
ring---see fig.\ref{4:fig:loop}. 
Using the results of the Section 5.1 (see
Eq.(\ref{4:1})) we conclude that any $M$-link ring subchain without volume
interactions not entangled with any of obstacles has the size
$R^{(0)}(M) \sim a M^{1/4}$. If $R^{(0)}$ is the size of an equilibrium
chain part in the lattice of obstacles, the entropy loss for ring chain,
$S$, as a function of its size, $R$, reaches its maximum for $R \simeq
R^{(0)}$ (see Eq.\ref{4:elastic1}) and the chain swelling for values of $R$
exceeding $R^{(0)}(M)$ is entropically unfavorable. At the same time in
the presence of excluded volume the following obvious inequality must be
fulfilled $R(M) \sim aM^{1/3}$, which follows from the fact that density of
the chain in the globular phase $\rho\sim R^3/N$ is constant. In connection
with the obvious relation $R(M) > R^{(0)}(M)$ we conclude that {\it the
swelling of chains in crumples due to their mutual inter-penetration with
the loops does not result in the entropy gain and, therefore, does not
occur in the system with finite density. It means that the size of crumple
on each scale is of order of its size in dense packing state and the
crumples are mutually segregated in space}. These questions are discussed
in details in the work \cite{4:grnesh}.

The system of densely packed globulized crumples corresponds to the chain
with the fractal dimension $D_f=3$ ($D_f=3$ is realized from the minimal
scale, $g^{*}$, up to the whole globule size). The value $g^{*}$ is of
order
\be \label{4.16}
g^{*} = N_{e}(\rho a^{3})^{-2},
\ee
where $\rho$ is the globule density. This estimation was obtained in
\cite{4:grnesh} using the following arguments: $g=(\rho a^{3})^{-2}$ is the
mean length of the chain part between two neighboring (along the chain)
contacts with other parts; consequently $N_{e} g$ is the mean length of the
chain part between topological contacts (entanglements). Of course, as to
the phantom chains, Gaussian blobs of size $g$ are strongly overlapped with
others because pair contacts between monomers are screened (because of
so-called $\theta$-conditions \cite{4:lgkh}). However for nonphantom chains
these pair contacts are topologically essential because chain crossings are
prohibited for any value and sign of the virial coefficient.

The entropy loss connected with the crumpled state formation can be
estimated as follows:
\be \label{4.17}
S \simeq - \frac{N}{g^{*}}
\ee

Using Eq.(\ref{4.17}) the corresponding crumpled globule density, $\rho$,
can be obtained in the mean-field approximation via minimization of its
free energy. The density of the crumpled state is less than  that of usual
equilibrium state what is connected with additional topological
repulsive-type interactions between crumples:
\be \label{4.18}
\rho_{\rm crump} = \frac{\rho_{eq}}{1 + {\rm const}(a^{6}/CN_{e})} <
\rho_{\rm eq}
\ee
where $\rho_{\rm eq}$ is the density of the Lifshits' globule.

The direct experimental verification of the proposed self-similar fractal
structure of the unknotted ring polymer in the collapsed phase meets some
technical difficulties. One of the ways to justify the "crumpled globule"
(CG) concept comes from its indirect manifestations in dynamic and static
properties of different polymer systems. The following works should be
mentioned in that context:

1. The two-stage dynamics of collapse of the macromolecule after abrupt
changing of the solvent quality, found in recent light scattering
experiments by B. Chu and Q. Ying (Stony Brook) \cite{4:chu}.

2. The notion about the crumpled structure of the collapsed ring polymer
allowed to explain \cite{khone} the experiments on compatibility
enhancement in mixtures of ring and linear chains \cite{mck}, as well as to
construct the quantitative theory of a collapse of
$N$--isopropilacrylamide gel in a poor water \cite{4:frac}.

3. The paper \cite{shakh} where the authors claim the observation of the
crumpled globule in numerical simulations.

\subsubsection{Knot Formation Probability}

We can also utilize the CG-concept to estimate the trivial knot formation
probability for dense phase of the polymer chain. Let us repeat that the
main part of our modern knowledge about knot and link statistics has
been obtained with the help of numerical simulations based on the
exploiting of the algebraic knot invariants (Alexander, as a rule). Among
the most important results we should mention the following ones:

-- The probability of the chain self-knotting, $p(N)$, is determined as a
function of chain length $N$ under the random chain closure
\cite{volog,4:mish}. In the work \cite{4:muthu} (see also the recent
paper \cite{4:tsdeg}) the simulation procedure was extended up to chains of
order $N \simeq 2000$, where the exponential asymptotic of the type
$$
p_{0} \sim \exp (-N/N_{0}(T))
$$
has been found for trivial knot formation probability for chains in good
and $\theta$-solvents. A statistical study of random knotting probability
using the Vassiliev invariants has been undertaken in recent work
\cite{4:tsdeg}.

\begin{figure}
\centerline{\epsfig{file=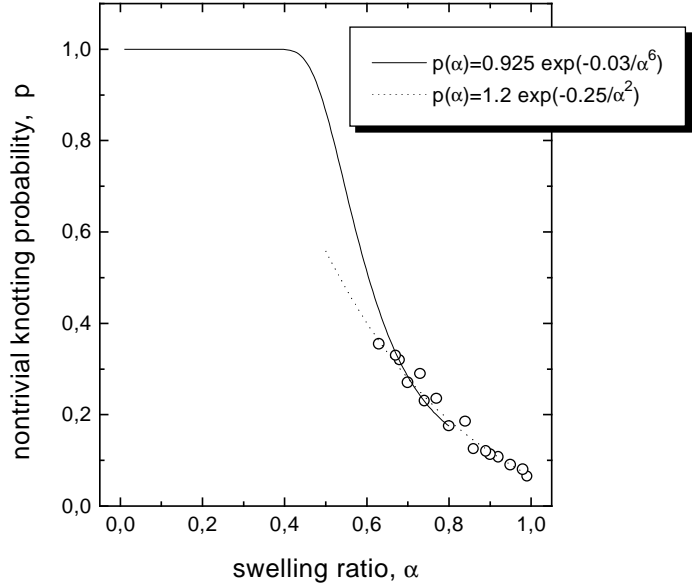,width=10cm}}
\caption{Dependence of non-trivial knot formation probability, $p$ on 
swelling parameter, $\alpha$, in globular state.}
\label{4:fig:swell}
\end{figure}

-- The knot formation probability $p$ is investigated as a  function
of swelling ratio $\alpha\; (\alpha<1)$ where $\disp \alpha =
\sqrt{\left<R_g^2\right>/\left<R_{g,0}^2\right>},\; \left<R_g^{2}\right>$
is the mean-square gyration radius of the closed chain and
$\left<R_{g,0}^2\right>=\frac{1}{12}Na^2$ is the same for unperturbed
$(\alpha=1)$ chain---see fig.\ref{4:fig:swell}, where points correspond to 
the data of Ref.\cite{volog}; dashed line gives approximation in weak 
compression regime and solid line---the approximation based on the
concept of crumpled globule. 

It has been shown that this probability 
decreases sharply when a coil contracts from
swollen state with $\alpha>1$ to the Gaussian one with $\alpha=1$
~\cite{33} and especially when it collapses to the globular state
\cite{volog,4:mish}.

-- It has been established that in region $\alpha>1$ the topological
constraints are screened by volume interactions almost completely \cite{33}.

-- It has been shown that two unentangled chains (of the same length) even
without volume interactions in the coil state repulse each other as
impenetrable spheres with radius of order $\sqrt{\left<R_{g,0}^2\right>}$
~\cite{volog,4:lebret}.

Return to fig.\ref{4:fig:swell}, where the knot formation probability $p$
is plotted as a function of swelling ratio, $\alpha$, in the globular
region $(\alpha<1)$. It can be seen that in compression region, especially
for $\alpha<0.6$ data of numerical experiment are absent. It is difficult
to discriminate between different knots in strongly compressed regime
because it is necessary to calculate Alexander polynomial for each
generated closed contour. It takes of order $O(l^{3})$ operations ($l$ is
the number of self-interactions in the projection). This value becomes as
larger as denser the system.

Let us present the theoretical estimations of the non-trivial knot
formation probability $p(\alpha)$ in dense globular state $(\alpha<0.6)$
based on the CG-concept. The trivial knot formation probability under
random linear chain closure, can be defined by the relation:
\be \label{4.6}
q(\alpha ) = \frac{Z(\alpha)}{Z_{0}(\alpha)}, \qquad
q(\alpha)= 1-p(\alpha)
\ee
where $Z(\alpha)$ is the partition function of unknotted closed chain with
volume interactions for fixed value of swelling parameter, $\alpha$, and
$Z_{0}(\alpha)$ is that of "shadow" chain without topological
constraints but with the same volume interactions. Both partition
functions can be estimated within the framework of the mean field theory.
To do so, let us write down the classic Flory-type representation for the
free energy of the chain with given $\alpha$ (in equations below we suppose
for the temperature $T\equiv 1$):
\be \label{4.7}
F(\alpha) = -\ln Z(\alpha) = F_{\rm int}(\alpha) + F_{\rm el}(\alpha)
\ee
where
\be
F_{\rm el}(\alpha)=-S(\alpha)
\ee
Here the contributions $F_{\rm int}(\alpha)$ from the volume interactions
to the free energies of unknotted and shadow chain of the same density
(i.e. of the same $\alpha$) are equivalent. Therefore, the only difference
concerns the elastic part of free energy, $F_{\rm el}$, or, in other words,
the conformational entropy. Thus, the equation (\ref{4.6}) can be
represented in the form:
\be \label{4.8}
q(\alpha)=\exp \Bigl(-F(\alpha)-F_{0}(\alpha)\Bigr) = \exp
\Bigl(S(\alpha)-S_{0}(\alpha) \Bigr)
\ee
According to Fixmann's calculations \cite{64} the entropy of phantom chain
$S_{0}(\alpha)$ ($S_{0}(\alpha)=\ln Z_{0}(\alpha)$) in region $\alpha <1$
can be written in the following form:
\be \label{4.9}
S_{0}(\alpha) \simeq - \alpha^{-2}
\ee

In the weak compression region $0.6<\alpha\le 1$ the probability of
nontrivial knotting, $p(\alpha)$, can be estimated from the expression of
the phantom ring entropy (Eq.(\ref{4.9})). The best fit of numerical data
\cite{volog} gives us
\be \label{4:weak}
p(\alpha)=1-A_1\exp\left(-B_1\alpha^{-2}\right) \qquad (0.6<\alpha \le 1)
\ee
where $A_1$ and $B_1$ are the numerical constant.

The nontrivial part of our problem is reduced to the estimation of the
entropy of strongly contracted closed unknotted ring $(\alpha\ll 1)$. Using
Eqs.(\ref{4.16}) and (\ref{4.17}) and the definition of $\alpha$ we find
\be \label{4.11}
S(\alpha)\simeq -\frac{1}{N_{e}}\alpha^{-6}
\ee
In the region of our interest $(\alpha<0.6)$ the $\alpha^{-2}$-term
can be neglected in comparison with $\alpha^{-6}$. Therefore, we the final
probability estimate has the form:

\be \label{4.12}
p(\alpha)=1-A_2\exp\left(-\frac{1}{N_{e}}\alpha^{-6}\right)
\qquad (\alpha<0.6)
\ee
where $A_2$ and $N_e$ are the numerical constants (their values are given
below).

The probabilities of the nontrivial knot formation, $p(\alpha)$, in weak
and strong compression regions are shown in fig.\ref{4:fig:swell} by the
dotted and solid lines respectively. The values of the constants are:
$A_1=1.2,\; B_1=0.25,\; A_2=0.925,\; N_e=34$; they are chosen by comparing
Eqs.(\ref{4:weak}) and (\ref{4.12}) with numerical data of
Ref.\cite{volog}.

\subsubsection{Quasi-Knot Concept in Collapsed Phase of Unknotted
Polymers}

Speculations about the crumpled structure of strongly contracted closed
polymer chains in the trivial topological state could be partially
confirmed by the results of Chapters 1 and 2. The crucial question is: {\it
why the crumples remain segregated in a weakly knotted topological state on
all scales in course of chain fluctuations}. To clarify the point we begin
by defining the topological state of a crumple, i.e. the unclosed part of
the chain. Of course, mathematically strict definition of a knot can be
formulated for closed (or infinite) contours exclusively. However the
everyday experience tells us that even unclosed rope can be knotted. Thus,
it seems attractive to construct a non-rigorous notion of a {\it quasi-knot}
for description of long linear chains with free ends.

Such ideas were expressed first in 1973 by I.M. Lifshits and A.Yu. Grosberg
\cite{53} for the globular state of the chain. The main conjecture was
rather simple: in the globular state the distance between the ends of the
chain is of order $R\sim aN^{1/3}$, being much smaller than the chain
contour length $L\sim Na$. Therefore, the topological state of closed loop,
consisting of the chain backbone and the straight end-to-end segment, might
roughly characterize the topological state of the chain on the whole. The
composite loop should be regarded as a quasi-knot of the linear chain.

The topological state of a quasi-knot can be characterized by the knot
complexity, $\eta$, introduced in Section 3 (see Eq.(\ref{1:complex})). It
should be noted that the quasi-knot concept failed for Gaussian chains where
the large space fluctuations of the end-to-end distance lead to the
indefiniteness of the quasi-topological state.

Our model of crumpled globule can be reformulated now in terms of
quasi-knots. Consider the ensemble of all closed loops of length $L$
generated with the right measure in the globular phase. Let us extract from
this ensemble the loops with $\eta(L)=0$ and find the mean quasi-knot
complexity, $\left<\eta(l)\right>$, of an arbitrary subpart of length $l$
($l/L=h=\mbox{const};\; 0<h<1$) of the given loop. In the globular state
the probability $\pi({\bf r})$ to find the end of the chain of length $L$
in some point ${\bf r}$ inside the globule of volume $R^3$ is of order
$\disp \pi({\bf r})\sim \frac{1}{R^3}$ being independent on ${\bf r}$ (this
relation is valid when $La \gg R^2$). So, for the globular phase we could
roughly suppose that the loops in the ensemble are generated with the
uniform distribution. Thus our system satisfies {\bf the "Brownian bridge"
condition} and according to conjecture of the Section 3 (Eq.(\ref{2:con2})
we can apply the following scale-invariant estimate for the averaged
quasi-knot complexity $\sqrt{\left<\eta^2(l)\right>}$
\be \label{4:quasi}
\sqrt{\left<\eta^2(l)\right>}\sim l^{1/2}=h^{1/2} L^{1/2}
\ee
This value should be compared to averaged complexity
$\sqrt{\left<\eta^2(l)\right>}$ of the part of the same length $l$ in the
equilibrium globule created by an open chain of length $L$, i.e. {\bf
without the Brownian bridge condition}
\be \label{4:quasi1}
\sqrt{\left<\eta^2(l)\right>}\sim l= h L
\ee

Comparing Eqs.(\ref{4:quasi}) and (\ref{4:quasi1}) we conclude that any
part of an unknotted chain in the globular state is far less knotted than
the same part of an open chain in the equilibrium globule, which supports
our mean-field consideration presented above. Let us stress that our
statement is thermodynamically reliable and is independent of kinetics of
crumpled globule formation.


\section {Some "tight" problems of the probability theory and
statistical physics}

Usually, in the conclusion it is accepted to overview the main results and
imperceptibly prepare the reader to an idea how important the work itself
is... We would not like go not by a usual way and to make a formal
conclusion, because the summary of received results together with
brief exposition of ideas and methods were indicated in the
introduction and some incompleteness of account could only stimulate the
fantasy of the reader.

On the contrary, we will try to pay attention to some hidden difficulties,
which we permanently met on our way, as well as to formulate possible, yet
unsolved problems, logically following from our consideration. Thus, we
shall schematically designate borders of given research and shall allow the
reader most to decide, whether a given subject deserves of further
attention or not.

\subsection{Remarks and comments to the section 2}

1. The derivation of Eqs.(\ref{1:28})--(\ref{1:29}) assumed the passage
from model with short--range interactions to the mean--field--theory,
in which all spins are supposed to interact with each others. From the
topological point of view such approximation is unphysical and requires the
additional verification. We believe, that the considered model could be
investigated with help of conformal theories and renormalisation group
technique in the case of "weak disorder", i.e. when exists the strong
asymmetry in choice of vertex crossings on a lattice.

2. As it was shown above, utilization of Jones topological invariant
with necessity results in the study of thermodynamic properties of a Potts
model. In the work \cite{2:kauf_sal} was mentioned, that Alexander
polynomials are naturally connected to a partition function of a free
fermion model and, hence, to an Ising model. Probably, the use of similar
functional representation of Alexander polynomials in the frameworks of our
disordered model would result in more simple equations, concerned with
statistical properties of the Ising spin glasses.

3. All results received in this work are sticked to a model, which is
effectively two--dimensional, since we are interested in statistical
properties of a planar projection of knot in which all space degrees of
freedom are thrown away and the topological disorder is kept only.
Thus, physically, the model corresponds to the situation of a globular
polymer chain located in a narrow two--dimensional slit. In connection with
that the following question is of significant interest: how the space
fluctuations of a trajectory in a three--dimensional space modify our
consideration and, in particular, the answer (\ref{1:24})?

\subsection{Remarks and comments to sections 3 and 4}

1. The investigation of topological properties of trajectories on
multiconnected manifolds (on planes with sets removed points) from
the point of view of the conformal field theory assumes a construction
of topological invariants on the basis of monodromy properties of
correlation functions of appropriate conformal theories. In connection
with that there is a question concerning the possibility of construction of
conformal theory with the monodromies of the locally--free group considered
in work.

2. Without any doubts the question about the relation between topological
invariants design and spectral properties of dynamic systems on hyperbolic
manifolds is of extreme importance. The nature of mentioned connection
consists in prospective dependence between knot invariants (in the simplest
case, Alexander polynomials), recorded in the terms of a trace of
products of elements of some hyperbolic group (see expression
(\ref{2:alex})) and trace formulae for some dynamic system on the
same group.

3. Comparing distribution function of primitive paths $\mu$
(\ref{2:3pseud}) with the distribution function of a knot complexity of
$\eta$ (\ref{1:complex})), we can conclude that both these invariants have
the same physical sense: a random walk in a covering space, constructed for
lattice of obstacles, is equivalent from the topological point of view to a
random walk on a Cayley tree. Thus, the knot complexity is proportional to
a length of the primitive (irreducible) word, written in terms of group
generators, i.e. is proportional to a geodesic length on some surface of
constant negative curvature. We believe, that the detailed study of this
interrelation will appear rather useful for utilization of algebraic
invariants in the problems concerning statistics of ensembles of
fluctuating molecules with a fixed topological state of each separate
polymer chain.

3. Questions, considered in Section 3 admit an interpretation in spirit
of spin glass theories, discussed at length of the Section 2. Let us
assume, that there is a closed trajectory of length $L$, which we randomly
drop on a plane with regular set of removed points. Let one point of a
trajectory is fixed. The following question appears: what is a probability
to find a random trajectory in a given topological state with respect to
the set of removed points? The topological state of a trajectory is a
typical example of the quenched disorder. According to the general concept,
in order to find an appropriate distribution function (statistical sum), it
is necessary to average the moments of topological invariant over a
Gaussian distribution (i.e. with the measure of trajectories on a plane).
The same assumptions are permitted us to assume, that the function
$g(r,\psi)$ (Jacobian of conformal transformation)---see
Fig.\ref{3:fig:poincare} has a sense of an ultrametric "potential", in
which the random walk takes place and where each valley corresponds to some
given topological state of the path. The closer $r$ is to 1, the higher are
the barriers between neighboring valleys. Thus, all reasonably long ($La\gg
c^2$) random trajectories in such potential will become "localized" in some
strongly entangled state, in the sense that the probability of spontaneous
disentanglement of a trajectory of length $La$ is of order of $\disp
\exp\left(-{\rm const}\frac{La}{c^2}\right)$. Probably this analogy could
be useful in a usual theory of spin glasses because of the presence
of explicit expression for the ultrametric Parisi phase space
(\cite{mezard}) in terms of a double-periodic analytic functions.

\subsection{Remarks and comments to section 5}

1. We would like to express the conjecture (see also \cite{4:grnech})
concerning the possibility of reformulation of some topological problems
for strongly collapsed chains (see Section 5.4) in terms of integration
over the set of trajectories with fixed fractal dimension but without any
topological constraints.

We have argued that in ensemble of strongly contracted
unknotted chains {\rm (}paths{\rm )} most of them have the fractal
dimension $D_f=3${\rm ;}

We believe that almost all paths in the ensemble of lines with fractal
dimension $D_f=3$ are topologically isomorphic to simple enough {\rm (}i.e.
close to the trivial {\rm )} knot.

Let us  remind, that the problem of the calculation of the partition
function for closed polymer chain with topological constraints can be
written as an integral over the set $\Omega$ of closed paths with fixed
value of topological invariant (see Chapter 1):
\be \label{5.1}
Z = \int\limits_{\Omega} D_{w}\{r\} e^{-H} = \int \ldots \int  D_{w}\{r\}
e^{-H} \delta [I-I_0],
\ee
where $D_{w}\{r\}$ means integration with the usual Wiener measure and
$\delta [I-I_{0}]$ cuts the paths with fixed value of topological invariant
($I_0$ corresponding to the trivial knots).

If our conjecture is true, then the integration over $\Omega$ in
Eq.(\ref{5.1}) for the chains in the globular phase (i.e. when $La\gg R^2$)
can be replaced by the integration over all paths without any topological
constraints, but with special new measure, $D_f\{r\}$:
\be \label{5.2}
Z = \int \ldots \int D_f\{r\} e^{-H}
\ee
The usual Wiener measure $D_{w}\{r\}$ is concentrated on trajectories with
the fractal dimension $D_f=2$. Instead of that, the measure $D_f\{r\}$ with
the fractal dimension $D_f=3$ for description of statistics of unknotted
rings should be used.

2. We believe, that the distribution of knot complexity found for some
model systems can serve as a starting point in construction of a
mean--field Ginsburg--Landau--type theory of fluctuating polymer chains
with a fixed topology. From a physical point of view it seems to be
important to rise the mean--field theory which takes into account
the influence of topological restrictions on phase transitions in bunches
of entangled directed polymers.

3. Let us note, that despite a number of experimental works, indirectly
testifying for the existence of a fractal globule (see section 5 and
references), the direct observation of this structure in real experiments
is connected to significant technical difficulties and is so far not
carried out. We believe, that the organization of an experiment on
determination of a microstructure of an entangled ring molecule in a
globular phase could introduce final clarity to a question on a crumpled
globule existence.

\end{document}